\documentclass[acmtog,table,authorversion]{acmart}

\usepackage{amsmath}
\usepackage{amsfonts}
\usepackage{graphicx}
\usepackage{tabularx}
\graphicspath{{./}{datadump/image_grid/}}

\usepackage{float}
\usepackage{listings}
\usepackage{verbatim}

\usepackage{multirow}
\usepackage{microtype}

\usepackage[capitalize]{cleveref}
\crefname{section}{Sec.}{Secs.}
\Crefname{section}{Sec.}{Secs.}
\crefname{paragraph}{Sec.}{Secs.}
\Crefname{paragraph}{Sec.}{Secs.}
\crefname{table}{Tab.}{Tabs.}
\Crefname{table}{Tab.}{Tabs.}
\crefname{figure}{Fig.}{Figs.}
\Crefname{figure}{Fig.}{Figs.}
\crefname{equation}{Eq.}{Eqs.}
\Crefname{equation}{Eq.}{Eqs.}
\acmSubmissionID{2062}
\definecolor{darkred}{rgb}{0.8, 0.0, 0.0}
\definecolor{darkgreen}{rgb}{0.0, 0.6, 0.0}

\newcommand{\notyet}[1]{\textcolor{orange}{[Not implemented in code yet]}}
\newcommand{\codematch}[1]{}
\newcommand{\mightchange}[1]{}

\newcommand{\added}[1]{#1}
\newcommand{\moved}[1]{#1}



\title{Editable Physically-based Reflections in Raytraced Gaussian Radiance Fields}

\author{Yohan Poirier-Ginter}
\orcid{0009-0001-5806-5136}
\affiliation{%
  \institution{Université Laval}
  \city{Quebec}
  \country{Canada}
}
\affiliation{%
  \institution{Inria, Université Côte d'Azur}
  \city{Nice}
  \country{France}
}
\email{yohan.poirier-ginter.1@ulaval.ca}

\author{Jeffrey Hu}
\orcid{0009-0003-4400-0862}
\affiliation{%
  \institution{Inria, Université Côte d'Azur}
  \city{Nice}
  \country{France}
}   
\email{hujh14@gmail.com}

\author{Jean-François Lalonde}
\orcid{0000-0002-6583-2364}
\affiliation{
  \institution{Université Laval}
  \city{Quebec}
  \country{Canada}
}
\email{jean-francois.lalonde@gel.ulaval.ca}

\author{George Drettakis}
\orcid{0000-0002-9254-4819}
\affiliation{%
  \institution{Inria, Université Côte d'Azur}
  \city{Nice}
  \country{France}
}
\email{george.drettakis@inria.fr}

\copyrightyear{2025}
\acmYear{2025}
\setcopyright{cc}
\setcctype{by}
\acmConference[SA Conference Papers '25]{SIGGRAPH Asia 2025 Conference Papers}{December 15--18, 2025}{Hong Kong, Hong Kong}
\acmBooktitle{SIGGRAPH Asia 2025 Conference Papers (SA Conference Papers '25), December 15--18, 2025, Hong Kong, Hong Kong}\acmDOI{10.1145/3757377.3763971}
\acmISBN{979-8-4007-2137-3/2025/12}

\begin{document}

\makeatletter
\def\@ACM@copyright@check@cc{}
\makeatother

\begin{teaserfigure}
  \centering
  \includegraphics[width=\textwidth]{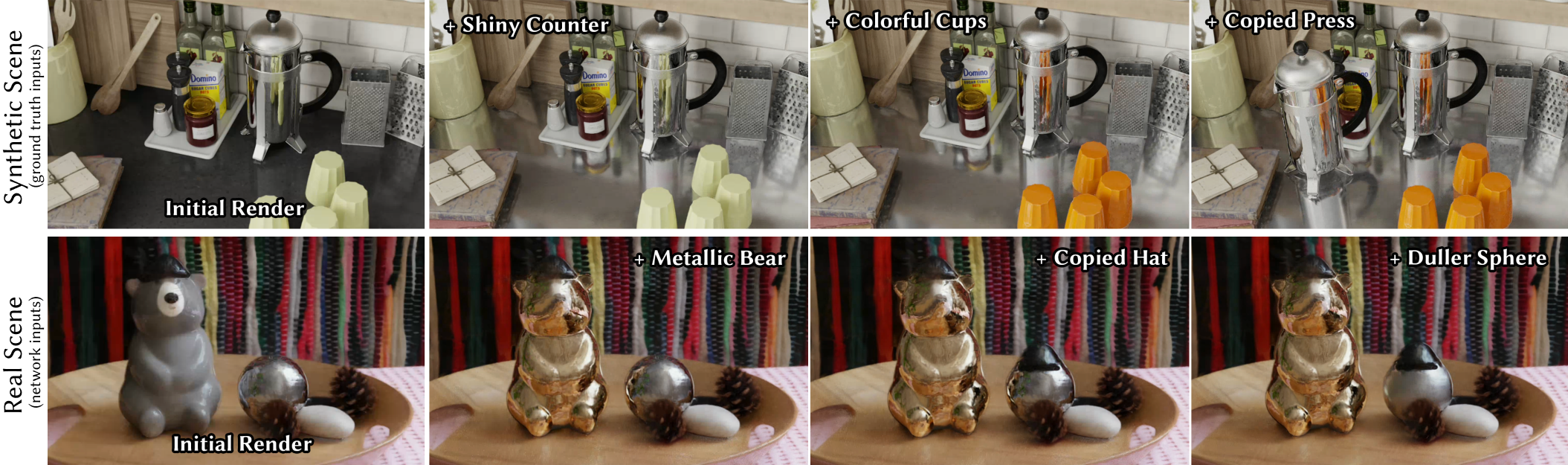}
  \caption{Our Gaussian-based radiance-field method allows interactive editing of \added{path traced} reflections, with consistent updates. We develop our proof-of-concept method in synthetic scenes with known ground truth inputs and show its extension to a real scene with network predicted inputs.}
  \label{fig:teaser}
\end{teaserfigure}

\begin{abstract}
Radiance fields such as 3D Gaussian Splatting allow real-time rendering of scenes captured from photos. They also reconstruct \added{most specular} reflections with high visual quality, but typically model them with ``fake'' reflected geometry, using primitives behind the reflector. 
Our goal is to correctly reconstruct the reflector and the reflected objects \added{such as to make specular reflections editable}; we present a \added{proof of concept} which exploits promising learning-based methods to extract diffuse and \added{specular} \added{buffers} from photos, as well as \added{geometry and BRDF \added{buffers}.} Our method builds on three key components. First, by using diffuse/\added{specular} \added{buffers} of \added{input training views}, we optimize a diffuse version of the scene and use \added{path tracing} to efficiently generate physically-based \added{specular} reflections. Second, we present a specialized training method that allows this process to converge. Finally, we present a fast \added{ray tracing} algorithm for 3D Gaussian primitives \added{that enables} efficient multi-bounce reflections. Our method reconstructs reflectors and reflected objects---including those not seen in the input images---in a \emph{unique} scene representation. Our solution allows real-time, consistent editing of captured scenes with \added{specular} reflections, including multi-bounce effects, changing roughness etc. 
We \added{mainly} show results using \added{ground truth} \added{buffers} from synthetic scenes, and \added{also preliminary} results in real scenes with currently imperfect learning-based \added{buffers}. Code and data are available at: \url{https://repo-sam.inria.fr/nerphys/editable-gaussian-reflections/}.
\end{abstract}

\maketitle

\begin{figure*}[t]
    \centering
    \includegraphics[width=1.0\linewidth]{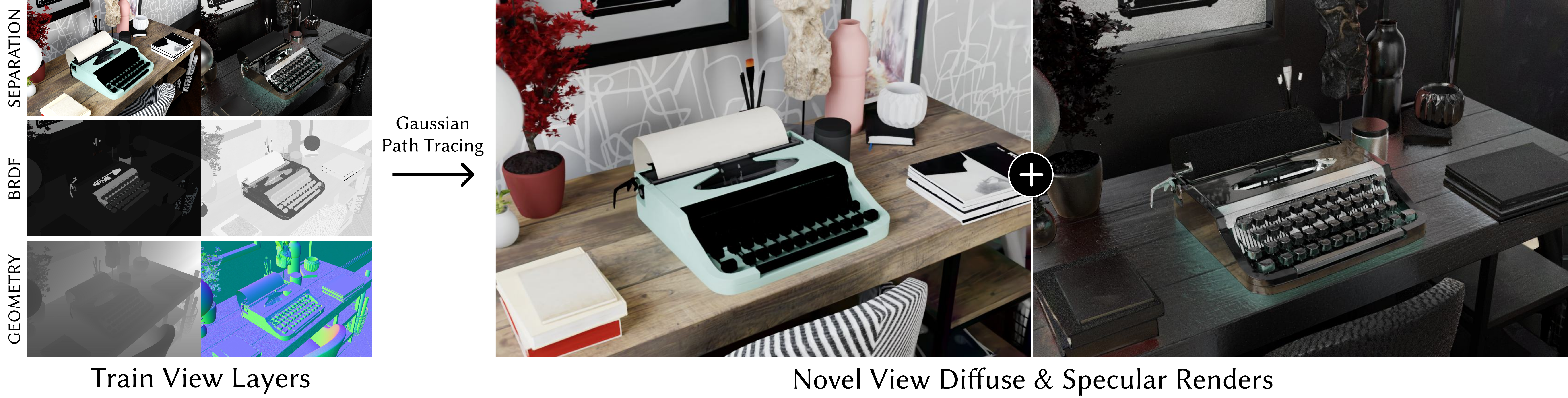}
    \vspace{-10pt}
    \caption{Our method takes as input several \added{buffers} for every training view (left), \added{and reconstructs a unique gaussian-based scene where reflections in novel views are computed with cached diffuse (middle) and physics-based path tracing (right).
    This figure shows ground truth input \added{buffers} from a synthetic scene, from top to bottom: separated targets (diffuse and non-diffuse), BRDF parameters (base reflectance and roughness), and geometry (depth and normals).}}
    \label{fig:method-overview}
\end{figure*}

\section{Introduction}
\label{sec:intro}

%
Neural Radiance Fields (NeRFs) \cite{mildenhall2020nerf} and 3D Gaussian Splatting (3DGS) \cite{kerbl20233d} allow high-quality novel-view synthesis with only posed photographs as input. One of their major strengths is that they render view-dependent effects such as \added{low- to medium-frequency specular} reflections with high visual quality. However, the representation of reflections is entangled \added{with the geometry}: they are baked into a directional radiance component, and often are actually ``fake mirror geometry'' situated behind the reflective surfaces~\cite{zhang2024refgaussian, meng2024mirror}. This precludes consistent editing of scenes with reflections, i.e., where reflections will be correctly updated when changing geometry or materials. 
We propose the first method that allows interactive reflection-consistent editing of radiance fields (\cref{fig:teaser}) by using distinct processes to reconstruct each of the diffuse and reflective components of the scene. Our method also allows reconstruction of reflected objects unseen in the input images.

%

In the original versions of NeRF and 3DGS, a directional radiance representation is modeled by a Multi-Layer Perceptron (MLP) and Spherical Harmonics (SH) respectively. However, these low-frequency representations cannot represent reflections accurately, and thus the optimization favors the creation of ``fake mirror geometry'' (see video).
Several improvements exist for better 
reflection in NeRFs, e.g., by changing parameterizations or encodings~\cite{verbin2022refnerf,ma2024specnerf}. Similarly, several methods  attempt to improve reflections in 3DGS, often for distant lighting~\cite{jiang2024gaussianshader,ye20243d}. These methods still suffer from the entangled reflection representation. Recently, NeRFcasting~\cite{verbin2024nerf} and EnvGS~\cite{xie2025envgs} use ray tracing but require a separate scene representation for reflections to allow stable optimization. 
Representing reflectors and reflected objects as separate scenes prevents consistent editing of the radiance field.


To address these problems, our key intuition is to have a unique scene with two distinct---but concurrent---optimizations: one for the diffuse,  and one for the \added{specular} components of the scene, supervised on diffuse and \added{specular} input \added{buffers} respectively. \added{In this paper, we use the term \emph{specular} to mean all non-diffuse reflections, from rough glossy to pure mirror.}
The diffuse \added{buffer} captures diffuse global illumination, including shadows and interreflections, while the \added{specular} \added{buffer} captures \added{mirror and glossy} reflections.
We reconstruct the diffuse component of the scene using ray traced Gaussian primitives and compute \added{specular} reflections with \added{path tracing} \added{while treating the diffuse component similarly to an irradiance cache}~\cite{irradiance-caching}.
Our reflections support surfaces of different roughness, and multi-bounce effects computed in a physically-based manner, i.e., exclusively with \added{path} tracing. They also allow interactive scene editing with physically-correct reflections. Note that transparency is more complicated, since \added{buffers} would have to deal with two rays---reflected and refracted; we thus leave it as future work.

\added{First, we present a \emph{proof-of-concept} of our method
on synthetic scenes, where rendered images are used in the place of photos, and
ground truth values are used for all buffers (diffuse, specular, materials etc.).
This solution} \moved{shows we can duplicate a reflective object in a scene, and all \emph{real, captured} objects will be correctly reflected in the duplicated object interactively, even with multiple bounces (see \cref{fig:teaser}).} 
\added{Second, we present an editable inverse rendering pipeline on synthetic
\emph{and real} images using buffers predicted by neural networks, albeit with lower quality for novel-view synthesis. In particular, }
\moved{recent intrinsic image decomposition methods can extract the required buffers from photographs~\cite{roberts2021hypersim,zeng2024rgbx}, by training on 
synthetic data where light transport is separated into diffuse and reflective (or non-diffuse) components at the first bounce.}
\added{That being said,} while these methods are improving at outstanding rates, the current quality of predicted \added{buffers} from existing methods (e.g., \cite{ke2024repurposing,zeng2024rgbx,diffusion-render}) is only partially sufficient for our goals.
\added{We thus focus mainly on our proof-of-concept} using synthetic data with perfect \added{buffers} to develop our method, and present \added{preliminary} results for scenes processed with the best possible existing learning-based methods \added{at the time of writing}.


\begin{figure*}[t]
    \centering
    \includegraphics[width=1.0\linewidth]{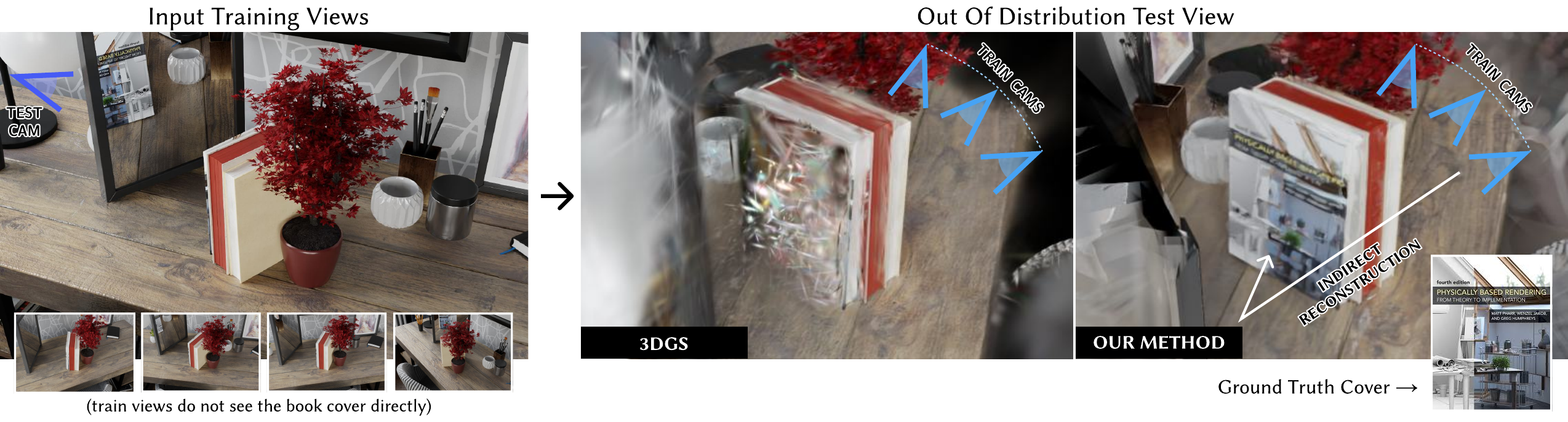}
    \vspace{-15pt}
    \caption{We showcase reconstruction of objects viewed indirectly by recovering the cover of a book only visible indirectly through a mirror in the training views (left). From these views alone, our method accurately recovers the cover (right), unlike standard methods like 3DGS (middle). Note that in these results, our method uses ground truth input buffers and additional Gaussians were initialized in a bounding box around the book (see supp.). Book cover by Matt Pharr, Wenzel Jakob, and Greg Humphreys, licensed under CC BY-NC-ND 4.0.}
    \label{fig:method-overview-2}
\end{figure*}

\added{In short, our method leverages intrinsic decomposition to separately reconstruct the diffuse and reflected components of a scene (see \cref{fig:method-overview}) to make editable specular reflections possible (see video).} \moved{We optimize a \emph{unique} representation of the scene, i.e., both the reflectors and reflected objects are represented with the same Gaussian primitives, by carefully designing separate losses and a training schedule.} \added{This unique representation makes it possible to reconstruct objects only viewed indirectly through specular reflections, as shown in~\cref{fig:method-overview-2}}. Finally, we present an efficient hardware-accelerated Gaussian primitive ray tracer, which uses efficient spatial data structures and other techniques to optimize Gaussian primitive traversal.

In summary our contributions are:
\begin{itemize}
\item A reconstruction method for radiance fields with distinct optimization for diffuse and \added{specular} components, using \added{path tracing} for the latter.
\item An efficient and accurate training method that reconstructs the diffuse and the \added{specular} components of the scene in a single representation. 
\item An efficient ray tracer for Gaussian primitives, that is fast enough to enable treatment of multiple bounces with minimal computational overhead. 
\end{itemize}
\added{
Given our advantage of additional input buffers, and notably diffuse/specular image decomposition, we achieve state-of-the-art (SOTA) diffuse/specular disentanglement for novel-view synthesis, even though overall visual quality is lower than previous methods.
Our disentanglement allows real-time reflection and material editing in both the proof-of-concept case with ground truth buffers and for real scenes. 
For the latter we present an experimental neural network that provides the \added{buffers} we require. While quality measured on our synthetic scenes given SOTA networks is low, it is sufficient for realistic, interactive scene manipulation in certain real scenes, showing the potential of the approach with future, better quality learning-based \added{buffer} predictions. 
}
Code and data are available at: \url{https://repo-sam.inria.fr/nerphys/editable-gaussian-reflections/}.

\section{Related work}
\label{sec:relwork}

Our goal is to provide physically-based \added{specular} reflections for radiance fields that allow consistent real-time editing and reconstruction of the unseen reflected objects. We review most closely related methods. 

\paragraph{Reflections in NeRF} Neural radiance fields, or NeRF~\cite{mildenhall2020nerf}, learn an implicit scene representation, parametrized with a multi-layer perceptron (MLP), from multi-view posed images. NeRF 
typically represents \added{specular} reflections with a combination of view-dependent color in the MLP and ``fake geometry'' representing the reflected objects \emph{behind} the reflector. 
Ref-NeRF~\cite{verbin2022refnerf} improves on this by using the reflected view direction, NeRFReN~\cite{guo2022nerfren} models the transmitted and reflected scene separately, similarly for MS-NeRF~\cite{yin2025ms}, and Mirror-NeRF~\cite{zeng2023mirror} assumes planar mirrors. Others~\cite{liang2023envidr,liu2023nero,wang2024unisdf} exploit signed distance fields~\cite{wang2021neus,li2023neuralangelo} to obtain more accurate scene geometries, or modify the directional encoding~\cite{ma2024specnerf} to better capture the spatially-varying nature of near-field lighting.
NeRF-casting~\cite{verbin2024nerf} integrates reflection features along reflection directions. However, the need for multiple MLP queries for each ray and the creation of a separate version of the scene makes it both impractical for real-time rendering and reflection-consistent scene editing. 




\paragraph{Reflections in 3D Gaussian Splatting} 3DGS~\cite{kerbl20233d} introduces a primitive-based representation for radiance fields, and 
uses efficient rasterization to render Gaussian primitives, achieving fast training and real-time rendering. 
3DGS-based approaches model reflections with a combination of view-dependent spherical harmonics (SH) and the creation of ``fake geometry'' representing reflected objects behind the reflected surface.
More recent works~\cite{jiang2024gaussianshader,ye20243d}, 
enhance reflection modeling by incorporating additional environment maps. However, these methods focus only on distant lighting.
3iGS~\cite{tang20243igs} integrates an illumination field through tensorial factorization (as in \cite{chen2022tensorf,jin2023tensoir}) and renders the final reflections with a neural renderer, but is constrained to bounded scenes. Ref-GS~\cite{zhang2025ref} models near-field reflection using a tensorial factorization within a 2DGS framework~\cite{huang20242d}, and models far-field illumination with a spherical feature grid. 
None of these approaches overcome the basic limitation of representing \added{specular} reflections with ``fake geometry'', making them unsuitable for consistent editing. Some approaches also assume knowledge of planar mirrors~\cite{meng2024mirror,liu2024mirrorgaussian}. \added{Most approaches are efficient for objects, but were not tested on full scenes~\cite{glossy-gs}.}  A predecessor of 3DGS~\cite{kopanas2022neural} models reflections with a separate point cloud rendered with an MLP and shows reflection editing, which is not truly consistent.

\paragraph{Ray Tracing with Gaussian Primitives} 3D Gaussian Ray Tracing (3DGRT)~\cite{moenne20243d} renders a 3D Gaussian representation using ray tracing instead of rasterization. They build a bounding volume hierarchy and cast a ray for each pixel using high-performance GPU ray tracing hardware~\cite{optix}. \added{3DGUT also showed that reflection rays are possible in a splatting framework~\cite{3dgut}.} A similar proposal is made in RaySplats~\cite{byrski2025raysplats}, more recently extended in REdiSplats~\cite{byrski2025redisplats} and in RayGauss~\cite{blanc2025raygauss}. After training, 3DGRT shows they can render secondary ray effects such as reflections. Critically, only mesh-Gaussian reflections can be handled by their framework; in contrast, we model Gaussian-Gaussian reflections \emph{during training}. We also improve over 3DGRT with several performance enhancements (see \cref{sec:fast-rt}). Inter-Reflective Gaussian Splatting (IRGS)~\cite{irgs} attempts to address inter-reflection modeling by extending the 2DGS framework with a differentiable \added{ray tracing} approach. EnvGS~\cite{xie2025envgs} trains two versions of the scene: a base scene (with rasterization), and ``environment Gaussians'' (with raytracing) which represents scene elements only visible through reflections. While this approach improves the separation between the diffuse and \added{specular} component, there are still residual reflections in the diffuse \added{buffer} (as we show in \cref{sec:evaluation}) and the lack of a BRDF model makes it unsuitable for editing. In contrast, we handle inter-reflections, and also reconstruct scene elements only visible in \added{specular} reflections, thanks to our  \emph{unique} scene representation. This allows consistent interactive reflection editing.

\paragraph{Inverse Rendering and Relighting} Scene editing could be achieved using full inverse rendering~\cite{marschner1998inverse,nimier2019mitsuba,li2018differentiable}. However, full physically-based inverse rendering is ill-posed~\cite{kouros2024unveiling} and expensive~\cite{li2018differentiable, nimier2019mitsuba,srinivasan2021nerv,zhang2021nerfactor}, and currently cannot handle complex scenes~\added{\cite{gir}}. \added{GS-IR~\cite{gs-ir} leverages 3DGS to accelerate it.} Relighting techniques~\cite{bi2024gs3,gao2024relightable,poirier2024diffusion,olat-gaussians} allow the \emph{lighting} to be edited, and are orthogonal to the kind of edits we show (moving reflective objects and changing materials). Combining them is interesting future work. 


\begin{figure}[t]
    \centering
    \includegraphics[width=1.0\linewidth]{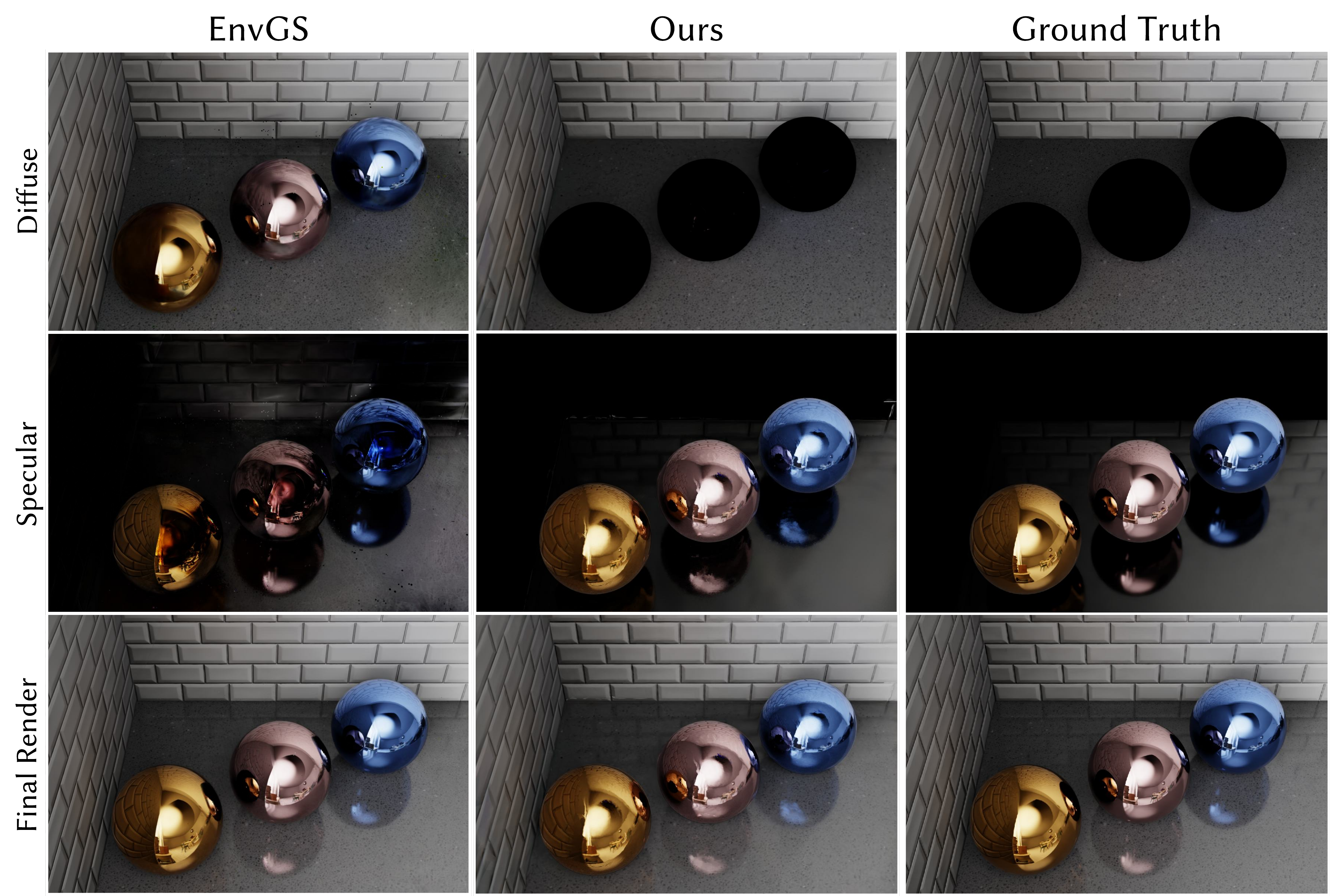}
    \caption[]{Radiance fields, recovered for example with EnvGS~\cite{xie2025envgs} (left), often rely on duplicate geometry to represent \added{specular} reflections; notice the replicas of the environment placed within the chrome spheres (top-left). Our method (middle) represents reflections with physically-based light transport \emph{only}, closely matching ground truth novel views (right). \added{Note that in these results, our method uses ground truth input buffers.} 
    }
    \label{fig:separation}
    \vspace{-0.2cm}
\end{figure}
\section{Method}
\label{sec:method}

At a high level, \added{our method starts with BRDF parameters from input views}, attaches these parameters as well as depth and \added{normals} to diffuse Gaussians using 3DGS-like optimization, and during optimization computes \added{specular} reflections by simulating full light transport of the \added{specular} component. At each intersection, accumulated parameters are used to perform proper ray tracing.


\subsection{Diffuse/\added{Specular} Separation for Physics-Based Reflections}
\label{sec:diff_n_diff}

%

\added{By computing and supervising diffuse and specular reflections separately}, our method makes it much harder for the optimization to ``cheat'' and create ``fake'' geometry to represent reflections.

To achieve this we make a key design choice by having separate supervision
signals: one for the diffuse and one for the \added{specular} component---both are computed with \added{ray tracing}. However, we create a \emph{unique} scene representation for both. 
The Gaussian primitives in this reconstruction will represent all geometry visible in the input views, but also the reflected objects \emph{at coherent positions} in space (see \Cref{fig:reconstruct-env}).

The RGB colors in the input images are the result of (real-world) physical light transport, expressed by the rendering equation:
\begin{equation}
L_o(\boldsymbol{x}, \boldsymbol{\omega}_o) = L_e(\boldsymbol{x}, \boldsymbol{\omega}_o) + \int_{\Omega} L_i(\boldsymbol{x}, \boldsymbol{\omega}_i) f(\boldsymbol{\omega}_i, \boldsymbol{n}, \boldsymbol{\omega}_o)  \cos \theta_i \, d\boldsymbol{\omega}_i \,,
\end{equation}
where $L_o$, $L_e$, $L_i$ are the outgoing, emitted, incoming radiance respectively, $\boldsymbol{\omega}_i$ and $\boldsymbol{\omega}_o$ are the incoming and outgoing directions, $\boldsymbol{n}$ is the surface normal at point $\boldsymbol{x}$,
$\cos \theta_i = \boldsymbol{\omega}_i \cdot \boldsymbol{n}$, and $f$ is the BRDF.
We use the approximation that \added{$f = f_d + f_s$ with $f_d$ and $f_s$ the diffuse and specular components.}
Ignoring the emissive term, our method approximates light transport by the sum of two terms:
\begin{equation}
\begin{split}
L_o(\boldsymbol{x}, \boldsymbol{\omega}_o) &= \int_{\Omega} L_i(\boldsymbol{x}, \boldsymbol{\omega}_i) f_d(\boldsymbol{\omega}_i, \boldsymbol{n})  \cos \theta_i \, d\boldsymbol{\omega}_i + \\
&\int_{\Omega} L_i(\boldsymbol{x}, \boldsymbol{\omega}_i) f_s(\boldsymbol{\omega}_i, \boldsymbol{n}, \boldsymbol{\omega}_o)  \cos \theta_i \, d\boldsymbol{\omega}_i \,.
\label{eq:split}
\end{split}
\end{equation}
Using Heckbert's light path notation~\cite{heckbert1990adaptive}, the first term computes all $\mathcal{L((D|S)*)DE}$ paths, where $\mathcal{
L}$ are the lights, $\mathcal{D}$ the diffuse vertices in a path, $\mathcal{S}$ the \added{specular} vertices and $\mathcal{E}$ is the eye or camera, while the second term computes all $\mathcal{L((D|S)*)SE}$ paths.   
When \added{path tracing} scenes, each term is computed with the same paths, but at the first intersection the first (diffuse) term is evaluated using only $f_d$, and the second (\added{specular}) term using only $f_s$.
\added{
We reconstruct the diffuse component by supervising on the diffuse \added{buffer}, resulting in a ``cached'' version of diffuse lighting $\hat{L}_d$ which we use like an irradiance cache already multiplied with albedo, i.e.,
\begin{equation}
L_o(\boldsymbol{x}, \boldsymbol{\omega}_o) = \hat{L}_d + \int_{\Omega} L_i(\boldsymbol{x}, \boldsymbol{\omega}_i) f_s(\boldsymbol{\omega}_i, \boldsymbol{n}, \boldsymbol{\omega}_o)  \cos \theta_i \, d\boldsymbol{\omega}_i \,.
\label{eq:cached}
\end{equation}
}
We thus avoid a large part of \added{the full} path tracing costs, \added{namely} the first term in \cref{eq:split}. This reconstruction contains ``baked'' shadows and diffuse color bleeding. 
Other quantities (depths, normals, material properties) are also supervised with corresponding \added{buffers}, while reflections (second term in \cref{eq:split}) are computed with path tracing.
Diffuse and \added{specular} components are shown in \cref{fig:method-overview,fig:separation}.

To develop our approach and evaluate it quantitatively, we first use synthetic scenes, rendering RGB images that serve the same purpose as real photos for radiance field reconstruction. We render exact image \added{buffers} per input view for diffuse and \added{specular} component of \cref{eq:split}. We also render \added{buffers} with all \added{depths}, normals, BRDF coefficients per pixel that are attached to Gaussians, and then used to perform proper \added{ray traced} reflections.
 For real images, we use a neural network trained on such \added{path traced} images (e.g., using the Hypersim~\cite{roberts2021hypersim} and InteriorVerse~\cite{zhu2022interiorverse} datasets) to extract diffuse and \added{specular} \added{buffers}, and \added{buffers} for the other parameters (\cref{sec:network}).
 
We use the Cook-Torrance BRDF~\cite{cook1982reflectance} and specifically, a subset of the Disney BRDF formulation~\cite{pbr-disney}. The parameters of this model are the roughness $\rho$ and the base reflectance or $F_0$ which is defined based on the metalness, specular and color parameters~\cite{pbr-disney}.

\paragraph{Reconstructing the Diffuse Component}

We reconstruct the diffuse component using a single diffuse RGB value per Gaussian primitive instead of spherical harmonics. 
This step could use the standard 3DGS splatting pipeline, but we prefer to use our \added{ray tracing} solution throughout to avoid the discrepancy induced by the affine approximation of splatting~\added{\cite{on-the-error-analysis}}, ensuring a single consistent solution. \added{Note that, like 3DGRT, our scenes are incompatible with 3DGS since we use per-pixel sorting to avoid popping artifacts.}

The diffuse component is supervised directly with the diffuse \added{buffer}, and reconstructs a version of the scene with diffuse global illumination and shadows. For the other quantities (normals, roughness and base reflectance), we attach values to each Gaussian primitive, and optimize with supervision from the per-input-view \added{buffers}. We also optimize the expected termination depth~\cite{kerbl20233d} with the depth \added{buffers}.


\paragraph{Ray Traced Physics-Based \added{Specular} Reflections for Radiance Fields.}

\begin{figure}[h]
    \centering
    \includegraphics[width=1.0\linewidth]{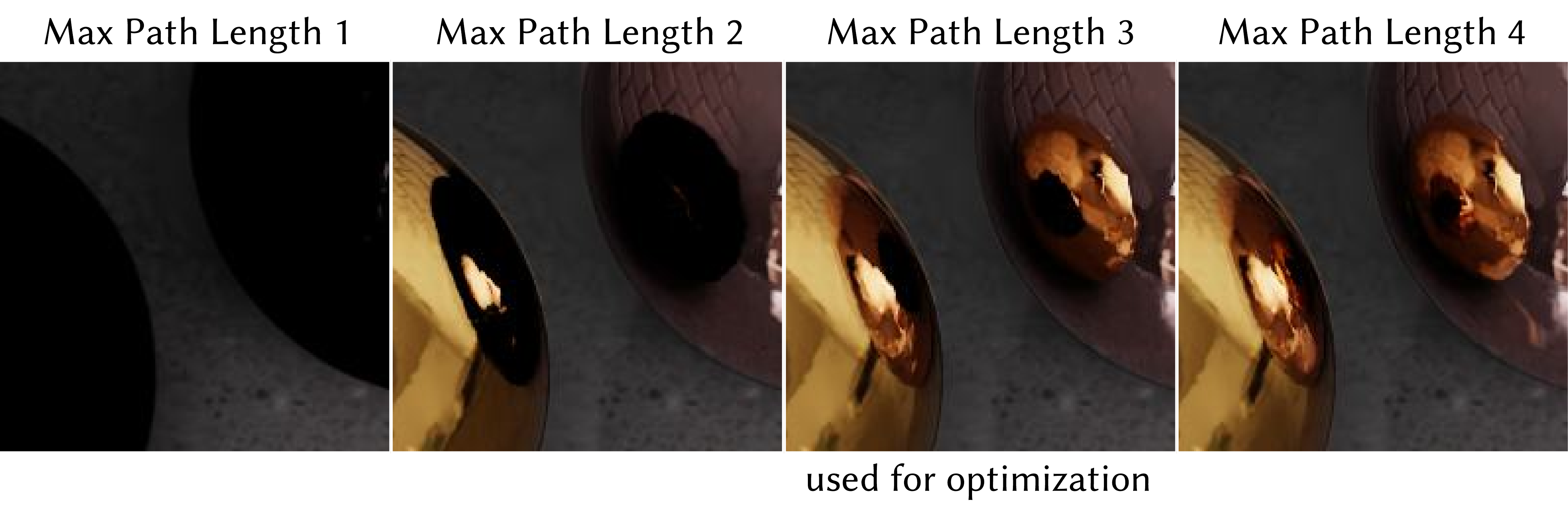}
    \vspace{-20pt}
    \caption{We support multi-bounce reflections during optimization. 
    The number of bounces can be adjusted, trading off performance vs. accuracy.}
    \label{fig:max_path_length}
    \vspace{-0.2cm}
\end{figure}

\begin{figure}[t]
    \centering
    \includegraphics[width=1.0\linewidth]{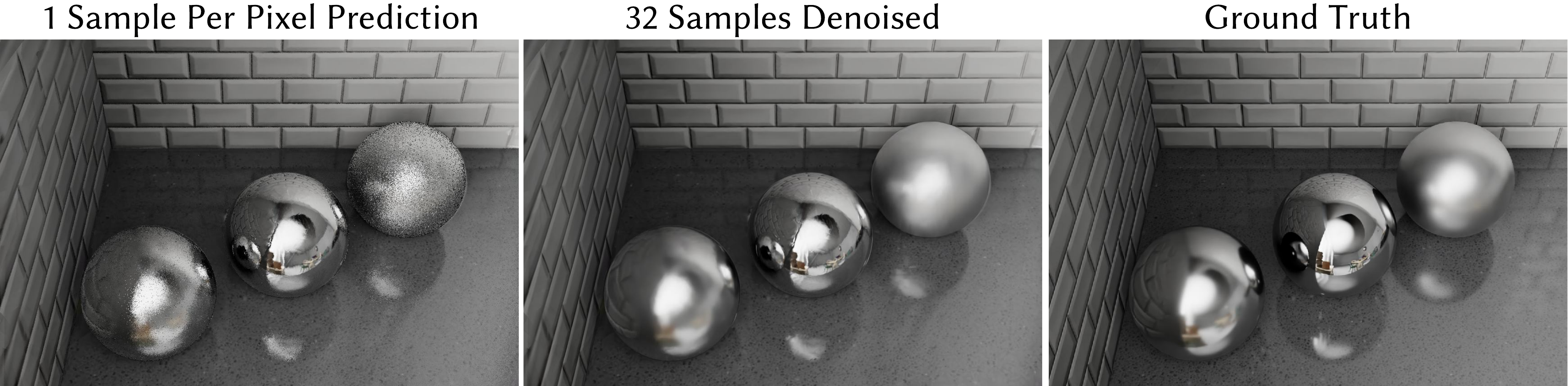}
    \caption{We support physically-based reflections with different levels of roughness in the same scene. At inference, we sample several samples per pixel (column 1) and denoise with the OptiX denoiser (column 2).}
    \label{fig:rough-surfaces}
    \vspace{-.2cm}
\end{figure}

We compute the $\mathcal{L((D|S)*)SE}$ paths with \added{path tracing} of the Gaussian primitives and cached diffuse. 
Each primitive is initialized with a base reflectance value $F_0$ initialized at $0.04$ and with roughness and normal values initialized at $0$.  
Each ray traverses Gaussians while accumulating values with alpha-blending as in 3DGS, for all quantities, including normals, roughness and $F_0$ values.
The accumulated ``normal'' value is then normalized to unit length. 
We compute the intersection at the expected termination depth
and use accumulated normal, roughness and $F_0$ values to importance sample a random reflection ray with the Cook-Torrance BRDF.  Our approach supports multiple ray bounces natively, see \cref{fig:max_path_length}. \added{In practice, we set the max path length to 3, resulting in good performance at the cost of some minor inaccuracies (i.e. the black spots in reflections in~\cref{fig:teaser})}. During these bounces we query the diffuse component representation for color, which is the (diffuse component of) radiance extracted from the input images (i.e., real-world lighting for photos). 
We train at 1 sample per pixel, but for offline rendering use $128$ samples and then apply the OptiX denoiser (see \cref{fig:rough-surfaces}). The \added{specular} component is supervised with the specular \added{buffer} provided as input.

Finally, we adapt dense initialization \cite{kotovenko2025edgs} instead of densification to further accelerate convergence; we found \added{ray tracing} effective at handling the large number of small Gaussians that dense initialization provides. \added{We do not densify further and aim for a number of gaussians comparable with the baselines;} for details refer to \cref{sec:sup-dense}.

\subsection{Optimizing a Unique Scene for Radiance Field Reflections}
\label{sec:method-optimizing}

Optimizing a representation that uses Gaussian primitive \added{ray tracing} is hard: previous approaches~\cite{xie2025envgs,verbin2024nerf} showed that flowing gradients naively through reflections damages geometric reconstruction. This can create a feedback effect where bad geometry skews reflection rays, leading to more damage. These previous approaches side-step this problem by optimizing two separate radiance fields, one for the primary scene and one for reflections (e.g., \cite{xie2025envgs}). However, such a choice precludes scene editing with multi-bounce reflections, and can hinder reconstruction of reflected objects unseen in the input views. We overcome this optimization instability and reconstruct a unique scene. 


Diffuse/\added{specular} separation resolves the ambiguity between texture and reflected objects, letting us maintain a unique scene while keeping training stable. Our training proceeds with two different losses: one for diffuse and one for \added{specular} components. It also recovers normals and BRDF directly through input \added{buffer} supervision and not via inverse rendering: we do \emph{not} differentiate normals and BRDF w.r.t. the RGB images, ensuring that the scene remains well-formed even with gradients flowing through reflections.


We cast a path through each pixel, and use the first segment of the path to
query diffuse global illumination (see \cref{sec:diff_n_diff}),
represented by the diffuse component reconstruction. We denote this quantity $d$ and
we supervise with the value $d^*$ in the diffuse \added{buffer}: 
\begin{equation}
\mathcal{L}_d = \lambda_d \ell(\hat{d}, d^*) \,,
\end{equation}
where $(\hat{\,})$ denotes the accumulated value, $(^*)$ the corresponding ground truth value from the \added{buffers}, and $\ell$ the L1 loss.
We then continue the path using the specular BRDF component $f_s$ and match the contribution of the remaining path segments $c_i,~i=2\ldots K$ to the \added{specular \added{buffer} $s$}:
\begin{equation}
\mathcal{L}_r = \lambda_r \ell(\sum_{i=2}^K \hat{c}_i, s^*) \,,
\end{equation}
where $K$ is the maximum path length. To this we add a loss matching the accumulated values of other attached properties namely depth $\hat{t}$, normal $\hat{n}$, roughness $\hat{\rho}$, and base reflectance $\hat{F_0}$ to their corresponding target \added{buffers}:
\begin{equation}
\mathcal{L}_a = \lambda_t \, \ell(\hat{t}, t^{*}) + \lambda_n \, \ell(\hat{n}, n^{*}) + \lambda_\rho \, \ell(\hat{\rho}, \rho^{*}) + \lambda_{F_0} \, \ell(\hat{F_0}, F_0^{*})
\end{equation}
This loss is only applied to the first path segment since these \added{buffers} are only known for this segment. Our final objective is:
\begin{equation}
\min_{\theta, d} \mathbb{E} \left[ \mathcal{L}_d + \mathcal{L}_r \right] + \min_{\theta,t,n,\rho,F_0} \mathbb{E}[ \mathcal{L}_a ] \,,
\label{eqn:all-losses}
\end{equation}
where $\theta$ contains the position, rotation, scale and alpha parameters for each Gaussian. \added{Note that both subobjectives are optimized at the same time.} We use values $\lambda_d = 5.0$, $\lambda_r = 3.0$, $\lambda_t = 2.5$, $\lambda_n = 2.5$, $\lambda_\rho = 1.0$ and $\lambda_{F_0} = 1.0$. We compute all losses in linear space.

Our training schedule starts with a warmup for 750 iterations with the first ray segment only, which reconstructs a first version of the diffuse component. We then enable all reflections and add $M=75,000$ far-field Gaussians initialized around the scene origin with positions randomly sampled from a normal distribution with $\sigma=4S$ truncated at $3\sigma$; the scene extent $S$ is estimated as in 3DGS by computing the radius of a sphere which bounds the input camera poses~\cite{kerbl20233d}. As optimization progresses, these primitives provide a (coarse) approximate reconstruction of the reflected objects in the environment (\cref{fig:reconstruct-env}).

\subsection{Efficient Ray Tracer for \added{Multi-Bounce} Radiance Field Reflections}
\label{sec:fast-rt}

To achieve our goals, we need \emph{efficient} Gaussian \added{ray tracing}, both for optimization and rendering (``inference''), beyond the speed of existing methods (Sec.~\ref{sec:relwork}). 
To do this we focus on four components: 1) Using oriented bounding boxes (OBBs) to bound Gaussians\footnote{The method of Moenne-Loccoz et al.~\shortcite{moenne20243d} describes the icosahedron approach, however a recent code release associated with the paper also uses OBBs. \cite{gscore} also proposed OBBs but for splatting.} 2) Avoiding multiple Bounding Volume Hierarchy (BVH) traversals, 3) Fused forward/backward pass and 4) Aggressive truncation. Numbers reported below use the configuration employed to compare with 3DGRT~\cite{moenne20243d} (see \cref{sec:sup-3dgrt}).

\begin{figure}[!h]
    \centering
    \includegraphics[width=1.0\linewidth]{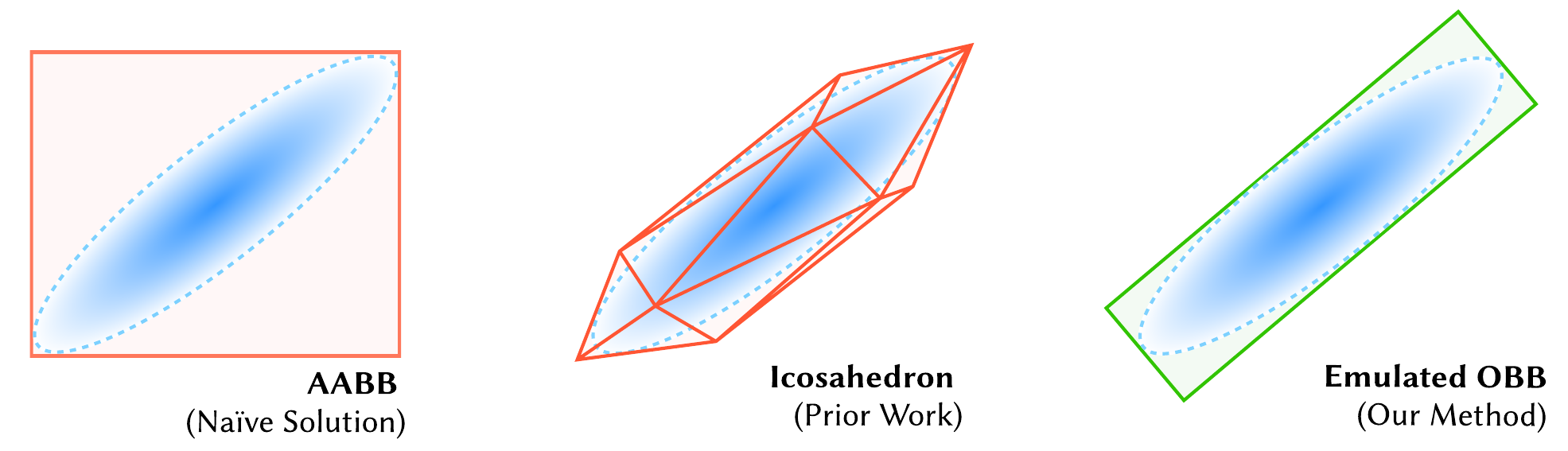}
    \caption{We encapsulate Gaussians with oriented bounding boxes (OBB) using instancing with hardware-accelerated OptiX transforms.}
    \label{fig:raytracing_shape}
    \vspace{-.2cm}
\end{figure}
OptiX only supports Axis-Aligned Bounding Boxes (AABBs) natively, thus naively emulating OBBs
would still result in slow traversal (\Cref{fig:raytracing_shape}). 

Instead, we exploit hardware accelerated transforms of AABBs provided by OptiX, by instancing an AABB in a two-level acceleration structure~\cite{hardware-ray-transforms}. This has the added benefit of letting us simplify the Gaussian evaluation since we avoid computing  covariance matrices explicitly and we do not invert the Gaussian transforms ourselves. On the \textsc{Kitchen} scene (\Cref{fig:synth-edit-1}), an incremental BVH update adds a minimal overhead of $9.5\%$. 



\paragraph{Avoiding Multiple BVH Traversals per Ray}
Per-pixel sorting a large number of Gaussian primitives is slow since it requires incoherent memory accesses. 3DGRT~\cite{moenne20243d} resolved this by collecting and sorting the nearest primitives in small (typically, 16) groups which fit in register memory, requiring multiple BVH traversals. To avoid this, we perform a single traversal and store all intersected Gaussians into a per-pixel linked-list (PPLL)~\cite{yang2010real} ``replay buffer'' which contains Gaussian id, alpha value, etc. We then loop over this buffer multiple times to collect, sort, and integrate the nearest 16 Gaussians. While this approach does increase memory consumption, our experiments show high-resolution scenes can still be handled on modern cards (e.g., RTX4090).
Our experiments show this results in speedups of $10$--$50\%$ depending on the scene and resolution.

\paragraph{Fused Forward/Backward Pass.}
To further accelerate optimization, we use a fused forward/backward pass where each pixel's backward pass starts immediately after the forward pass. We use a PPLL to store data for the backward pass: the intersected Gaussian ids, hit distances, precomputed alpha values, etc. Our backward pass is quite fast, a combined forward/backward is $2.23$ times slower than just the forward pass while 3DGS is $3.31$ times slower.

\paragraph{Aggressive Primitive Truncation.}

\begin{figure}[t]
    \centering
    \includegraphics[width=1.0\linewidth]{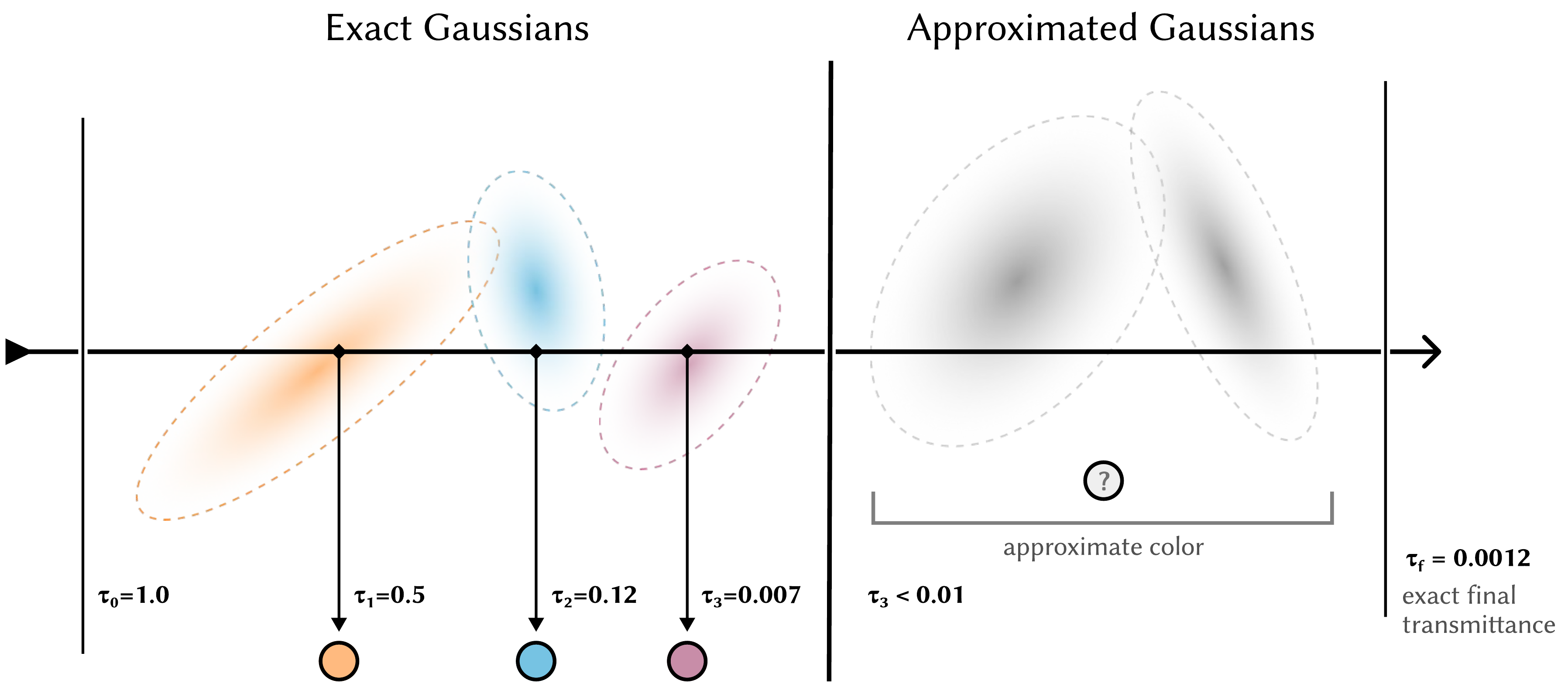}
    \caption{We accelerate integration by approximating Gaussians of low contribution, enabling aggressive transmittance thresholding during optimization.}
    \vspace{-.3cm}
    \label{fig:raytracing_skip}
\end{figure}

Since transmittance decays rapid\-ly along a ray, all Gaussians past a certain point have minimal contribution to the final color. 3DGRT~\cite{moenne20243d} handled this by reducing the transmittance threshold from $\tau=0.001$ to $\tau=0.03$ at inference. Doing so during optimization fails since discarding even seemingly insignificant Gaussians biases gradients of the others in front, making them grow to mask the background. 
To avoid this instability, we discard farthermost Gaussians, and scale the gradients of those remaining using an approximation of their color, similar to  \cite{byrski2025raysplats,hahlbohm2025efficient}, see \Cref{fig:raytracing_skip}. See \cref{sec:sup-truncation} for more details.


\subsection{A Dedicated Network to Infer Disentangled Layers}
\label{sec:network}

Recent learning-based methods have made incredible advances in inferring different intrinsic quantities such as depth, normals or reflectance properties directly from images~\cite{zeng2024rgbx,ke2024repurposing,garcia2025finetuning,diffusion-render,intrinsic-anything,idarb}. Many of these methods fine-tune a diffusion model (typically StableDiffusion2 (SD2)~\cite{rombach2022high}). Marigold~\cite{ke2024repurposing} and RGB$\leftrightarrow$X~\cite{zeng2024rgbx} require many denoising steps; more recent work \cite{xu2025matters,garcia2025finetuning} showed that for strongly conditioned tasks, a single step model can be equally accurate and cheaper at inference. \added{Additionally, concurrent work~\cite{diffusion-render} leverages video models to improve quality and temporal stability; we leave extension to video models as future work, and base our model on SD2.}

\begin{figure}[t]
    \centering
    \includegraphics[width=1.0\linewidth]{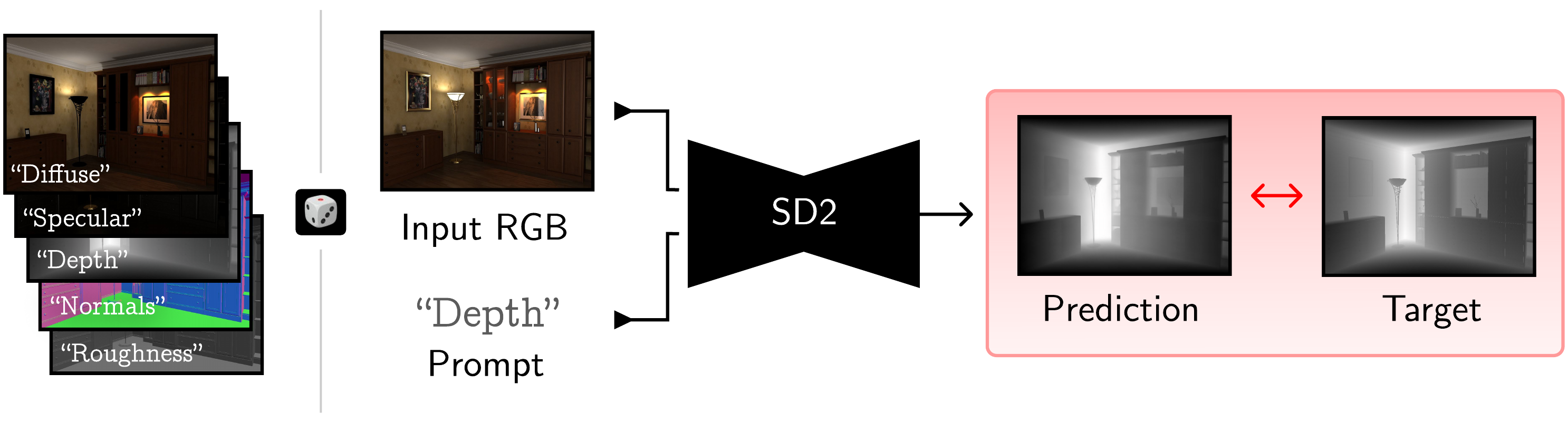}
    \caption{\added{We fine-tune a pretrained Stable Diffusion 2 network into a one-step inverse rendering network.}}
    \label{fig:network-training}
\end{figure}

\begin{figure}[h!]
    \centering
    \includegraphics[width=1.0\linewidth]{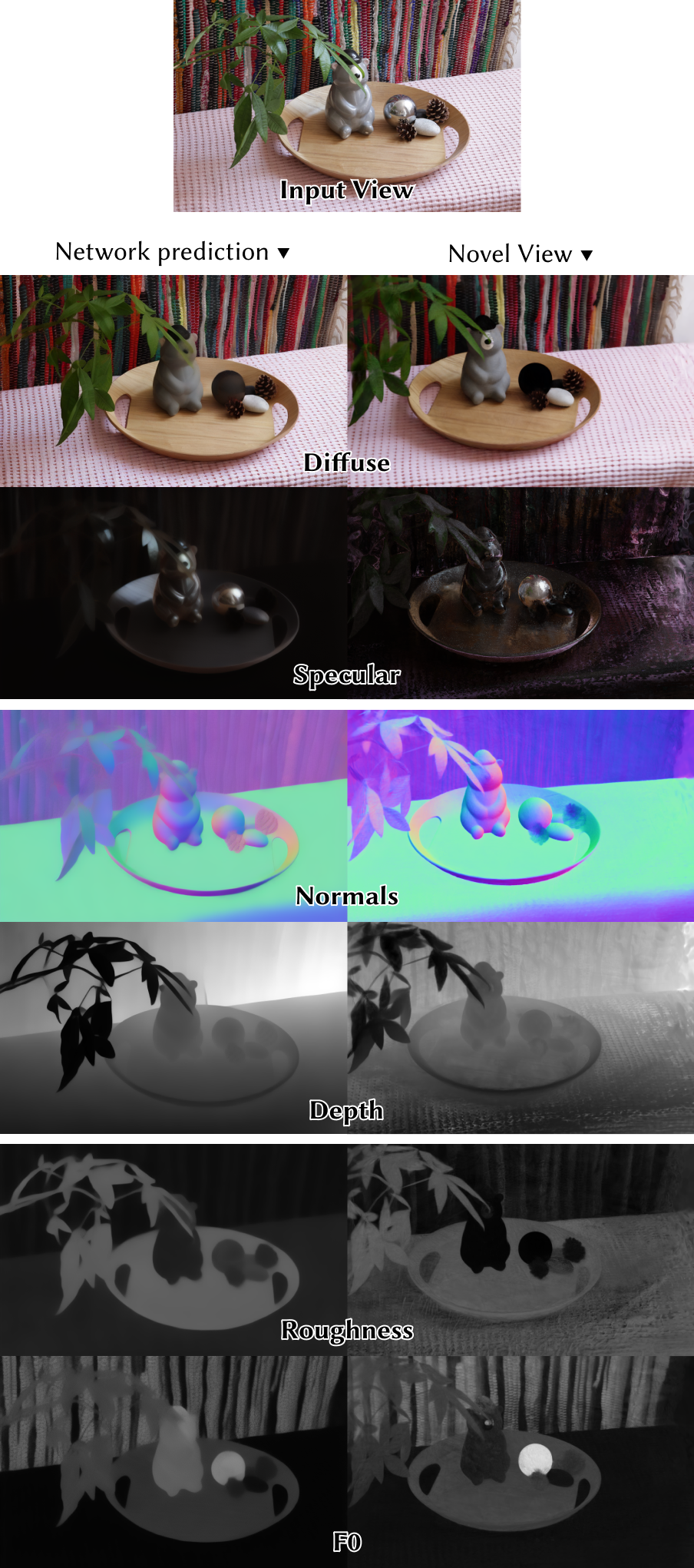}
    \caption{Example network predictions (left) for a given input view (top), and novel view renderings (right) of the different layers used in our approach obtained after optimization. }
    \label{fig:real-scene-layers}
\end{figure}

Based on these recent results, and since none predict the non-diffuse \added{buffer} needed by our method, we fine-tune SD2 into a single step model that predicts all the \added{buffers} (diffuse, \added{specular}, normals, see \cref{sec:method-optimizing}) and switches between them with text prompts, using the approach of RGB$\leftrightarrow$X~\cite{zeng2024rgbx} (\cref{fig:network-training}). For training, we use a mix of the Hypersim~\cite{roberts2021hypersim} and InteriorVerse~\cite{zhu2022interiorverse} datasets and provide as input the RGB image and a prompt designating the desired buffer. We use a latent mean squared error (MSE) loss for fast training. \Cref{fig:real-scene-layers} shows that our network yields plausible predictions of all target \added{buffers}.


\section{Results and Evaluation}
\added{We first present results of real-time reflection editing of radiance fields. Since no previous method can perform such editing, we present our best-effort evaluation of diffuse/specular disentanglement, and compare our ray-tracer with that of 3DGRT~\cite{moenne20243d}.} Finally, in \cref{sec:sup-sparse} we show that we perform better than 3DGS for a sparse set of views.

\label{sec:evaluation}


\begin{figure}[t]
    \centering
    \includegraphics[width=0.96\linewidth]{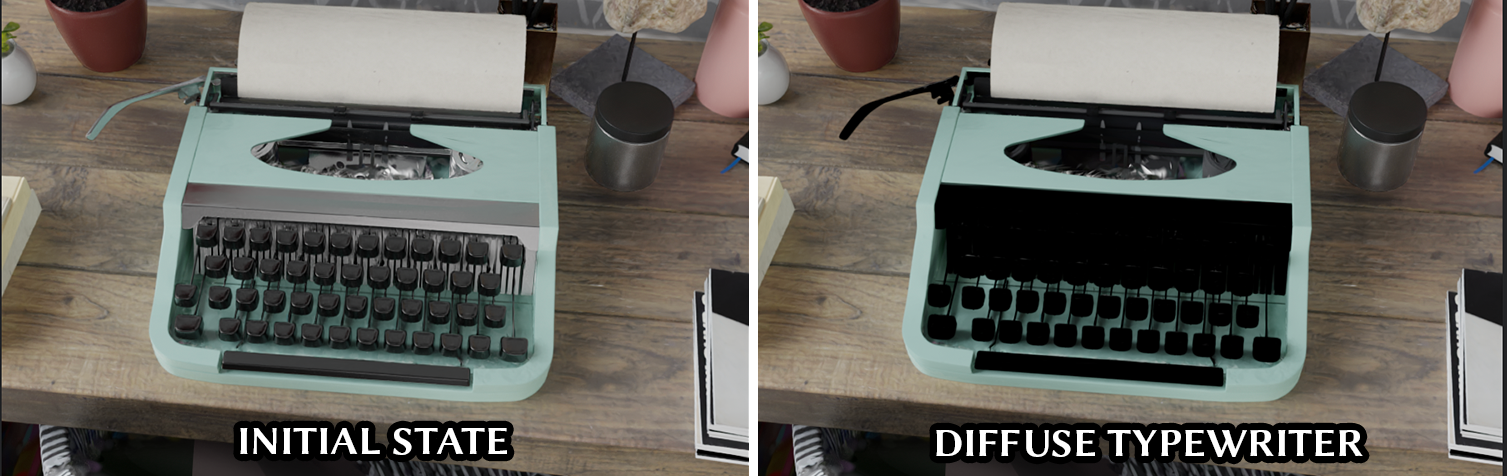} \\
    \includegraphics[width=0.96\linewidth]{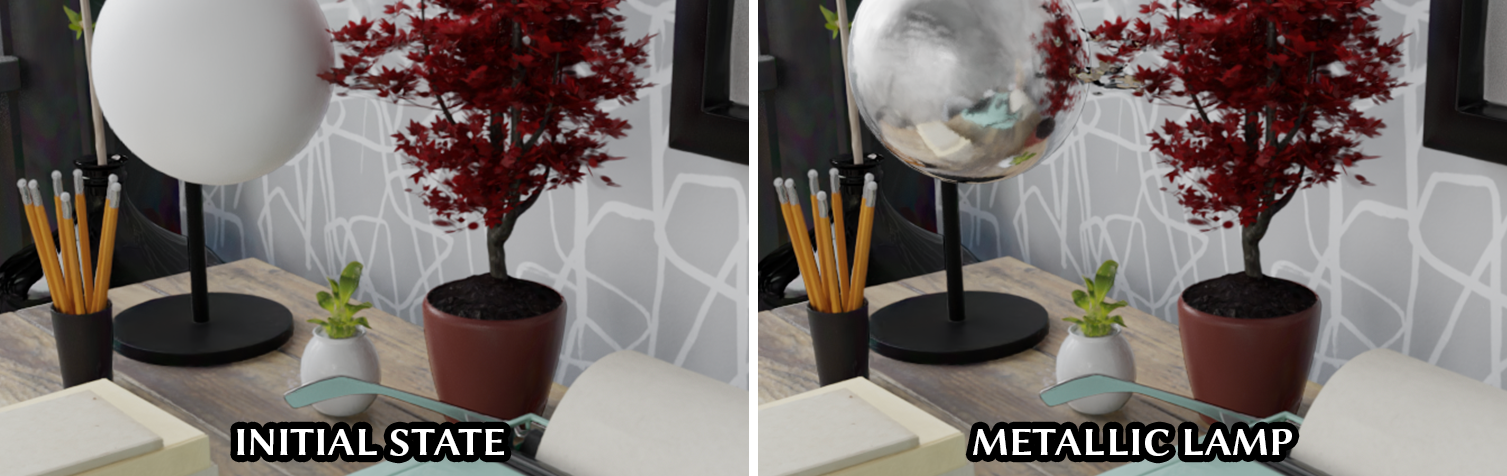} \\
    \includegraphics[width=0.96\linewidth]{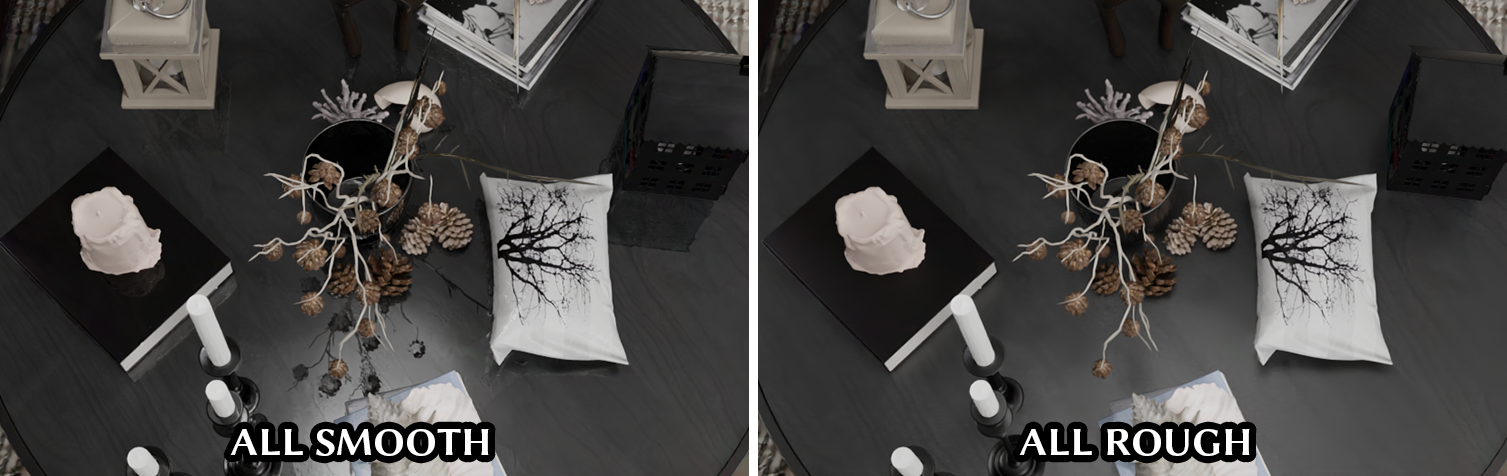} \\
    \includegraphics[width=0.96\linewidth]{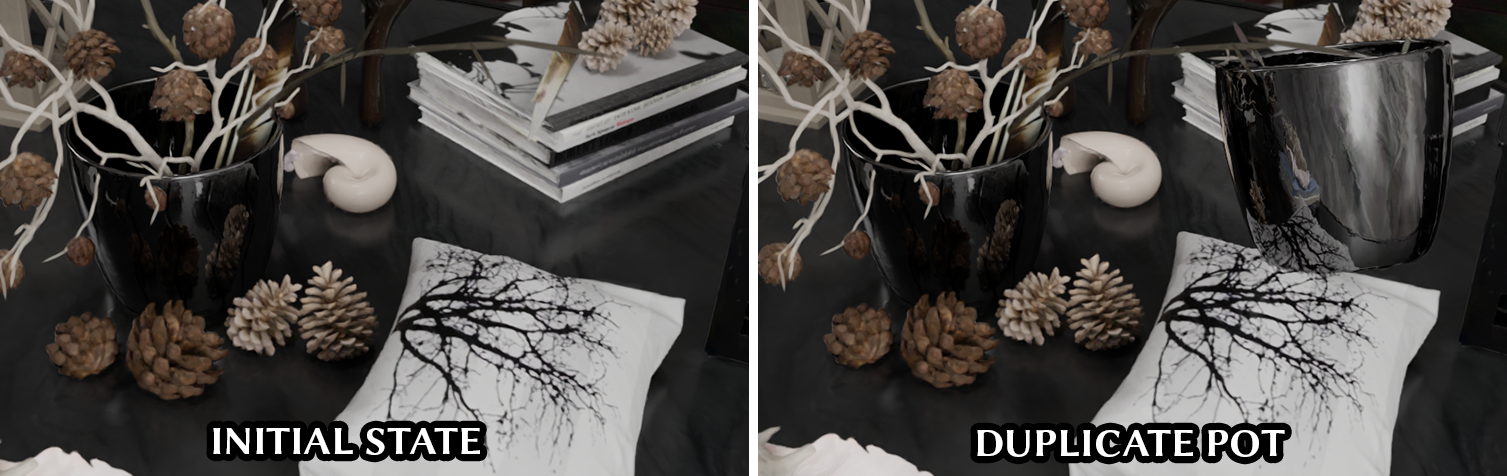}
    \caption{Editing two synthetic scenes: making a typewriter diffuse (first row), making a lamp metallic (second row), changing every object to be smooth or diffuse (third row), duplicating a reflective pot (fourth row). Observe how reflections are consistent with the scene edits.}
    \label{fig:synth-edit-2}
\end{figure}

\begin{figure}[t]
    \centering
    \includegraphics[width=1.0\linewidth]{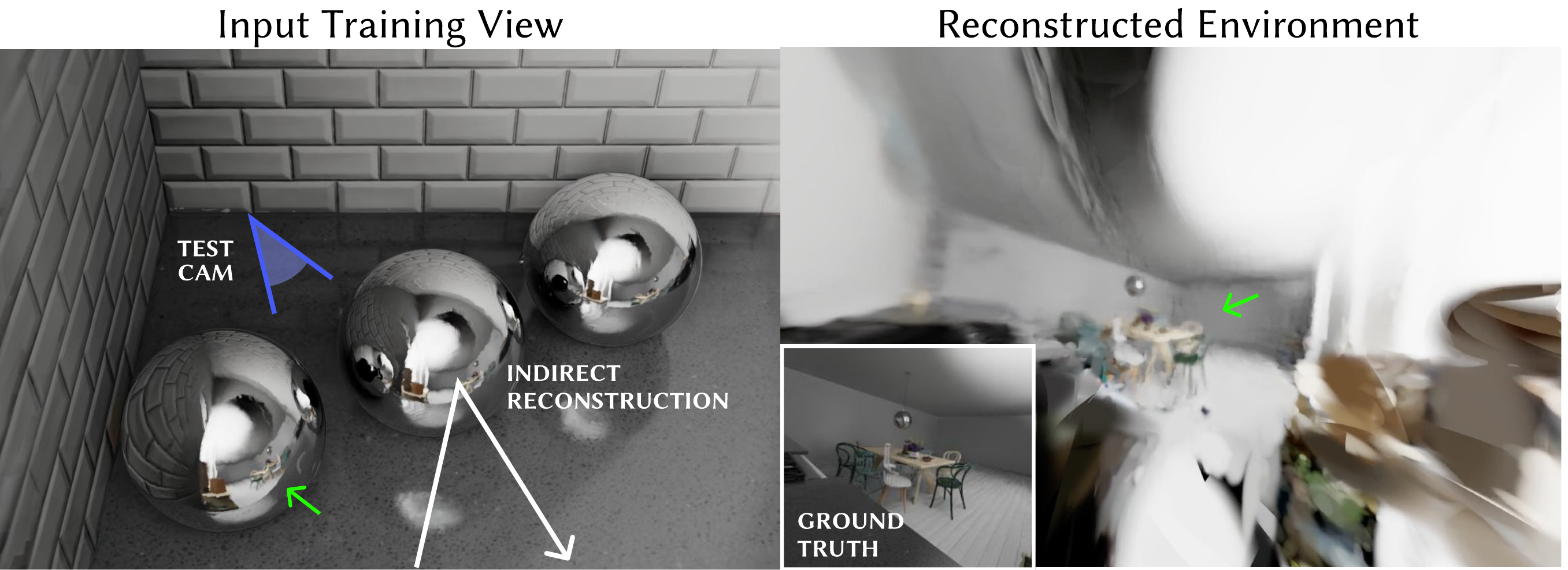}
    \caption{Our method reconstructs the environment behind the camera with a high enough degree of fidelity to produce realistic novel views of parts of the scene \emph{never observed in the input images}. This reconstruction is part of the same scene and can be observed by simply turning the camera around.}
    \vspace{-.3cm}
    \label{fig:reconstruct-env}
\end{figure}

\added{\subsection{Real-Time Editing}}
We show results of real-time editing of radiance field with \added{specular} reflections in the supplemental video and in 
\cref{fig:teaser,fig:synth-edit-1,fig:synth-edit-2}. We can consistently move reconstructed reflectors: reflections are correctly updated in \emph{other} objects in the radiance field. Similarly, we can modify material properties with consistent updates. These operations are only possible because of our full disentanglement and \emph{multiple bounces} we enable. Note however that shadows are ``baked'' into the diffuse layer, and are not updated. In \cref{fig:method-overview-2,fig:reconstruct-env}, we see that our method reconstructs the \emph{unseen} reflected objects, with accuracy that is sufficient for \added{specular} reflection computation. 

\subsection{Evaluation on Synthetic Scenes}

\added{We cannot directly compare our method with previous work, since no other solution allows consistent editing of specular reflections in radiance fields. As a best-effort evaluation, we compare quality of diffuse/specular disentanglement which is required for reflection editing. We show results both for the proof-of-concept case with ground truth (g.t.) buffers, as well as the inverse rendering approach using inferred buffers. In all cases our method has the advantage of using these additional buffers---in particular diffuse/specular image decomposition---that previous methods cannot exploit.}

To allow precise evaluation, we use three synthetic scenes adapted from~\cite{poirier2024diffusion} at $1152 \times 768$ resolution and changed material properties to be shinier. We also removed all transparency from material properties of scene objects. We render RGB images and additional layers, used as input to our method.

\begin{figure}[t]
    \centering
    \includegraphics[width=0.96\linewidth]{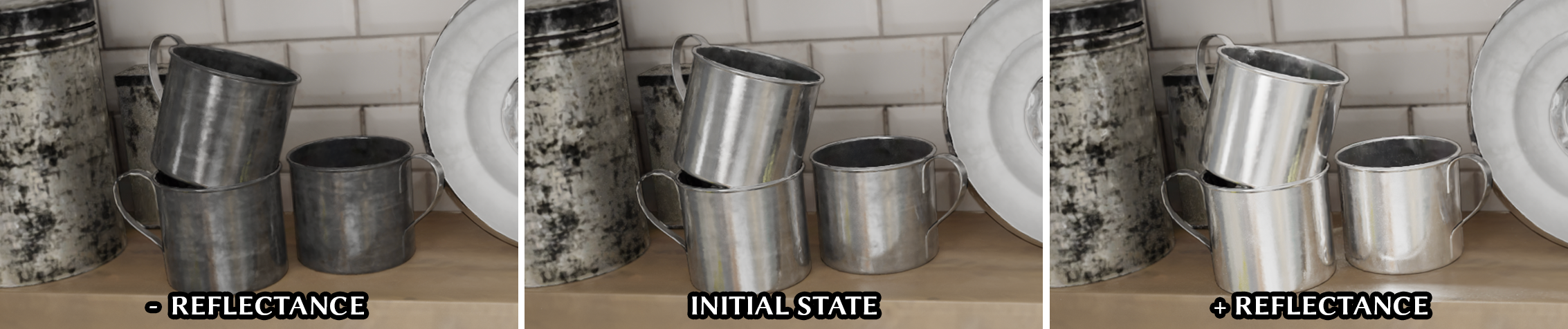}
    \includegraphics[width=0.96\linewidth]{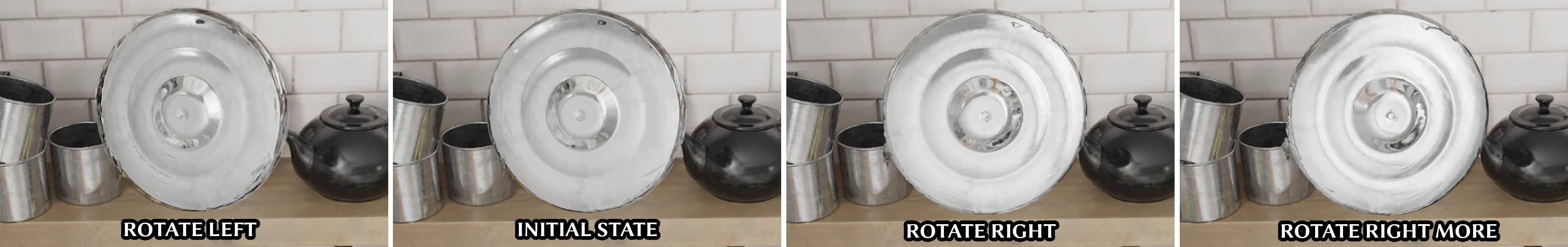}
    \includegraphics[width=0.96\linewidth]{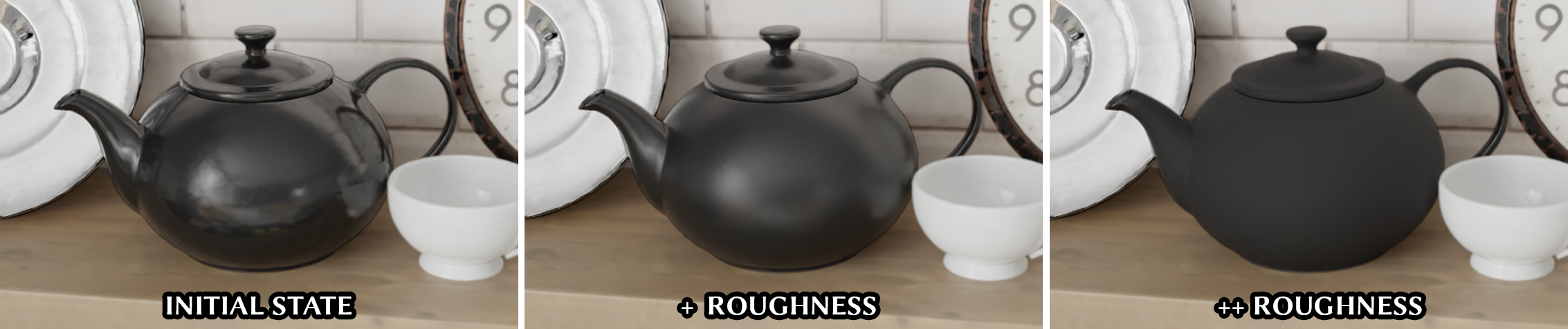}

    \caption{Editing a synthetic scene: changing the base reflectance ($F_0$) of the cups (top), rotating the plate (middle), changing the roughness of the teapot (bottom). Observe how reflections are consistent with the scene edits.}
    \label{fig:synth-edit-1}
\end{figure}

We compare to methods based on 3DGS-based solutions that claim to model diffuse, in particular: Gaussian Shader~\cite{jiang2024gaussianshader}, 3DGS-DR~\cite{ye20243d}, Reflective-GS~\cite{yao2024reflective} and the recent (to appear) EnvGS~\cite{xie2025envgs} which uses ray tracing and is the closest method to ours. 
\added{Again, our method has the advantage of additional buffers that these previous approaches are not designed to use.}

\begin{figure*}[t]
    \centering
    \renewcommand{\arraystretch}{0.5}
    \begin{tabular}{@{\hskip 2pt}c@{\hskip 2pt}c@{\hskip 2pt}c@{\hskip 2pt}c@{\hskip 2pt}c@{\hskip 2pt}c@{\hskip 2pt}c@{\hskip 2pt}}
     & 3DGS & GaussianShader & 3DGS-DR & EnvGS\added{\textsubscript{(g.t. normals)}} & Ours\added{\textsubscript{(g.t. inputs)}} & G.T. \\
    \rotatebox{90}{\hspace{1em}Diffuse} &
    \includegraphics[width=0.145\linewidth]{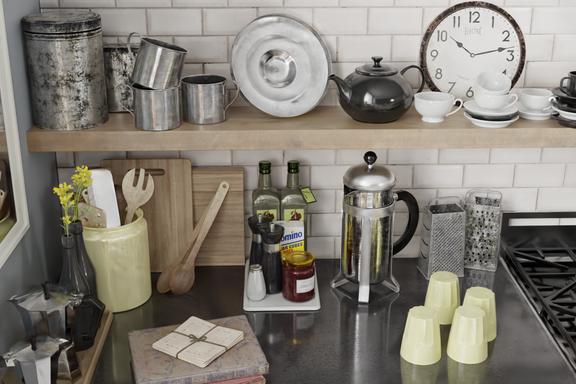} &
    \includegraphics[width=0.145\linewidth]{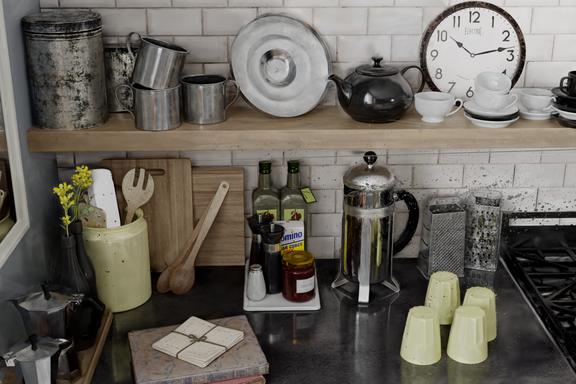} &
    \includegraphics[width=0.145\linewidth]{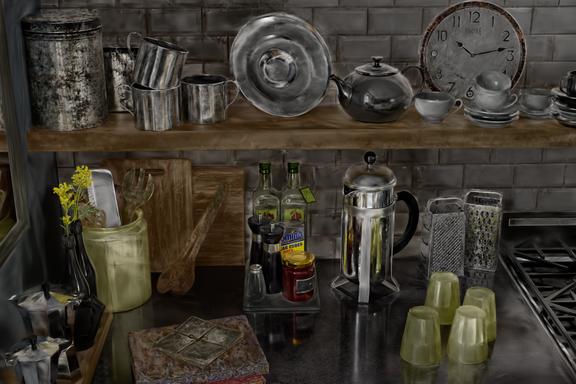} &
    \includegraphics[width=0.145\linewidth]{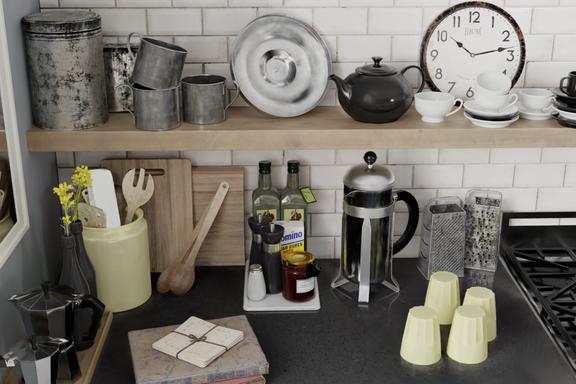} &
    \includegraphics[width=0.145\linewidth]{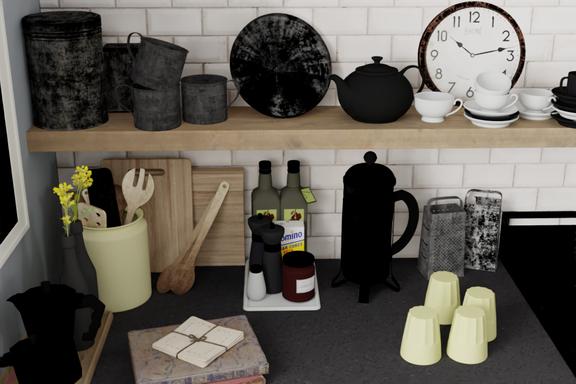} &
    \includegraphics[width=0.145\linewidth]{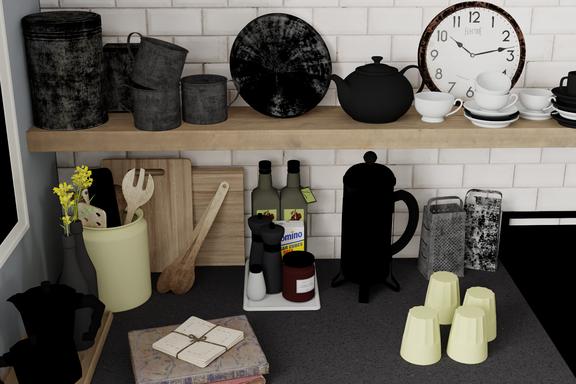} \\
    \rotatebox{90}{\hspace{0.5em}\added{Specular}} &
    \includegraphics[width=0.145\linewidth]{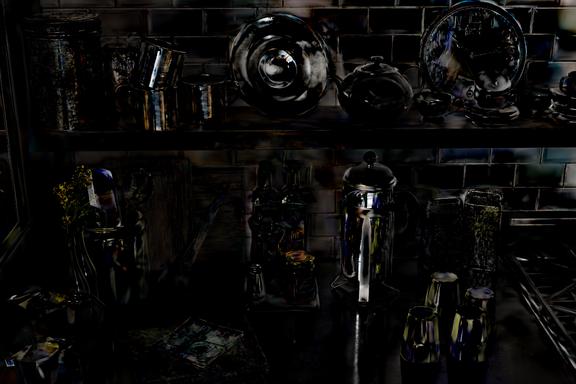} &
    \includegraphics[width=0.145\linewidth]{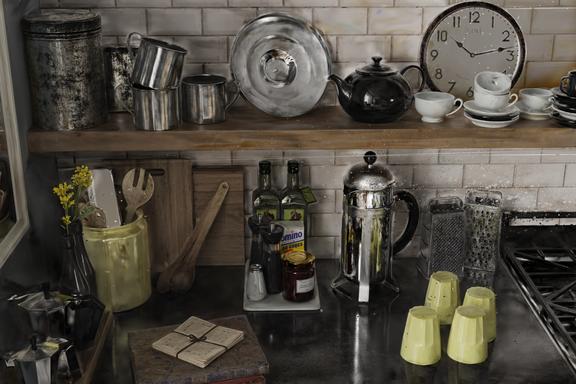} &
    \includegraphics[width=0.145\linewidth]{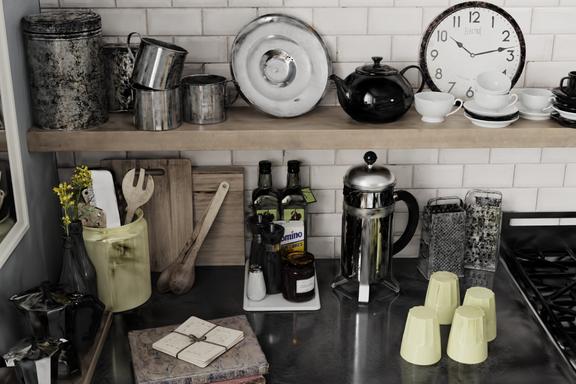} &
    \includegraphics[width=0.145\linewidth]{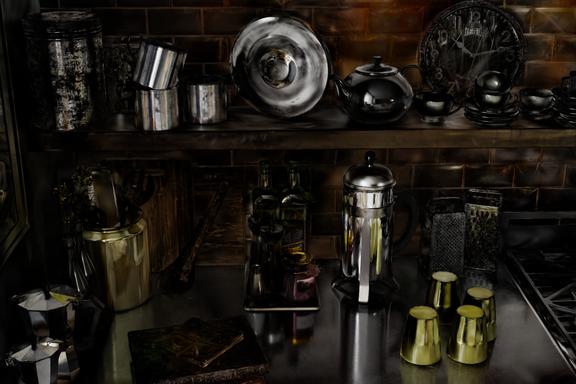} &
    \includegraphics[width=0.145\linewidth]{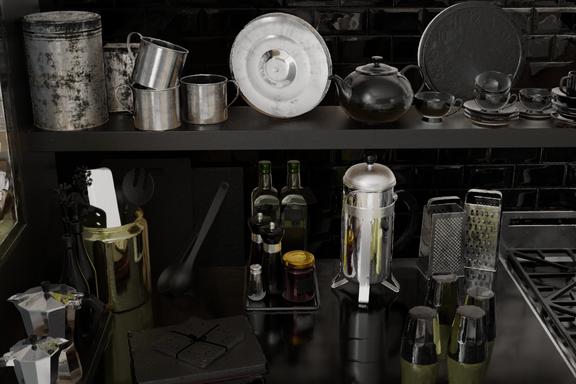} &
    \includegraphics[width=0.145\linewidth]{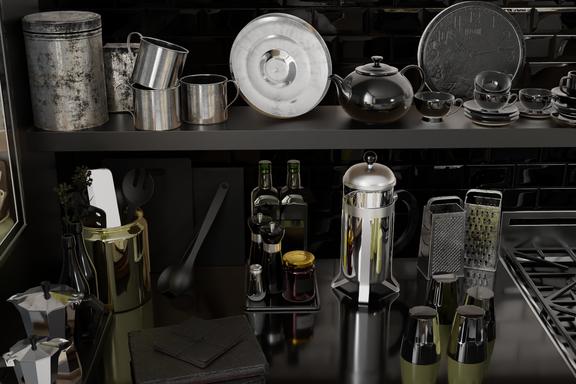} \\
    \rotatebox{90}{\hspace{1.5em}Final} &
    \includegraphics[width=0.145\linewidth]{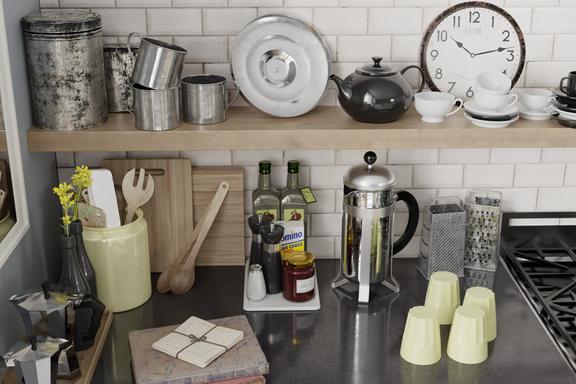} &
    \includegraphics[width=0.145\linewidth]{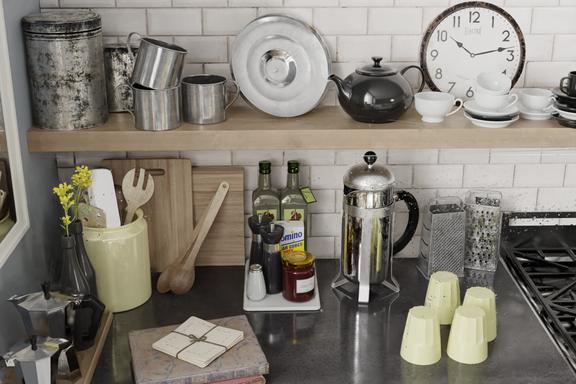} &
    \includegraphics[width=0.145\linewidth]{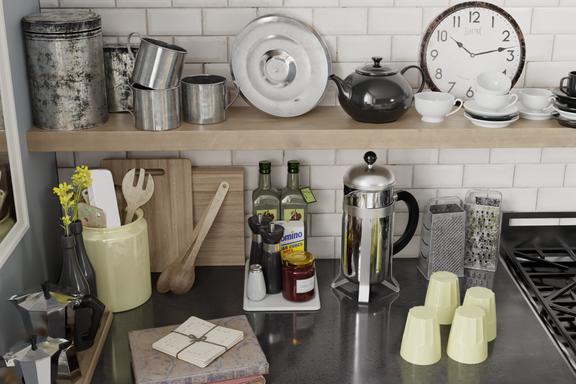} &
    \includegraphics[width=0.145\linewidth]{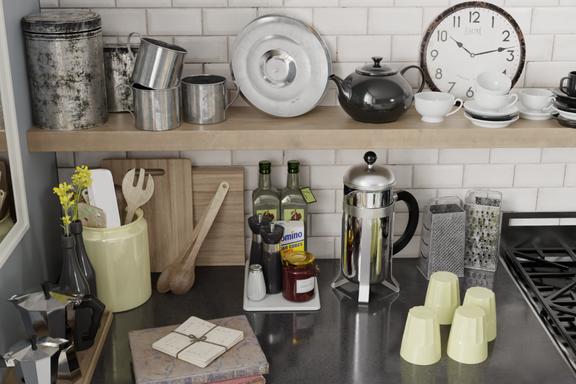} &
    \includegraphics[width=0.145\linewidth]{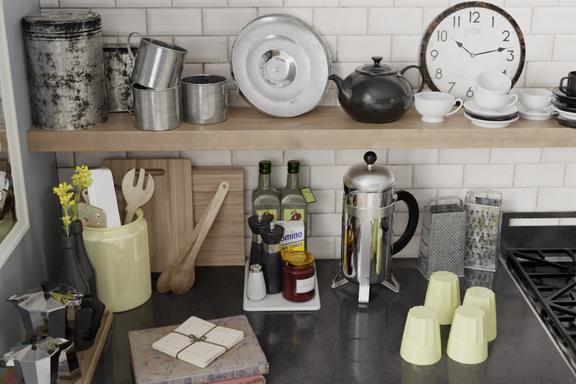} &
    \includegraphics[width=0.145\linewidth]{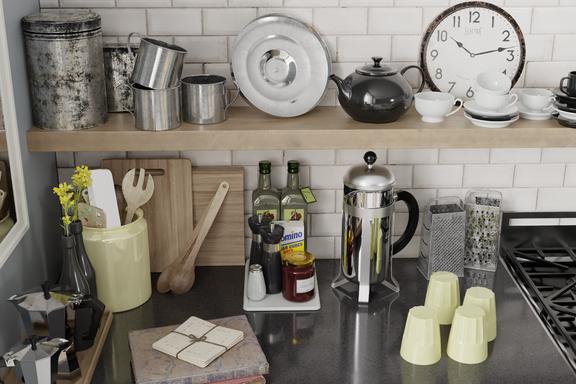} \\
    \end{tabular}

    \vspace{0.4em}
    
    \begin{tabular}{@{\hskip 2pt}c@{\hskip 2pt}c@{\hskip 2pt}c@{\hskip 2pt}c@{\hskip 2pt}c@{\hskip 2pt}c@{\hskip 2pt}c@{\hskip 2pt}}
    \rotatebox{90}{\hspace{1em}Diffuse} &
    \includegraphics[width=0.145\linewidth]{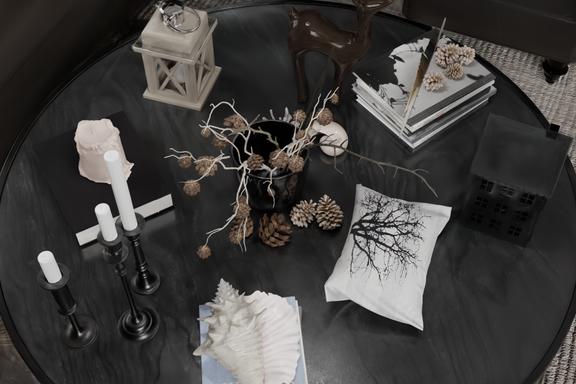} &
    \includegraphics[width=0.145\linewidth]{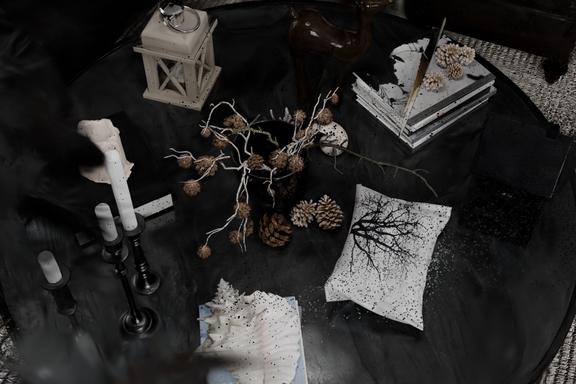} &
    \includegraphics[width=0.145\linewidth]{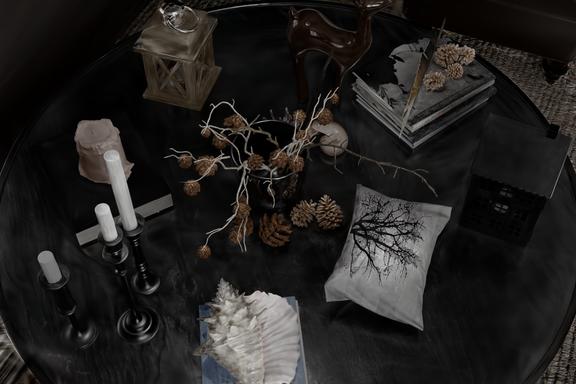} &
    \includegraphics[width=0.145\linewidth]{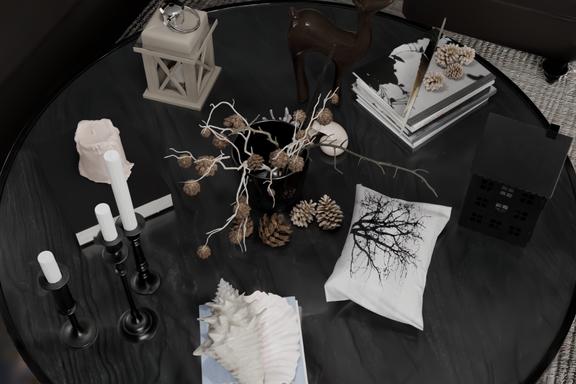} &
    \includegraphics[width=0.145\linewidth]{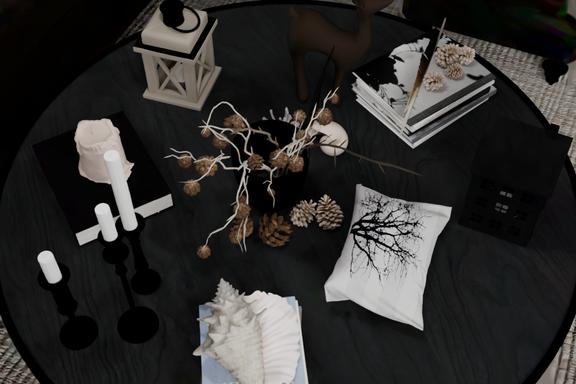} &
    \includegraphics[width=0.145\linewidth]{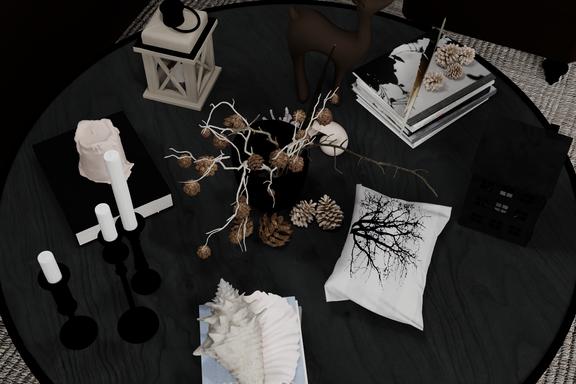} \\
    \rotatebox{90}{\hspace{0.5em}\added{Specular}} &
    \includegraphics[width=0.145\linewidth]{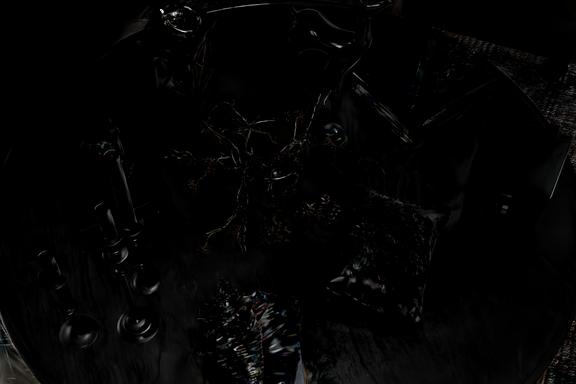} &
    \includegraphics[width=0.145\linewidth]{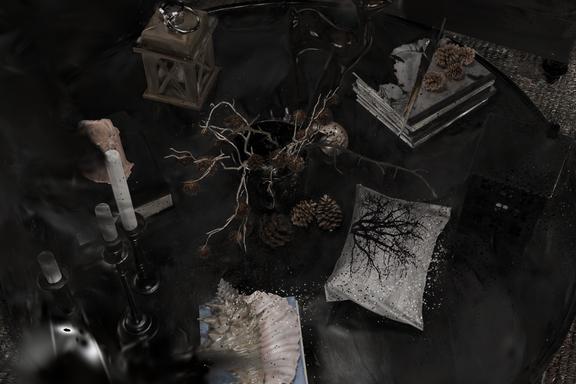} &
    \includegraphics[width=0.145\linewidth]{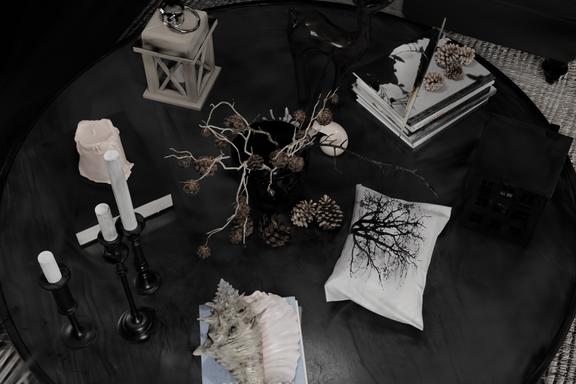} &
    \includegraphics[width=0.145\linewidth]{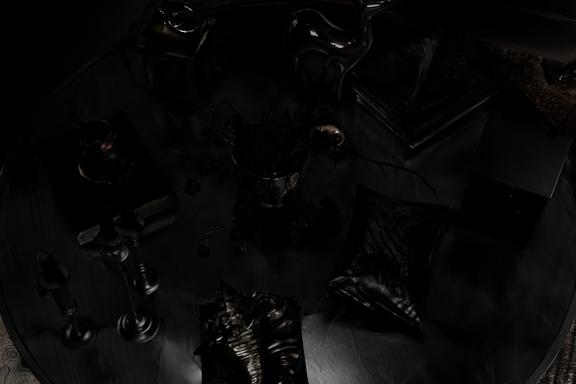} &
    \includegraphics[width=0.145\linewidth]{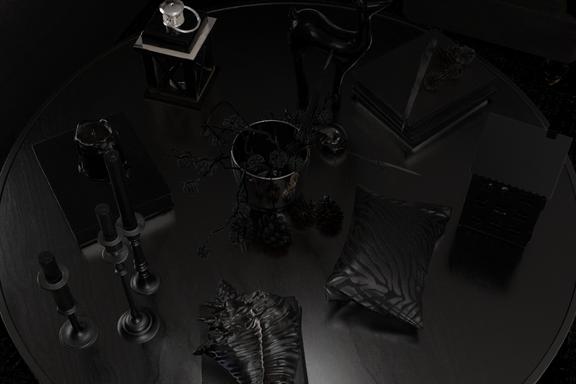} &
    \includegraphics[width=0.145\linewidth]{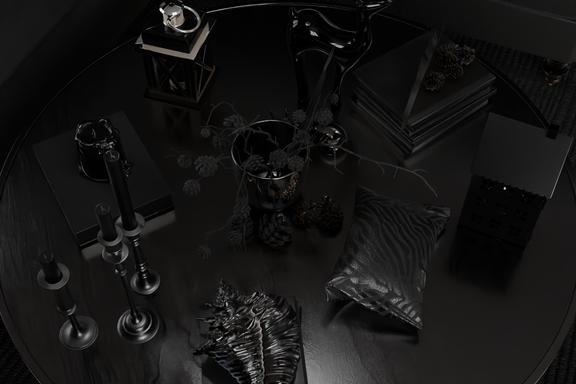} \\
    \rotatebox{90}{\hspace{1.5em}Final} &
    \includegraphics[width=0.145\linewidth]{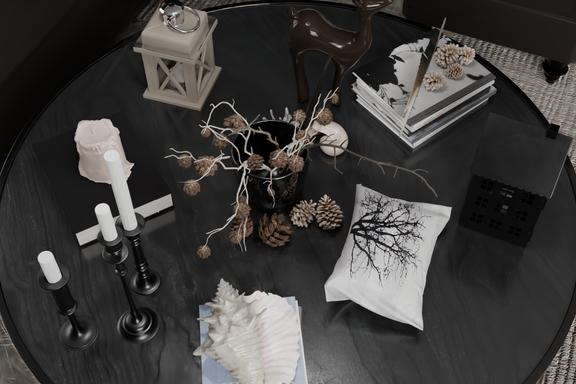} &
    \includegraphics[width=0.145\linewidth]{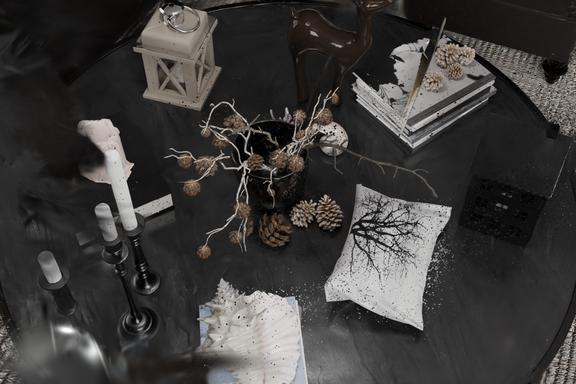} &
    \includegraphics[width=0.145\linewidth]{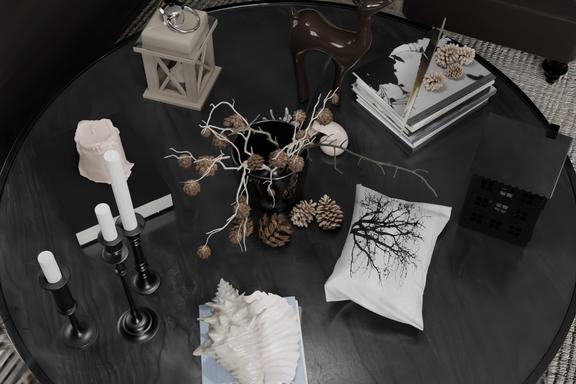} &
    \includegraphics[width=0.145\linewidth]{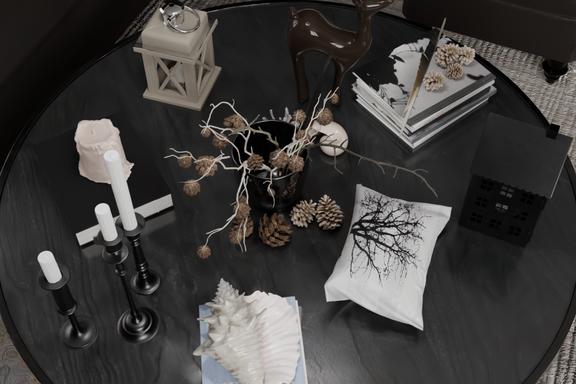} &
    \includegraphics[width=0.145\linewidth]{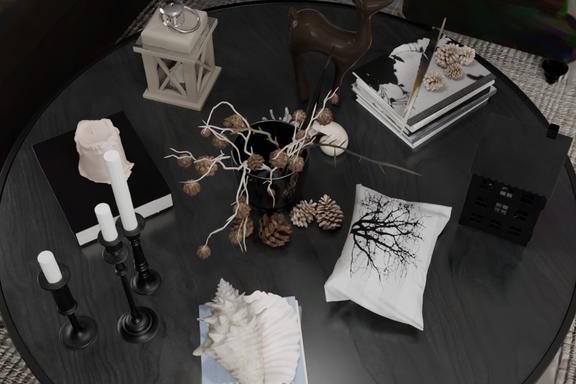} &
    \includegraphics[width=0.145\linewidth]{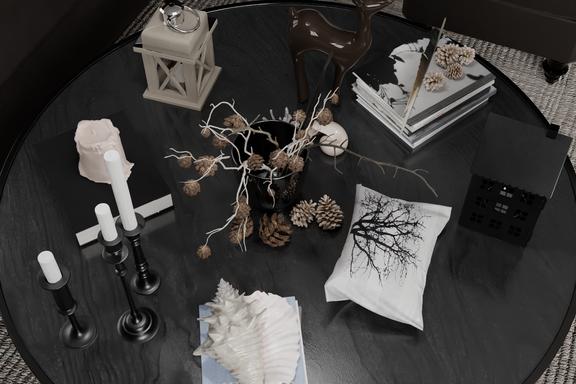} \\
    \end{tabular}
    
    \caption{Disentanglement results of different methods on \textsc{Shiny kitchen} and \textsc{Shiny Livingroom}. Inlays show that our method does not mix reflections into diffuse. \added{Note that in these results, our method uses ground truth input buffers and EnvGS uses ground truth normals.}}
    \label{fig:image-grid}
\end{figure*}

\Cref{fig:image-grid} shows that, in complex scenes, existing approaches incorrectly separate the diffuse from \added{specular} components of the scene---reflections are represented as ``fake mirror'' geometry in the diffuse, making them unusable for editing. In contrast, assuming an optimal network \added{(given \emph{ground truth} inputs)}, our method reconstructs a clean reflection-free diffuse pass. 

In \cref{tbl:disentanglement-psnr} we show the PSNR for the 3 synthetic scenes for each method (more details in \cref{sec:sup-disentangle}). Metrics are computed on a test path separate from input views. The top part of the table does not use any ground truth values, only network predictions; our method (Ours\added{\textsubscript{(net. inputs)}}) achieves on average better disentanglement for diffuse/\added{specular} compared to all other methods, including the concurrent \added{EnvGS, using network predicted normals (EnvGS\textsubscript{(net. normals)}), despite its lower overall quality}.  \added{As a proof-of-concept, in the last two rows we show results for our method using} all the ground truth layers (Ours\textsubscript{(g.t. inputs)})\added{. We also show EnvGS using} ground truth normals \added{(EnvGS\textsubscript{(g.t. normals)})}. Since our method does not create ``fake'' geometry to reproduce the input RGB images, reconstructing the final image is much harder, and thus PSNR is lower. Note however that all other methods have lower final PSNR than 3DGS and do not improve much on the diffuse nor \added{specular} layers, showing their inability to properly model \added{specular} reflections in complex scenes. In addition, qualitatively our image quality is sufficient for editing (see video and figures).
Achieving higher PSNR requires solving the hard problem of accurate reconstruction of the reflected part of the scene, as well as improved optimization (see \cref{sec:concl}).

\begin{table}
\centering
\small
\caption{Disentanglement performance comparison. \added{Note that our method has the advantage of using additional buffers unavailable to other methods. The second part of the table is shown as a proof-of-concept, using g.t. train view inputs where possible.} }

\setlength{\tabcolsep}{2.0pt}

\resizebox{\columnwidth}{!}{
\begin{tabular}{llllllllll}
\toprule
 & 
\multicolumn{3}{l}{Shiny Kitchen} & \multicolumn{3}{l}{Shiny Livingroom} & \multicolumn{3}{l}{Shiny Office} \\
 & \makebox[0pt][l]{\scriptsize Diffuse} & \makebox[0pt][l]{\scriptsize Spec.} & \makebox[0pt][l]{\scriptsize Final} & \makebox[0pt][l]{\scriptsize Diffuse} & \makebox[0pt][l]{\scriptsize Spec.} & \makebox[0pt][l]{\scriptsize Final} & \makebox[0pt][l]{\scriptsize Diffuse} & \makebox[0pt][l]{\scriptsize Spec.} & \makebox[0pt][l]{\scriptsize Final} \\
\midrule
3DGS & 13.11 & 13.17 & \textbf{32.83} & 17.01 & 17.35 & \textbf{32.72} & 19.10 & 15.74 & 
\textbf{34.91} \\
Gauss. Shader & 
14.88 & 12.15 & 30.13 & 20.57 & 16.91 & 24.26 & \textbf{21.77} & 15.56 & 32.75 \\
3DGS-DR & 
10.10 &  8.59 & 32.79 & 16.08 & 12.97 & 31.39 &  9.74 &  8.96 & 34.66 \\
Refl. GS & 
13.50 & 13.45 & 32.17 & 20.43 & 20.46 & 29.85 & 20.63 & 17.24 & 33.68 \\
EnvGS\added{\textsubscript{(net. normals)}} & 
14.33 & 14.48 & 32.82 & 22.74 & \textbf{21.22} & 30.65 & 20.92 & 16.33 & 34.16 \\
Ours\added{\textsubscript{(net. inputs)}} & 
\textbf{20.36} & \textbf{16.95} & 20.41 & \textbf{23.77} & 20.35 & 21.21 & 20.60 & \textbf{17.40} & 17.75 \\
\midrule
EnvGS\added{\textsubscript{(g.t. normals)}} & 
14.51 & 15.15 & \textbf{32.96} & 22.64 & 21.46 & \textbf{30.31} & 20.81 & 15.90 & \textbf{34.34} \\
Ours\added{\textsubscript{(g.t. inputs)}} & 
\textbf{33.20} & \textbf{24.30} & 26.96 & \textbf{29.68} & \textbf{26.46} & 26.96 & \textbf{31.74} & \textbf{24.48} & 27.54 \\
\bottomrule

\end{tabular}
}

\label{tbl:disentanglement-psnr}
\end{table}

We also show optimization times (\Cref{tbl:time}), which shows that our method optimizes significantly faster than all previous methods and $\sim$$5\times$ faster than the second best method EnvGS.
In \cref{sec:sup-perf} we show statistics of number of primitives and FPS: as expected, our method is slower but still interactive, since we trace multiple rays per pixel. 

\subsection{Evaluation on Real Scenes}
On real scenes, we use the network (\cref{sec:network}) to compute layers from input images. The prediction quality, seen for example in \cref{fig:real-scene-layers}, is far from perfect, but we do manage to obtain sufficiently good results to demonstrate similar editing examples (see video and \cref{fig:teaser}). 
We compare to EnvGS in \cref{fig:ours_vs_envgs} where we show renderings of  diffuse and specular layers, and the final novel view (left EnvGS, right ours). We can clearly see that the EnvGS diffuse layer contains \added{specular} content, especially visible in the \textsc{Bear} scene (far right).

In \cref{sec:sup-network-ablations}, we ablate the quality of network predictions, progressively replacing GT layers with predicted versions. The most important layers are diffuse/\added{specular} that have a significant impact on PSNR; all other layers are less important.



\begin{table}[h!]
\small
\centering
\caption{Training times for different methods.}



\centering

\begin{tabular}{lccc}
\toprule
 & Shiny Kitchen & Shiny Livingroom & Shiny Office \\
\midrule
GShader & 2:24:26 & 2:28:28 & 2:31:09 \\
3DGS-DR & 0:44:09 & 0:45:45 & 0:51:41 \\
ReflGS & 1:28:57 & 1:28:38 & 1:43:22 \\
EnvGS\added{\textsubscript{(g.t. norm.)}} & 3:34:27 & 2:40:52 & 3:13:21 \\
Ours\added{\textsubscript{(net. inputs)}} & 0:48:52 & 0:23:56 & 0:37:05 \\
Ours\added{\textsubscript{(g.t. inputs)}} & \textbf{0:23:16} & \textbf{0:20:48} & \textbf{0:24:13} \\
\bottomrule
\end{tabular}
\vspace{-.4cm}
\label{tab:time}

\label{tbl:time}
\end{table}

\subsection{Comparing our Raytracer to 3DGRT}
We compared our raytracer's performance to 3DGRT when used as a drop-in replacement for regular 3DGS, by swapping it for the splatting rasterizer while limiting to 1 ray per pixel and integrating RGB color only. \added{We trained for 7k iterations and our method does not have spherical harmonics;} for details of the exact configuration used, please see \cref{sec:sup-3dgrt}.
\cref{tbl:raytracing-perf} shows that our raytracer improves training times and FPS performance over 3DGRT.

\begin{table}[h]
\caption{Performance comparison between our raytracer and 3DGRT when used as a drop-in replacement for 3DGS. Average across all MipNerf scenes at different resolutions. Our method was run without spherical harmonics.
}
\centering
\small
\resizebox{\columnwidth}{!}{
\begin{tabular}{lccccccccc}
\toprule
\multirow{2}{*}{Downsampl.} & \multicolumn{3}{c}{Training Time} & \multicolumn{3}{c}{FPS} & \multicolumn{2}{c}{PSNR (dB)} \\
\cmidrule(lr){2-4} \cmidrule(lr){5-7} \cmidrule(lr){8-9}
& 3DGRT & Ours & Speedup & 3DGRT & Ours & Speedup & 3DGRT & Ours \\
\midrule

\textsc{2} & 00:29:24 & \textbf{00:06:51} & 4.35$\times$ & 26.77 & \textbf{49.93} & 2.34$\times$ & \textbf{25.44} & 25.01 \\
\textsc{4} & 00:10:03 & \textbf{00:02:27} & 4.08$\times$ & 75.49 & \textbf{143.93} & 2.43$\times$ & \textbf{25.83} & 25.35 \\
\textsc{8} & 00:05:28 & \textbf{00:01:08}  & 4.77$\times$ & 139.09 & \textbf{248.75} & 2.24$\times$ & \textbf{26.27} & 26.25 \\
\bottomrule
\end{tabular}
}

\label{tbl:raytracing-perf}
\vspace{-.5cm}
\end{table}

\section{Limitations and Conclusion}
\label{sec:concl}
We make an important step forward to allow truly disentangled, physically-based reflections for radiance fields. The main limitation of our method is the performance of the network used to extract layers: we are confident that such approaches will keep improving and provide the quality needed.
Another limitation is lack of support for transparency. This requires determining how to estimate transparency from images (including producing training data and corresponding networks), and an efficient rendering method. 
Full scene editing with shadow updates requires a solution with (at least partial) relighting.
Finally, improving image quality even with perfect layers requires more accurate reconstruction of the reflected scene. These are all hard problems and exciting new research directions. 

\added{ In conclusion, we presented a new approach that allows real-time, consistent specular reflection editing in radiance fields, enabled by our diffuse/specular disentanglement and our support for multi-bounce reflections.}
The key to our solution is the use of separate supervision of the diffuse and \added{specular} layers of images, building on learning-based predictors of such layers. We also show how to provide stable training and several performance improvements to Gaussian ray tracing that, taken together, allow real-time, consistent, physically-based editing of reflections, and reconstruction of unseen reflected objects.

\begin{acks}
Thanks to Ishaan Shah for the \href{https://github.com/graphdeco-inria/graphdecoviewer}{Gaussian Viewer}. This research was co-funded by the European Union (EU) ERC Advanced grant FUNGRAPH No 788065 and ERC Advanced Grant NERPHYS No 101141721. Views and opinions expressed are however those of the author(s) only and do not necessarily reflect those of the EU or the European Research Council. Neither the EU nor the granting authority can be held responsible for them. This research was also supported by NSERC grant RGPIN-2020-04799 and the Digital Research Alliance Canada. The authors are grateful to Adobe and NVIDIA for generous donations, and the OPAL infrastructure from Université Côte d’Azur.
\end{acks} 

\bibliographystyle{ACM-Reference-Format}
\bibliography{references}


\clearpage
\newcommand{\mywidth}{0.96\linewidth}

\begin{figure*}[t]
    \centering
    \renewcommand{\mywidth}{0.15\linewidth}
    \begin{tabular}{@{\hskip 2pt}c@{\hskip 2pt}c@{\hskip 2pt}c@{\hskip 2pt}c@{\hskip 2pt}c@{\hskip 2pt}c@{\hskip 2pt}c@{\hskip 2pt}}
     & EnvGS\added{\textsubscript{(net. normals)}} & Ours\added{\textsubscript{(net. inputs)}} & EnvGS\added{\textsubscript{(net. normals)}} & Ours\added{\textsubscript{(net. inputs)}} & EnvGS\added{\textsubscript{(net. normals)}} & Ours\added{\textsubscript{(net. inputs)}} \\
    \rotatebox{90}{\hspace{1em}Diffuse} &
    \includegraphics[width=\mywidth]{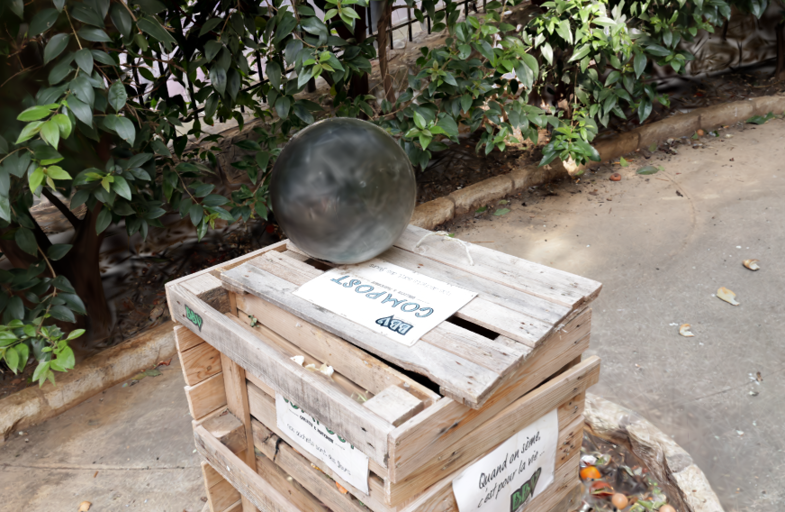} &
    \includegraphics[width=\mywidth]{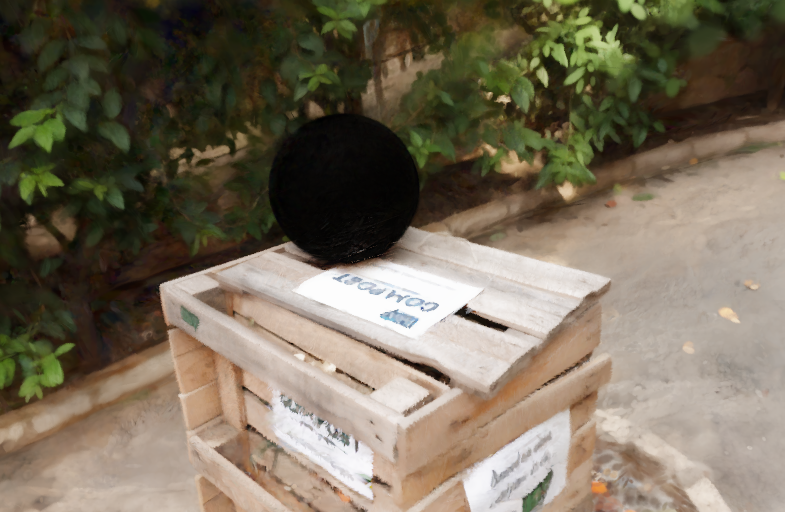} &
    \includegraphics[width=\mywidth]{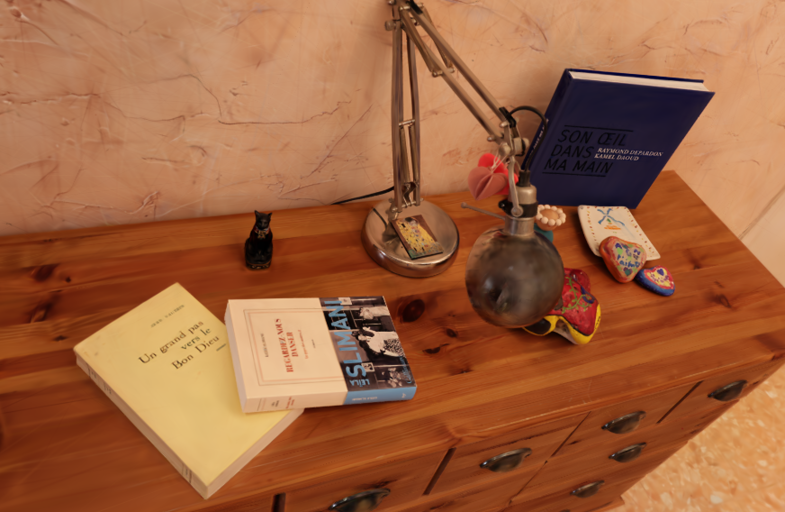} &
    \includegraphics[width=\mywidth]{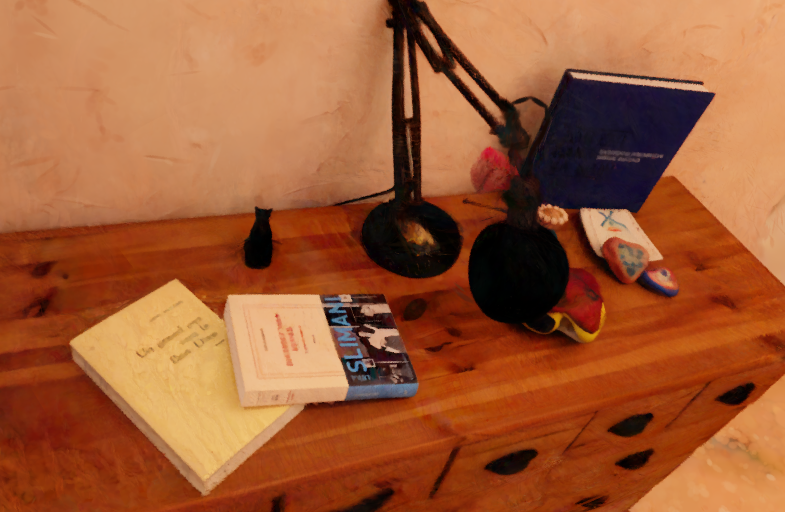} &
    \includegraphics[width=\mywidth]{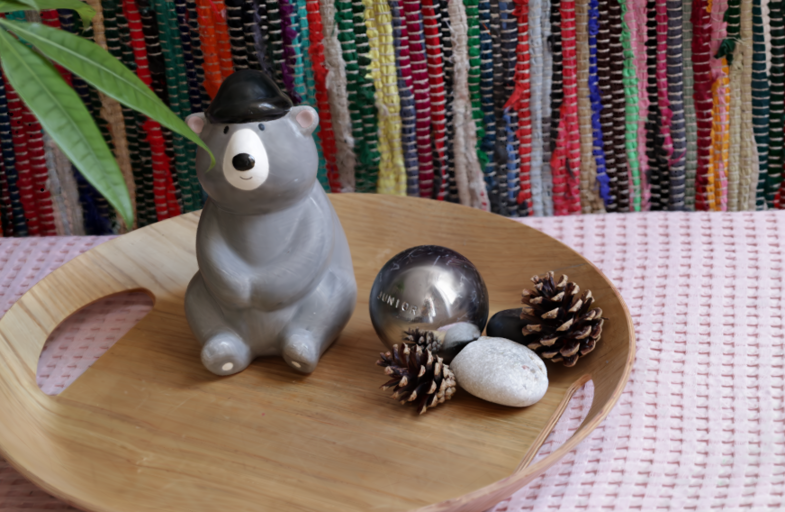} &
    \includegraphics[width=\mywidth]{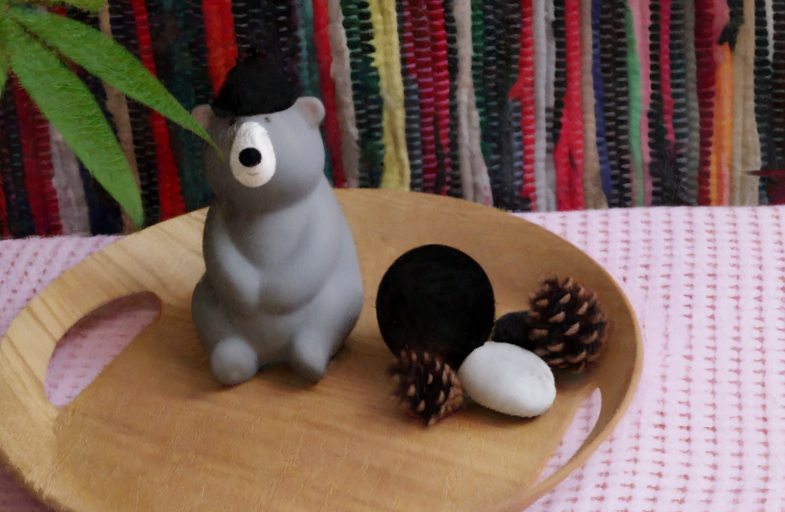} \\
    \rotatebox{90}{\hspace{0.5em}\added{Specular}} &
    \includegraphics[width=\mywidth]{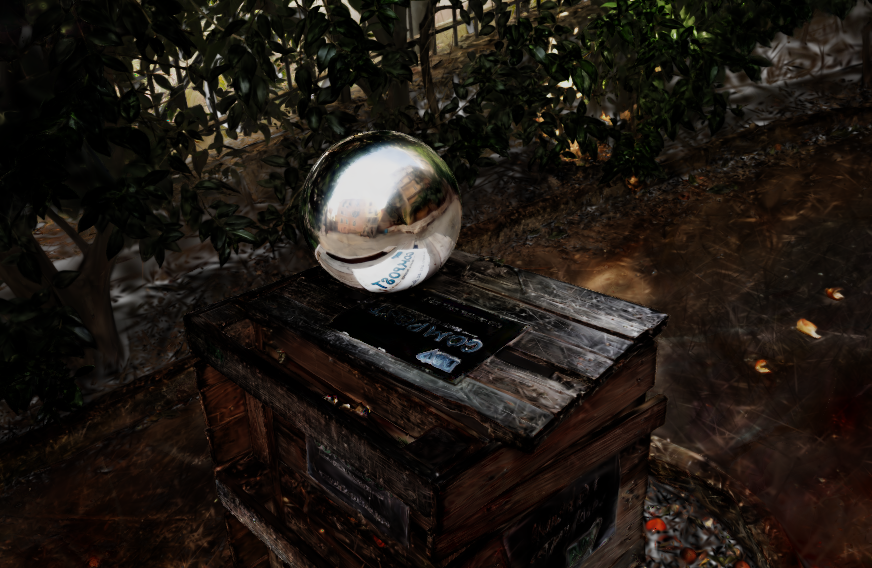} &
    \includegraphics[width=\mywidth]{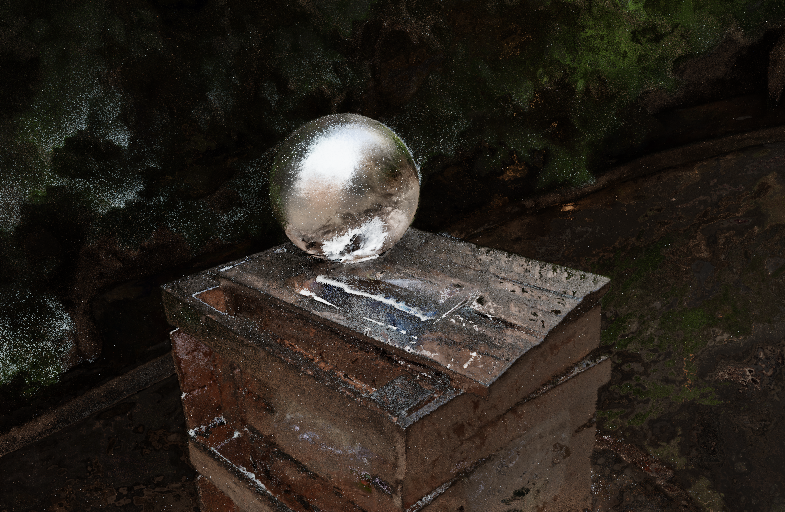} &
    \includegraphics[width=\mywidth]{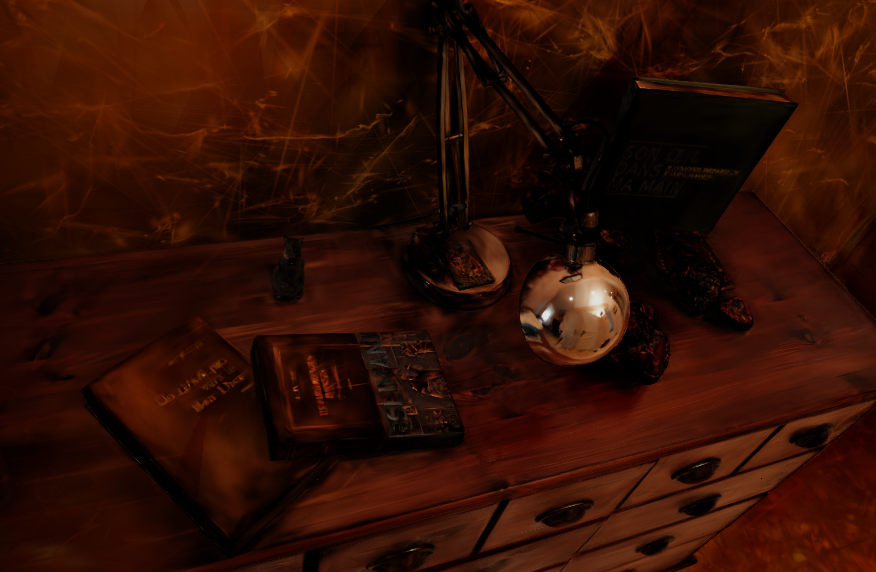} &
    \includegraphics[width=\mywidth]{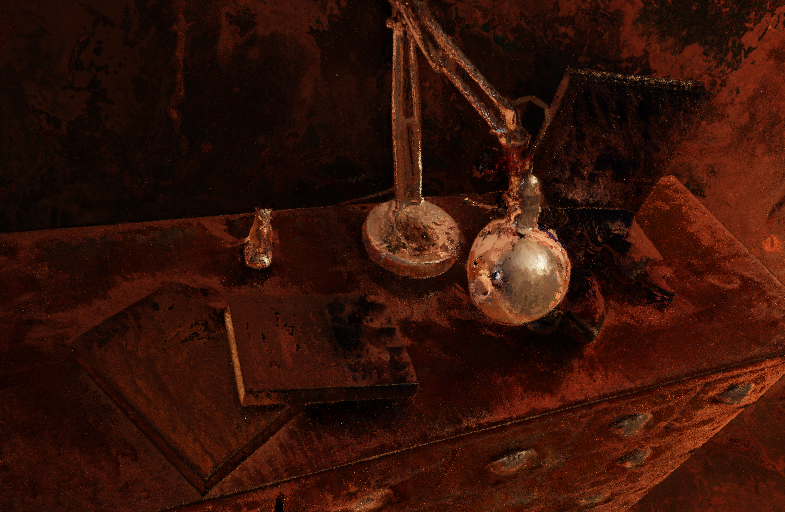} &
    \includegraphics[width=\mywidth]{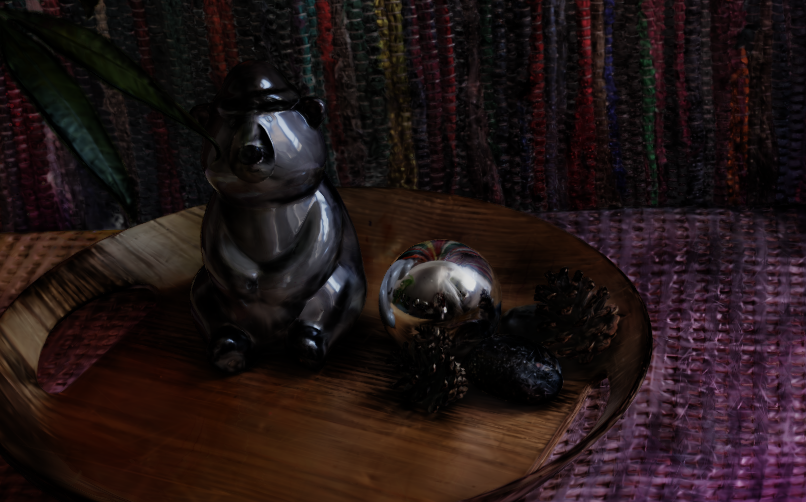} &
    \includegraphics[width=\mywidth]{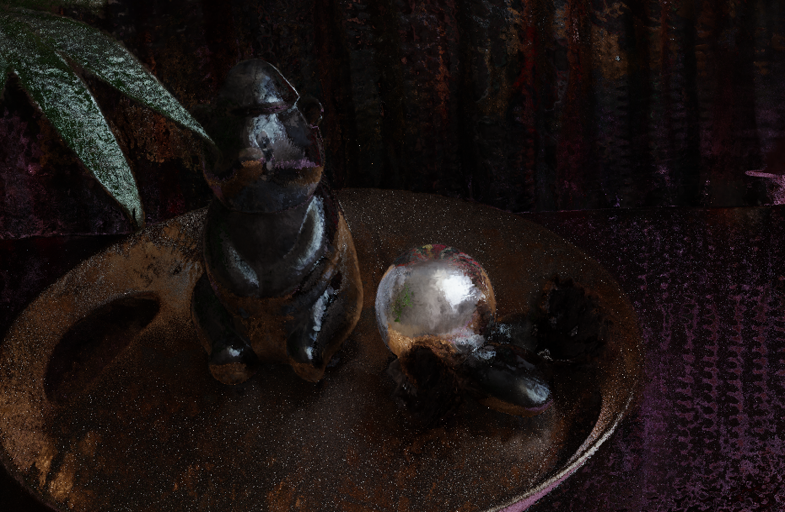} \\
    \rotatebox{90}{\hspace{1.5em}Final} &
    \includegraphics[width=\mywidth]{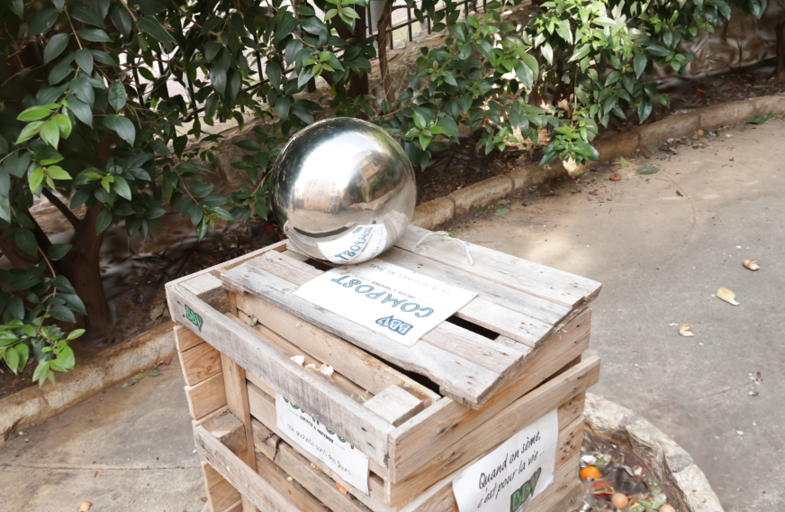} &
    \includegraphics[width=\mywidth]{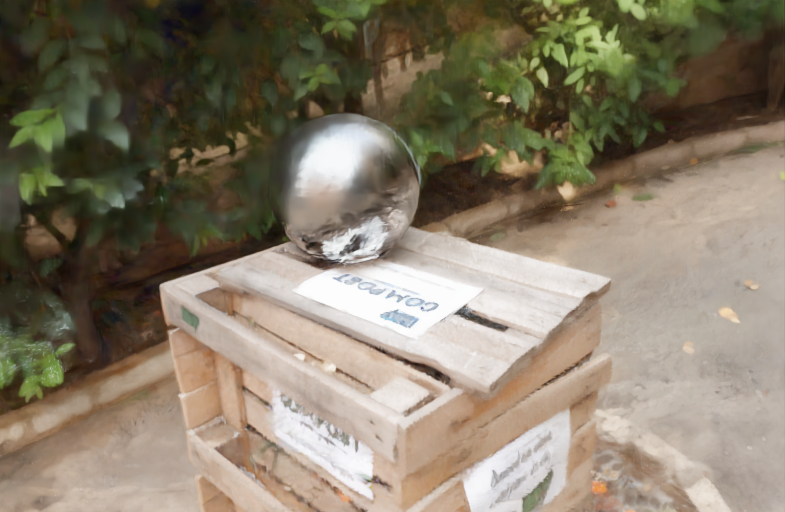} &
    \includegraphics[width=\mywidth]{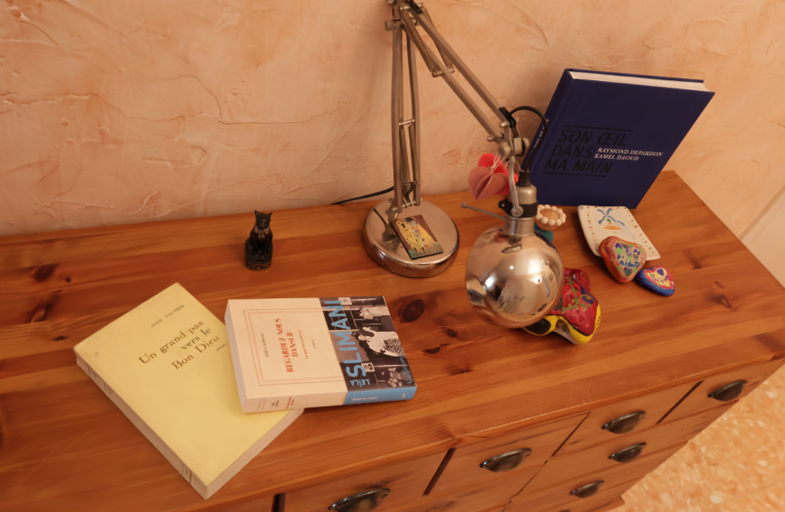} &
    \includegraphics[width=\mywidth]{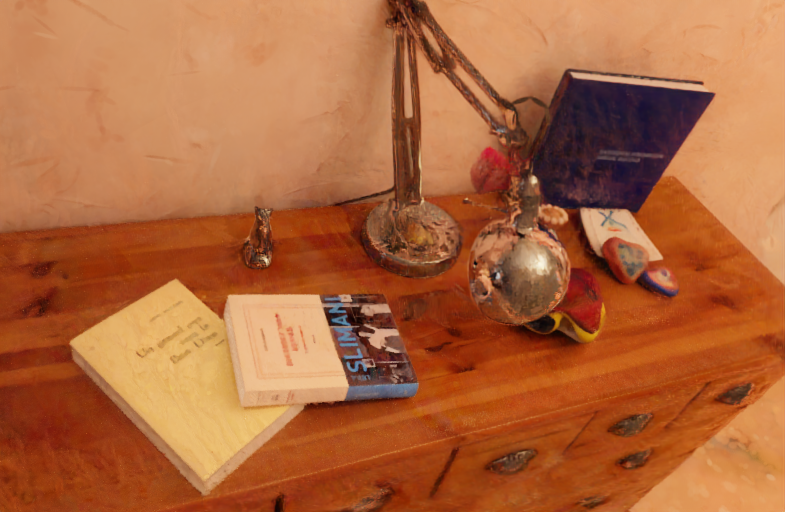} &
    \includegraphics[width=\mywidth]{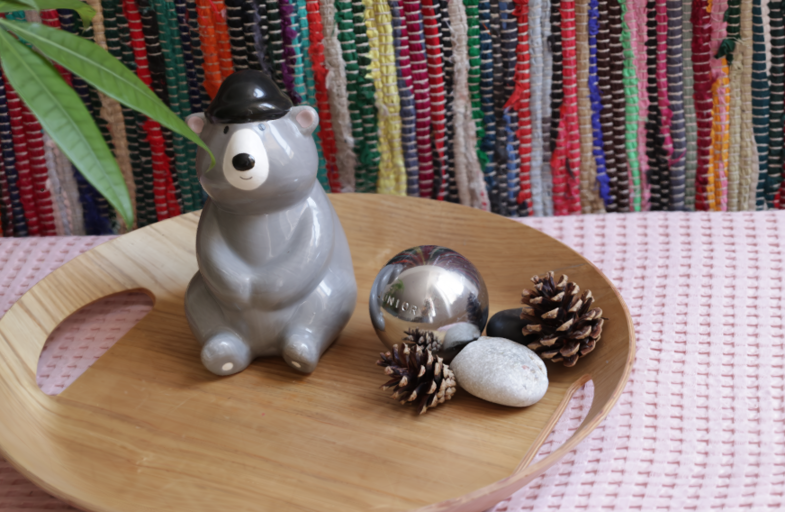} &
    \includegraphics[width=\mywidth]{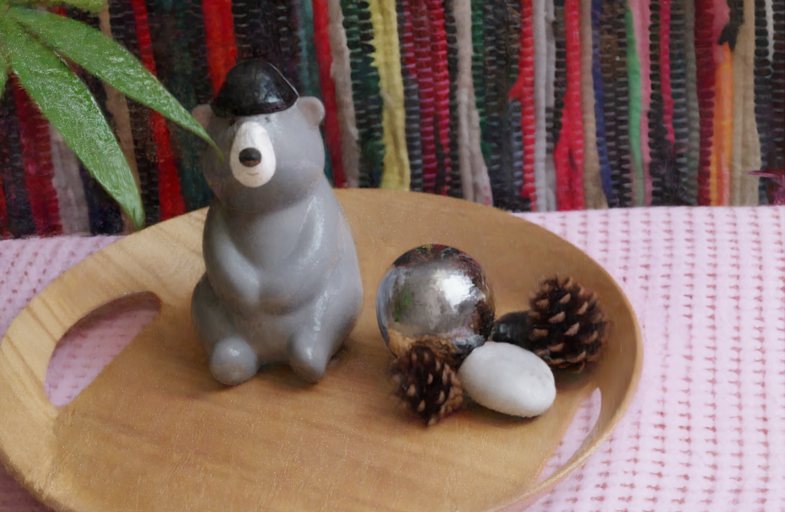} \\
    \end{tabular}
    \vspace{-10pt}
    \caption[]{Our method compared against EnvGS~\cite{xie2025envgs} on real scenes from the Neural Catacaustics dataset~\cite{kopanas2022neural}. Qualitatively, we achieve better disentanglement at the cost of lower visual fidelity in the final render. \added{Note that the denoiser was only applied to the final image.}}
    \label{fig:ours_vs_envgs}
\end{figure*}

\clearpage
\appendix
\section*{Supplemental Material}
In this document, we present additional details of the algorithm and complete
sets of results for the various experiments and ablations. Curated versions of some of these results were
presented in the main paper.

\section{Method details}

In this section we present additional details of the method.

\subsection{Gaussian Evaluation using OBBs}

We use oriented bounding boxes instead of polygons to bound Gaussians. The OBBs are emulated using instancing in a two-level acceleration structure~\cite{hardware-ray-transforms}: the bottom-level contains just a single unit AABB while the top-level contains one copy of this box per gaussian instanced with its local-to-world matrix. To compute intersections, Optix maps world-space rays to local coordinates, letting us reuse the local coordinate rays for easy gaussian evaluation; we never form covariance matrices. 

\begin{figure}[!h]
    \centering
    \includegraphics[width=0.8\linewidth]{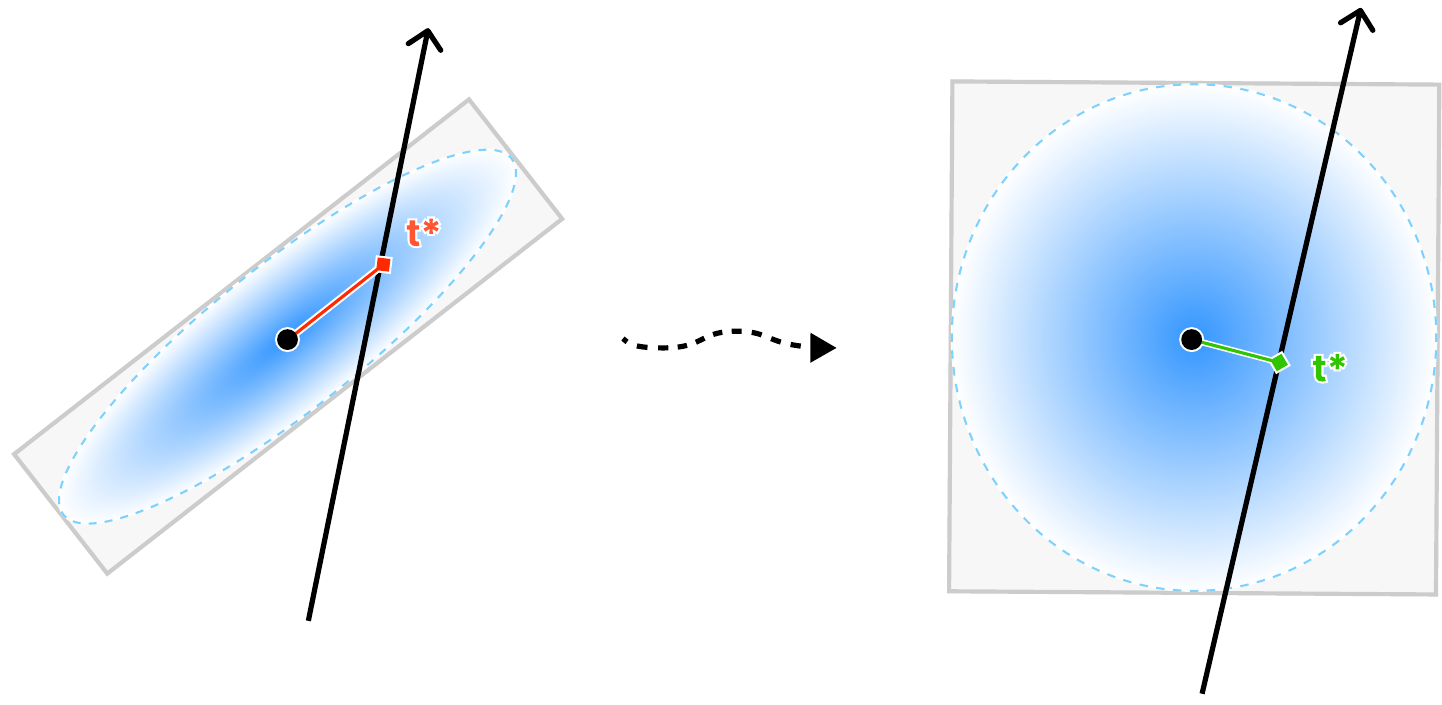}
    \caption{Our method leverages hardware-accelerated transforms, providing us with rays in local coordinates ``for free''. We can reuse these transformed rays to cheaply evaluate Gaussians by finding the same point $t^\star$ in local coordinates.}
    \label{fig:raytracing_eval}
\end{figure}

OptiX performs the required inversion natively in hardware and maps the ray to local coordinates to perform the bounding-box test. This has the added benefit of letting us simplify the Gaussian evaluation since we can reuse the local-space ray origin $\boldsymbol{o}_L$ and direction $\boldsymbol{d}_L$. As shown in \cref{fig:raytracing_eval}, we evaluate the Gaussians at the same point in local coordinates where finding the peak simplifies to 
\begin{equation}
    t^{\star}_L = \boldsymbol{o}_L^T \hat{\boldsymbol{d}}_L \,,
\end{equation}
and with the local-coordinate peak $\boldsymbol{x}^\star_L = \boldsymbol{o}_L + t^{\star}_L \hat{\boldsymbol{d}}_L$, our final expression for the alpha value simplifies to  
\begin{equation}
\alpha = o \cdot \exp(-\frac{1}{2}||\boldsymbol{x}^\star_L||^{2p})\,,
\end{equation}
with $p$ the exponent of a generalized Gaussian (see Eq.~(9) from \cite{moenne20243d}). 
In doing so, we never form covariance matrices explicitly nor do we invert the Gaussian transforms ourselves in the forward pass.

\subsection{Aggressive Primitive Truncation}
\label{sec:sup-truncation}
Even though a Gaussian's color is only dependent on the other ones in front of it, its alpha gradients are dependent on all Gaussians behind it since changing the alpha value can reveal or hide them. To make training with higher transmittance thresholds possible, we use an estimate of the contribution of all truncated Gaussians. 

Specifically, we remark that the final transmittance $\tau_f$ can always be evaluated cheaply even during an out-of-order BVH traversal since multiplication commutes. Using this value, the final accumulated color $c = c_{front}$ can instead be rendered more accurately as $c' = c_{kept} + (\tau_i - \tau_f) \hat{c}_{back}$ where $\tau_i$ is the transmittance value after passing the transmittance threshold,  $c_{front}$ is the color accumulated from the Gaussians before the threshold is kept, and  $\hat{c}_{back}$ estimates the color that would have been accumulated from the remaining truncated Gaussians. We differentiate through this expression but not through the color estimation i.e. $\hat{c}_{back}$ is detached.

We tried different more complicated estimators for $\hat{c}_{back}$ in our initial experiments (like stochastic sampling and out-of-order transparency) but found that surprisingly, setting $\hat{c}_{back} = c_{front}$ worked well enough to stabilize training while also providing a small improvement to the final pixel color. 

\subsection{Dense Initialization}
\label{sec:sup-dense}
Inspired by EDGS~\cite{kotovenko2025edgs}, rather than use SFM-based initialization, we opt to leverage our predicted depth maps to create a dense initialization. These are aligned with COLMAP points, downscaled, backprojected, and binned in 3D. We use all bins with counts greater than some threshold for initialization. We aim for 250k to 2M points, depending on scene size, \added{to try and roughly match the baselines}. \added{Dense initialization provides a small (<1 dB average) PSNR improvement as shown in~\cref{tbl:dense-init}.}
\begin{table}[h]
\caption{\added{PSNR (dB) comparison between dense initialization and SfM (no densification).}}
\centering
\added{
\begin{tabular}{lcc}
\toprule
Scene & Dense & SfM (no densification) \\
\midrule
Shiny Kitchen    & 27.01 & 26.25 \\
Shiny Livingroom & 26.87 & 26.47 \\
Shiny Office     & 27.49 & 26.39 \\
\midrule
Average          & 27.12 & 26.37 \\
\bottomrule
\end{tabular}
}
\label{tbl:dense-init}
\end{table}

\subsection{Weight-based Pruning}
We create far-field Gaussians randomly, many of which are never seen by any training view directly or indirectly. To ensure these Gaussians are pruned, we replace 3DGS' opacity pruning with weight-based pruning, by accumulating the total per-gaussian integration weight at each step. Gaussians that have a total weight (summed across all pixels) of less than $0.1$ on average are pruned every 500 steps. 

\section{Implementation Details}

In this section, we present details of the implementation.

\subsection{Interactive Rendering}
We implement temporal accumulation by reusing samples across frames when the scene remains static, followed by denoising to produce stable images. To reduce artifacts such as surface acne, reflection rays ignore backfacing Gaussians (that have an assigned normal pointing an opposite direction from the ray) for a small initial distance along the ray (0.1 meters).

\subsection{Floater Removal}
Our naive initialization for far-field Gaussians can result in floaters since many of these are initialized very close to cameras. In the synthetic scenes, we resolved this by placing a sphere around each camera and culling Gaussians when their centroid moves inside this sphere. We perform this culling every 500 iterations. The sphere radii are set to $0.8$ times the distance to the nearest point computed from the input depth map. At inference, we also set a znear plane at a fixed distance across all cameras (1 meter in the synthetic scenes).

\subsection{Scale Decay}
Raytracing performance is strongly dependent on reducing the total number of ray-gaussian intersections; to favor smaller Gaussians whenever possible we decay every Gaussian's scale by a factor of $0.999$ every iteration.

\subsection{Avoiding Invalid Reflections}
The Gaussian primitive reconstruction of the scene is not always well adapted to ray-tracing. In particular, the accumulated normal can sometimes be incorrect since the orientation of Gaussians traversed by a given ray may not be consistent, or simply the ray traverses object edges. The latter case results in flickering during rendering. Consequently, we consider normals with length smaller than $0.6$ before normalization as invalid and terminate the path.

\subsection{Reconstruction through Rough Reflections}
We found reconstructing the environment through rough reflectors challenging. Since random sampling leads to gradients of high variance when roughness values are very high, for reflection rays we found it beneficial to scale down gradients by a factor of $(1-\rho)^3$ where $\rho$ is the roughness of the surface the ray originates from in order to prioritize gradients of higher quality. 

\subsection{Tone Mapping and Ranges in Network Training}

For tone mapping in network training, we define a pair of functions: \texttt{tonemap(x)} maps values from $[0, \infty)$ to $[0, 1]$, and \texttt{untonemap(x)} reverses the mapping. We use a filmic tone mapping function rather than the standard Reinhard function ($\text{reinhard}(x) = x / (1 + x)$). All predicted RGB values are modeled using the equation $\text{RGB} = A \cdot I + R$, where $A$ is albedo, $I$ is irradiance, and $R$ is the residual. Because linear EXR files contain a wide range of values, we first normalize them with scaling factor $k$ prior to tone mapping. Additionally, we randomly vary $k$ for data augmentation. 
\[
k \cdot \text{RGB} = A \cdot (k \cdot I) + (k \cdot R),
\]

When interfacing with models like SD2, which require input and output in the range $[-1, 1]$, we first map all buffers into $[0, 1]$ using the tone mapping described above, then apply a linear transformation $x \mapsto 2x - 1$ to match the expected range. RGB, irradiance, diffuse, and glossy components are all tone-mapped. In contrast, albedo, roughness, metalness, and normals are kept in their native format, without tone mapping. For depth, we predict affine-invariant values. These are accurate up to an unknown scale and offset, which are normalized using the 2nd and 98th percentiles.

\subsection{Tone Mapping in Comparisons}

Our method operates directly in linear space. For all baselines, we undo the filmic tonemapping curve in the diffuse and final images, and compute the reflection pass as the (clamped) residual in linear space. All reflections are presented with tonemapping reapplied. 

\subsection{Scene Editing}

We compute selections by placing bounding boxes manually. In the synthetic scenes, we use the blend files to position these boxes, and in the real scenes, place them by using the point cloud as references. The points within the bounds are then filtered based on attached properties e.g. retaining all points with $F_0 \in [0.2, 0.4]$. In the real scene certain materials were edited prior to starting recording (removed excessive reflections from the background and moved the plant out of view). In real scenes, we also make the assumption of white reflectors.

\subsection{Indirect Reconstruction}

In the scene showcasing indirect reconstruction of a book cover (\cref{fig:method-overview-2}), $5000$ small Gaussians (scale $0.001$) were randomly initialized in a bounding box coarsely placed around the book, and added at the same time as the other far-field points. These points were assigned a random initial color and this was done for both our method and 3DGS.

\section{Additional Results}

We present extensive results and comparisons.

\subsection{Tinted Chrome Spheres}

\added{We compare to EnvGS on the scene with three tinted chrome spheres shown in \cref{fig:separation}. As expected, performances are higher across the board on this simple scene with large, smooth reflectors which provide clear view of the reflected scene.}

\begin{table}[h]
\caption{\added{PSNR comparison on the tinted chrome spheres scene.}}
\centering
\added{
\begin{tabular}{lccc}
\toprule
Method & Diffuse & Specular & Render \\
\midrule
EnvGS\textsubscript{(g.t. normals)} & 15.32 & 16.93 & \cellcolor{orange!25}{35.24} \\
Ours\textsubscript{(g.t. inputs)}     & \cellcolor{orange!25}{37.86} & \cellcolor{orange!25}{30.51} & 30.29 \\
\bottomrule
\end{tabular}

}
\label{tab:chromeball-scene}
\end{table}

\subsection{Full Disentanglement Results}
\label{sec:sup-disentangle}

Although previous methods achieve high image quality, they exhibit poor disentanglement between diffuse and \added{specular} components; results are presented in \Cref{tab:disentanglement}. With ground truth \added{buffers}, our method maintains the disentanglement while achieving good final image quality, which is critical for real-time editing applications. The next best method, EnvGS, cannot achieve a good disentanglement for highly reflective objects even with ground truth normals. However, when rendering using only our network predicted \added{buffers}, the final PSNR is low, highlighting a limitation of our approach.

\subsection{Full Performance Statistics}
\label{sec:sup-perf}

In \Cref{tab:time-all} we show optimization times for all methods on an A6000,
\Cref{tab:num_gaussians} the final number of Gaussian primitives, and in
\Cref{tab:fps} the frames per second for rendering of each method. All results were measured at resolution $1024 \times 768$.

Note that not all methods train for the same number of iterations. We trained all baselines as-is, and our method for 8k iterations; ours benefits from faster convergence due to dense initialization and from the absence of spherical harmonics. Since our metrics plateau rapidly we trained for 8k iteration only, aiming to approximately match 3DGS's training times.

\subsection{Comparisons with 3DGRT}
\label{sec:sup-3dgrt}

We provide details of the configuration used for the comparison.
We did not implement spherical harmonics and used just the L1 loss at a weight of 1.0 since our method does not support SSIM. Like 3DGRT we used an alpha threshold of 0.0113, a generalized kernel exponent of 2 and a transmittance threshold of $0.01$ at both training and inference. We implemented their densification based on scaled 3D positional gradients but only split gaussians without cloning. We reset opacity to 0.1 instead of 0.01 and trained for 7k iterations on the Shiny (synthetic) scenes at resolution 768x512 and on the MipNeRF scenes. We adjusted densification thresholds to roughly match 3DGRT's number of Gaussians; we used 0.002 for the Shiny scenes and 0.0015 for MipNerf scenes, leaving all other hyperparameters untouched from 3DGS. Results are presented in \Cref{tab:3DGRT} and \Cref{tab:shiny}.

\subsection{Sparse Views}
\label{sec:sup-sparse}

\begin{figure}[!h]
    \centering
    \includegraphics[width=1.0\linewidth]{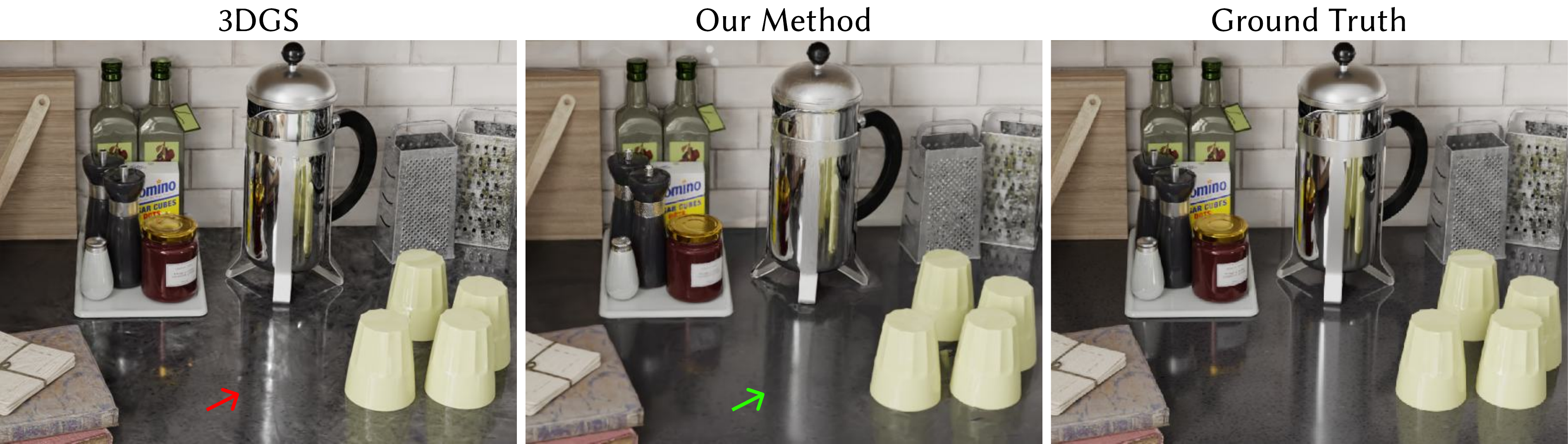}
    \label{fig:sparse-views}
    \caption{Training with just 4 views still results in clean reflections provided a good point cloud initialization.}
\end{figure}

In \Cref{fig:sparse-views} we show that with a small number of input views (just 4 views), our method produces better quality results for reflections than 3DGS. Note that these results use the point cloud obtained from all 200 views at initialization. 

\subsection{Network Ablations}
\label{sec:sup-network-ablations}
In this section we show a set of ablations on the performance of our method by replacing GT \added{buffers} with network \added{buffers} to see the effect. The results are shown in \Cref{tab:network-ablations}.
From top to bottom, NONE is all GT, DISENTANG. is with network diffuse/specular \added{buffers}, BRDF with network BRDF parameters, and combinations thereof.

\begin{table*}
\caption{
\label{tab:network-ablations}
We show results of our method where we replace different GT \added{buffers} with network prediction and show the result on quantitative metrics. From top to bottom, NONE is all GT, DISENTANG. is with network diffuse/\added{specular} \added{buffers}, BRDF with network BRDF parameters, and combinations thereof. }

\setlength{\tabcolsep}{8.0pt}
\subsubsection*{PSNR}
\begin{tabular}{llllllllll}
\toprule
 & 
 \multicolumn{3}{l}{Shiny Kitchen} & \multicolumn{3}{l}{Shiny Livingroom} & \multicolumn{3}{l}{Shiny Office} \\
 & \makebox[0pt][l]{\scriptsize Diffuse} & \makebox[0pt][l]{\scriptsize Reflective} & \makebox[0pt][l]{\scriptsize Final} & \makebox[0pt][l]{\scriptsize Diffuse} & \makebox[0pt][l]{\scriptsize Reflective} & \makebox[0pt][l]{\scriptsize Final} & \makebox[0pt][l]{\scriptsize Diffuse} & \makebox[0pt][l]{\scriptsize Reflective} & \makebox[0pt][l]{\scriptsize Final} \\
\midrule
NONE & 
\cellcolor{orange!25}31.66 & \cellcolor{orange!25}24.02 & \cellcolor{orange!25}26.69 & 28.35 & \cellcolor{orange!25}26.25 & \cellcolor{orange!25}26.21 & \cellcolor{orange!25}30.65 & \cellcolor{orange!25}24.08 & \cellcolor{orange!25}26.90 \\
DISENTAG.  & 
20.36 & 19.51 & 22.88 & 24.09 & 23.77 & 23.38 & 20.83 & 20.69 & 20.93 \\
BRDF & 
31.46 & 18.23 & 19.93 & \cellcolor{orange!25}28.36 & 22.09 & 22.80 & 30.55 & 20.87 & 23.24 \\
GEOM. & 
30.28 & 19.62 & 22.43 & 27.56 & 22.49 & 23.45 & 29.09 & 21.85 & 24.59 \\
DISENTAG.+GEOM. & 
20.33 & 17.85 & 20.75 & 23.76 & 21.62 & 21.75 & 20.62 & 20.05 & 20.44 \\
DISENTAG.+BRDF & 
20.32 & 16.40 & 18.25 & 24.06 & 20.56 & 20.74 & 20.84 & 19.03 & 19.52 \\
GEOM.+BRDF & 
29.90 & 16.21 & 17.83 & 27.43 & 19.88 & 20.82 & 28.84 & 20.55 & 22.40 \\
DISENTAG.+GEOM.+BRDF & 
20.33 & 15.18 & 17.04 & 23.75 & 19.25 & 19.67 & 20.57 & 19.35 & 20.19 \\
\bottomrule
\end{tabular}

\subsubsection*{SSIM}
\begin{tabular}{llllllllll}
\toprule
 & 
 \multicolumn{3}{l}{Shiny Kitchen} & \multicolumn{3}{l}{Shiny Livingroom} & \multicolumn{3}{l}{Shiny Office} \\
 & 
 \makebox[0pt][l]{\scriptsize Diffuse} & \makebox[0pt][l]{\scriptsize Reflective} & \makebox[0pt][l]{\scriptsize Final} & \makebox[0pt][l]{\scriptsize Diffuse} & \makebox[0pt][l]{\scriptsize Reflective} & \makebox[0pt][l]{\scriptsize Final} & \makebox[0pt][l]{\scriptsize Diffuse} & \makebox[0pt][l]{\scriptsize Reflective} & \makebox[0pt][l]{\scriptsize Final} \\
\midrule
NONE & 
\cellcolor{orange!25}0.936 & \cellcolor{orange!25}0.772 & \cellcolor{orange!25}0.877 & \cellcolor{orange!25}0.922 & \cellcolor{orange!25}0.747 & \cellcolor{orange!25}0.849 & \cellcolor{orange!25}0.924 & \cellcolor{orange!25}0.750 & \cellcolor{orange!25}0.874 \\
DISENTAG. & 
0.722 & 0.624 & 0.814 & 0.718 & 0.663 & 0.792 & 0.751 & 0.627 & 0.802 \\
BRDF & 
0.932 & 0.559 & 0.782 & 0.918 & 0.682 & 0.806 & 0.921 & 0.680 & 0.838 \\
GEOM. & 
0.920 & 0.552 & 0.772 & 0.900 & 0.624 & 0.760 & 0.895 & 0.640 & 0.815 \\
DISENTAG.+GEOM. & 
0.712 & 0.492 & 0.732 & 0.711 & 0.591 & 0.729 & 0.735 & 0.567 & 0.760 \\
DISENTAG.+BRDF & 
0.719 & 0.493 & 0.741 & 0.717 & 0.598 & 0.750 & 0.751 & 0.579 & 0.777 \\
GEOM.+BRDF & 
0.915 & 0.407 & 0.711 & 0.891 & 0.588 & 0.735 & 0.892 & 0.595 & 0.793 \\
DISENTAG.+GEOM.+BRDF & 
0.709 & 0.399 & 0.690 & 0.709 & 0.554 & 0.696 & 0.732 & 0.532 & 0.746 \\
\bottomrule
\end{tabular}

\subsubsection*{LPIPS}
\begin{tabular}{llllllllll}
\toprule
 & 
 \multicolumn{3}{l}{Shiny Kitchen} & \multicolumn{3}{l}{Shiny Livingroom} & \multicolumn{3}{l}{Shiny Office} \\
 & 
\makebox[0pt][l]{\scriptsize Diffuse} & \makebox[0pt][l]{\scriptsize Reflective} & \makebox[0pt][l]{\scriptsize Final} & \makebox[0pt][l]{\scriptsize Diffuse} & \makebox[0pt][l]{\scriptsize Reflective} & \makebox[0pt][l]{\scriptsize Final} & \makebox[0pt][l]{\scriptsize Diffuse} & \makebox[0pt][l]{\scriptsize Reflective} & \makebox[0pt][l]{\scriptsize Final} \\
\midrule
NONE &  
\cellcolor{orange!25}0.090 & \cellcolor{orange!25}0.201 & \cellcolor{orange!25}0.123 & \cellcolor{orange!25}0.119 & \cellcolor{orange!25}0.255 & \cellcolor{orange!25}0.206 & \cellcolor{orange!25}0.093 & \cellcolor{orange!25}0.245 & \cellcolor{orange!25}0.126 \\
DISENTAG. & 
0.208 & 0.274 & 0.170 & 0.239 & 0.286 & 0.272 & 0.202 & 0.322 & 0.195 \\
BRDF & 
0.096 & 0.351 & 0.188 & 0.127 & 0.286 & 0.227 & 0.096 & 0.307 & 0.145 \\
GEOM. & 
0.108 & 0.389 & 0.210 & 0.149 & 0.398 & 0.270 & 0.118 & 0.385 & 0.177 \\
DISENTAG.+GEOM. & 
0.221 & 0.428 & 0.242 & 0.261 & 0.407 & 0.322 & 0.223 & 0.433 & 0.237 \\
DISENTAG.+BRDF & 
0.210 & 0.363 & 0.214 & 0.242 & 0.327 & 0.289 & 0.203 & 0.360 & 0.210 \\
GEOM.+BRDF & 
0.112 & 0.490 & 0.249 & 0.163 & 0.412 & 0.289 & 0.122 & 0.425 & 0.192 \\
DISENTAG.+GEOM.+BRDF & 
0.225 & 0.494 & 0.268 & 0.266 & 0.433 & 0.338 & 0.226 & 0.462 & 0.250 \\
\bottomrule
\end{tabular}

\end{table*}

\subsection{Full Comparisons with Other Methods for Disentanglement}

\begin{figure*}[t]
\centering
\renewcommand{\arraystretch}{0.5}
\begin{tabular}{@{\hskip 2pt}c@{\hskip 2pt}c@{\hskip 2pt}c@{\hskip 2pt}c@{\hskip 2pt}}
& \textbf{Diffuse} & \textbf{Residual} & \textbf{Final Render} \\
\rotatebox{90}{GShader} &
  \includegraphics[width=0.20\linewidth]{shiny_kitchen/gaussian_shader_diffuse.jpg} &
  \includegraphics[width=0.20\linewidth]{shiny_kitchen/gaussian_shader_glossy.jpg} &
  \includegraphics[width=0.20\linewidth]{shiny_kitchen/gaussian_shader_render.jpg} \\
\rotatebox{90}{3DGS-DR} &
  \includegraphics[width=0.20\linewidth]{shiny_kitchen/3dgs_dr_diffuse.jpg} &
  \includegraphics[width=0.20\linewidth]{shiny_kitchen/3dgs_dr_glossy.jpg} &
  \includegraphics[width=0.20\linewidth]{shiny_kitchen/3dgs_dr_render.jpg} \\
\rotatebox{90}{ReflGS} &
  \includegraphics[width=0.20\linewidth]{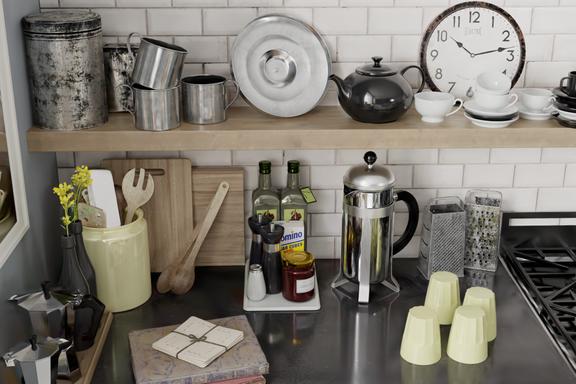} &
  \includegraphics[width=0.20\linewidth]{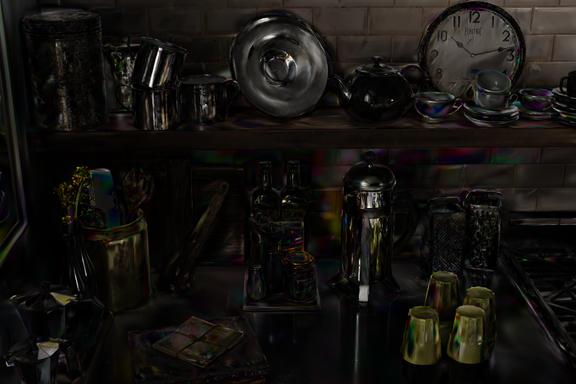} &
  \includegraphics[width=0.20\linewidth]{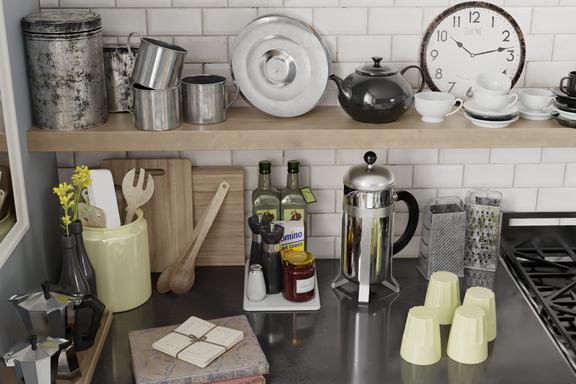} \\
\rotatebox{90}{EnvGS\added{\textsubscript{(net. normals)}}} &
  \includegraphics[width=0.20\linewidth]{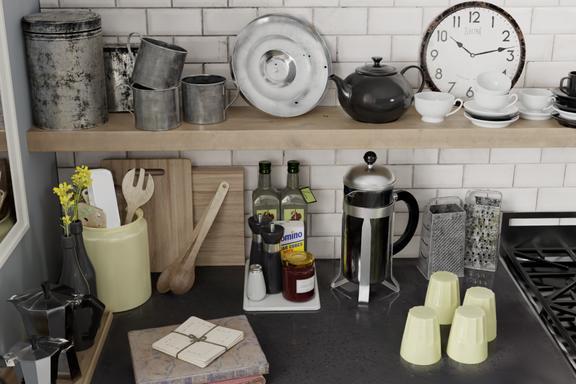} &
  \includegraphics[width=0.20\linewidth]{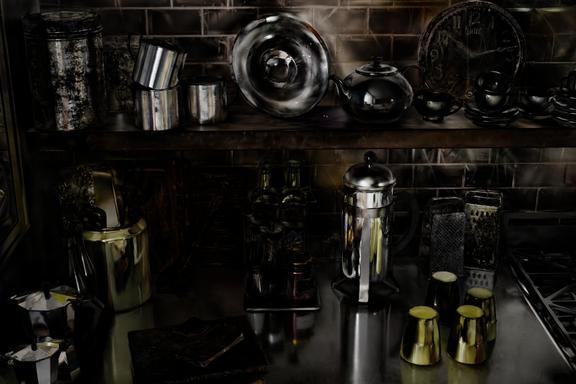} &
  \includegraphics[width=0.20\linewidth]{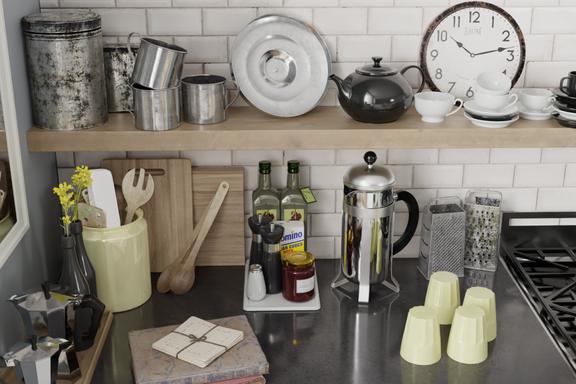} \\
\rotatebox{90}{EnvGS\added{\textsubscript{(g.t. normals)}}} &
  \includegraphics[width=0.20\linewidth]{shiny_kitchen/envgs_gt_diffuse.jpg} &
  \includegraphics[width=0.20\linewidth]{shiny_kitchen/envgs_gt_glossy.jpg} &
  \includegraphics[width=0.20\linewidth]{shiny_kitchen/envgs_gt_render.jpg} \\
\rotatebox{90}{Ours\added{\textsubscript{(net. inputs)}}} &
  \includegraphics[width=0.20\linewidth]{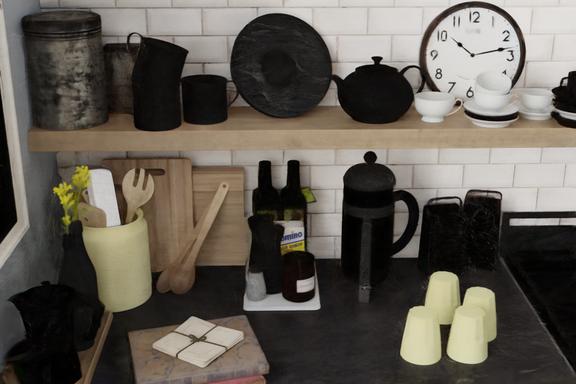} &
  \includegraphics[width=0.20\linewidth]{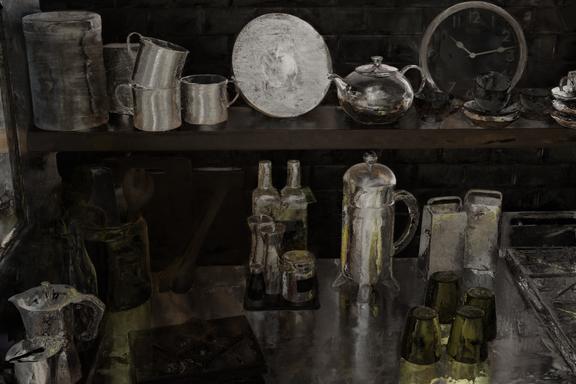} &
  \includegraphics[width=0.20\linewidth]{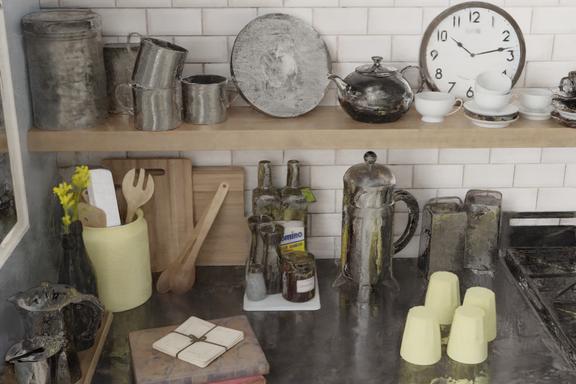} \\
\rotatebox{90}{Ours\added{\textsubscript{(g.t. inputs)}}} &
  \includegraphics[width=0.20\linewidth]{shiny_kitchen/ours_diffuse.jpg} &
  \includegraphics[width=0.20\linewidth]{shiny_kitchen/ours_glossy.jpg} &
  \includegraphics[width=0.20\linewidth]{shiny_kitchen/ours_render.jpg} \\
\rotatebox{90}{G.T.} &
  \includegraphics[width=0.20\linewidth]{shiny_kitchen/ground_truth_diffuse.jpg} &
  \includegraphics[width=0.20\linewidth]{shiny_kitchen/ground_truth_glossy.jpg} &
  \includegraphics[width=0.20\linewidth]{shiny_kitchen/ground_truth_render.jpg} \\
\end{tabular}
\caption{Disentanglement results for the \textsc{Shiny Kitchen} scene.}
\label{tab:image_grid_kitchen}
\end{figure*}

\begin{figure*}[t]
\centering
\renewcommand{\arraystretch}{0.5}
\begin{tabular}{@{\hskip 2pt}c@{\hskip 2pt}c@{\hskip 2pt}c@{\hskip 2pt}c@{\hskip 2pt}}
& \textbf{Diffuse} & \textbf{Residual} & \textbf{Final Render} \\
\rotatebox{90}{GShader} &
  \includegraphics[width=0.20\linewidth]{shiny_livingroom/gaussian_shader_diffuse.jpg} &
  \includegraphics[width=0.20\linewidth]{shiny_livingroom/gaussian_shader_glossy.jpg} &
  \includegraphics[width=0.20\linewidth]{shiny_livingroom/gaussian_shader_render.jpg} \\
\rotatebox{90}{3DGS-DR} &
  \includegraphics[width=0.20\linewidth]{shiny_livingroom/3dgs_dr_diffuse.jpg} &
  \includegraphics[width=0.20\linewidth]{shiny_livingroom/3dgs_dr_glossy.jpg} &
  \includegraphics[width=0.20\linewidth]{shiny_livingroom/3dgs_dr_render.jpg} \\
\rotatebox{90}{ReflGS} &
  \includegraphics[width=0.20\linewidth]{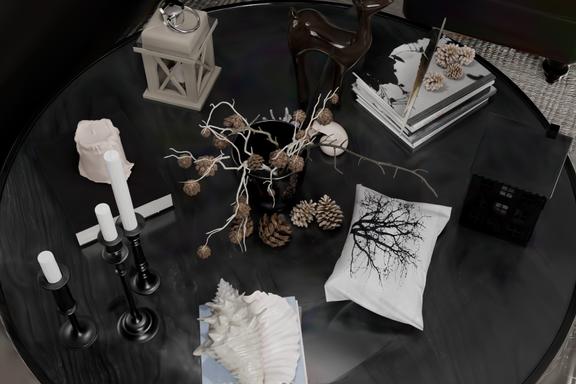} &
  \includegraphics[width=0.20\linewidth]{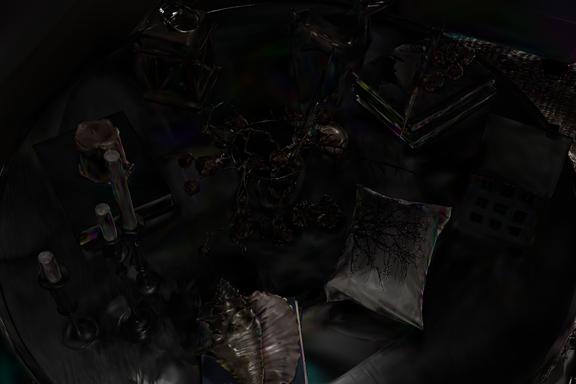} &
  \includegraphics[width=0.20\linewidth]{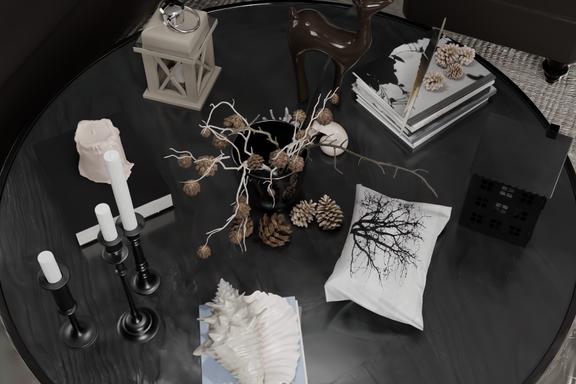} \\
\rotatebox{90}{EnvGS\added{\textsubscript{(net. normals)}}} &
  \includegraphics[width=0.20\linewidth]{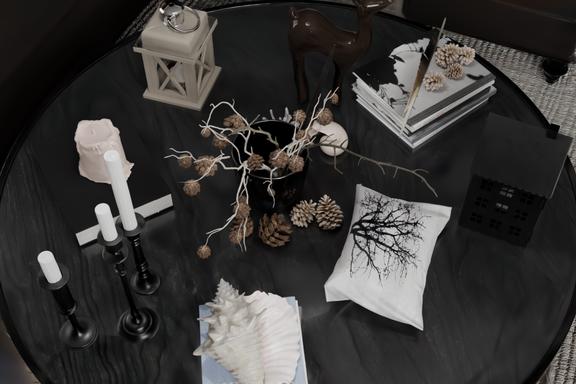} &
  \includegraphics[width=0.20\linewidth]{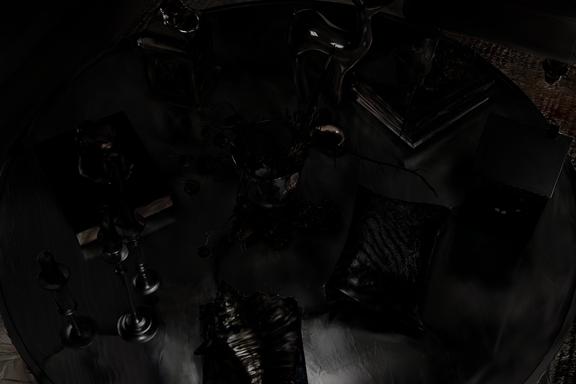} &
  \includegraphics[width=0.20\linewidth]{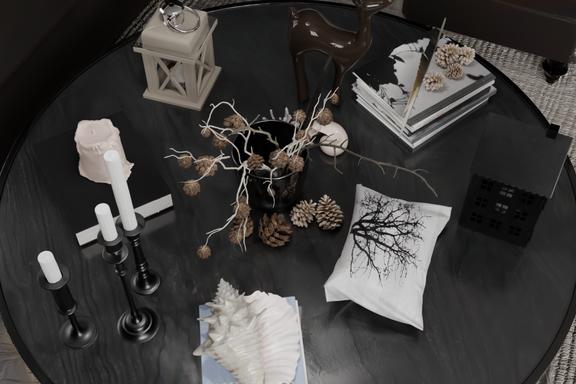} \\
\rotatebox{90}{EnvGS\added{\textsubscript{(g.t. normals)}}} &
  \includegraphics[width=0.20\linewidth]{shiny_livingroom/envgs_gt_diffuse.jpg} &
  \includegraphics[width=0.20\linewidth]{shiny_livingroom/envgs_gt_glossy.jpg} &
  \includegraphics[width=0.20\linewidth]{shiny_livingroom/envgs_gt_render.jpg} \\
\rotatebox{90}{Ours\added{\textsubscript{(net. inputs)}}} &
  \includegraphics[width=0.20\linewidth]{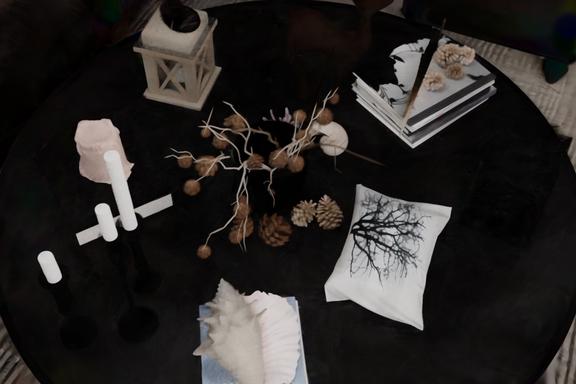} &
  \includegraphics[width=0.20\linewidth]{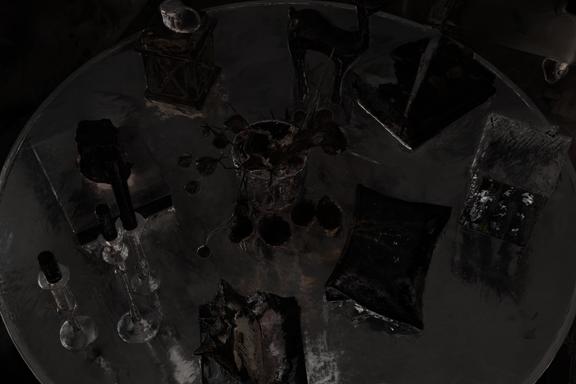} &
  \includegraphics[width=0.20\linewidth]{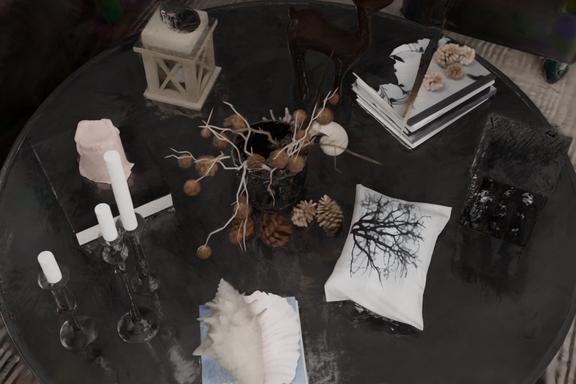} \\
\rotatebox{90}{Ours\added{\textsubscript{(g.t. inputs)}}} &
  \includegraphics[width=0.20\linewidth]{shiny_livingroom/ours_diffuse.jpg} &
  \includegraphics[width=0.20\linewidth]{shiny_livingroom/ours_glossy.jpg} &
  \includegraphics[width=0.20\linewidth]{shiny_livingroom/ours_render.jpg} \\
\rotatebox{90}{G.T.} &
  \includegraphics[width=0.20\linewidth]{shiny_livingroom/ground_truth_diffuse.jpg} &
  \includegraphics[width=0.20\linewidth]{shiny_livingroom/ground_truth_glossy.jpg} &
  \includegraphics[width=0.20\linewidth]{shiny_livingroom/ground_truth_render.jpg} \\
\end{tabular}
\caption{Disentanglement results for the \textsc{Shiny Livingroom} scene.}
\label{tab:image_grid_livingroom}
\end{figure*}

\begin{figure*}[t]
\centering
\renewcommand{\arraystretch}{0.5}
\begin{tabular}{@{\hskip 2pt}c@{\hskip 2pt}c@{\hskip 2pt}c@{\hskip 2pt}c@{\hskip 2pt}}
& \textbf{Diffuse} & \textbf{Residual} & \textbf{Final Render} \\
\rotatebox{90}{GShader} &
  \includegraphics[width=0.20\linewidth]{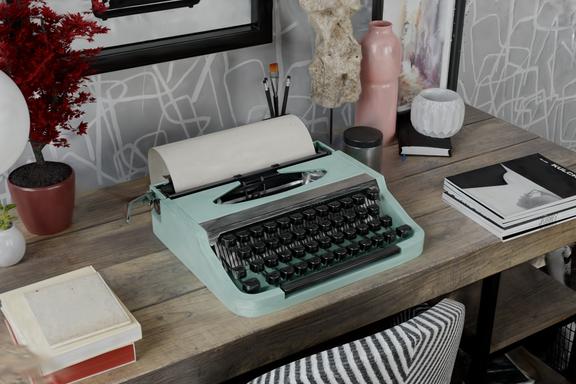} &
  \includegraphics[width=0.20\linewidth]{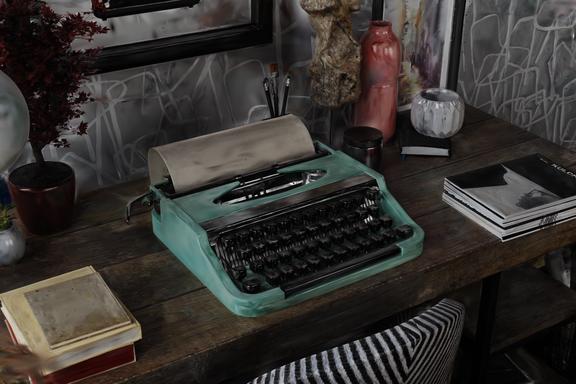} &
  \includegraphics[width=0.20\linewidth]{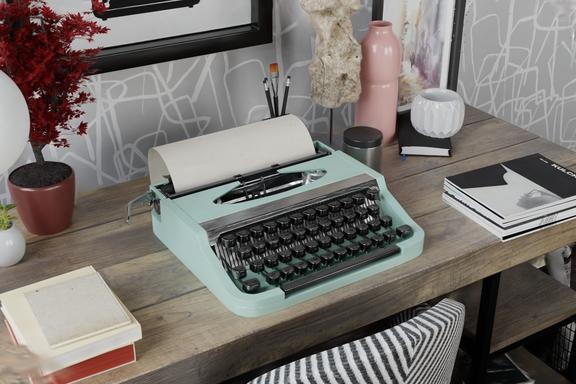} \\
\rotatebox{90}{3DGS-DR} &
  \includegraphics[width=0.20\linewidth]{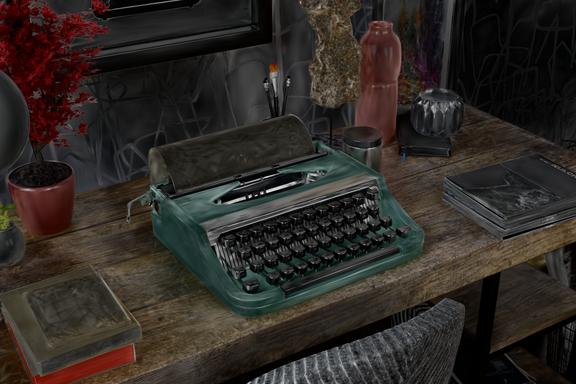} &
  \includegraphics[width=0.20\linewidth]{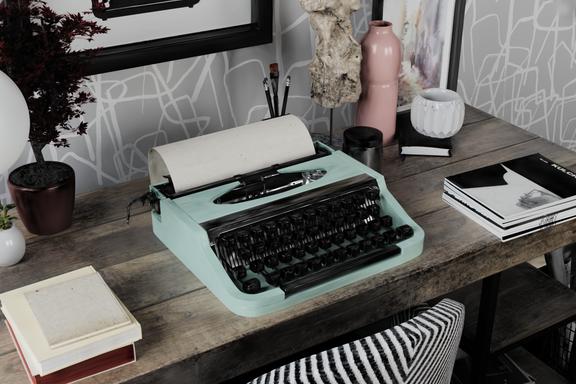} &
  \includegraphics[width=0.20\linewidth]{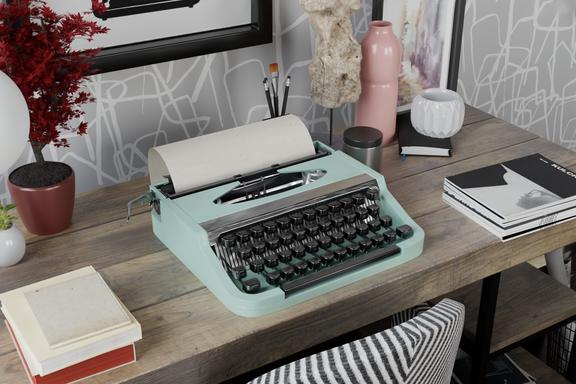} \\
\rotatebox{90}{ReflGS} &
  \includegraphics[width=0.20\linewidth]{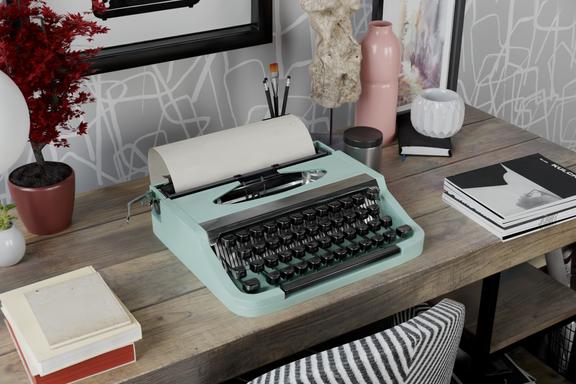} &
  \includegraphics[width=0.20\linewidth]{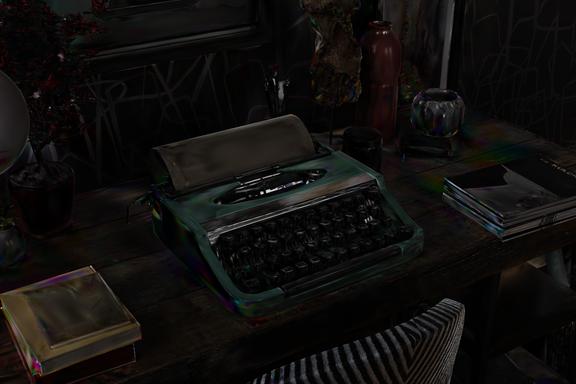} &
  \includegraphics[width=0.20\linewidth]{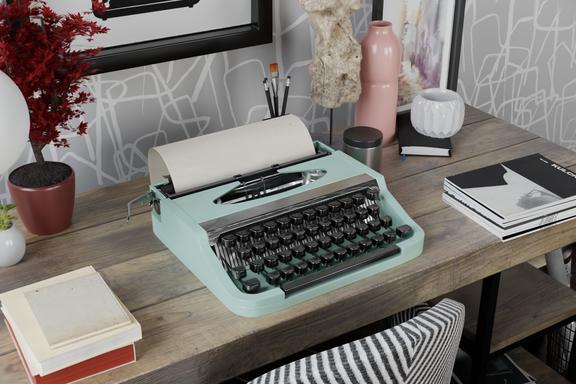} \\
\rotatebox{90}{EnvGS\added{\textsubscript{(net. normals)}}} &
  \includegraphics[width=0.20\linewidth]{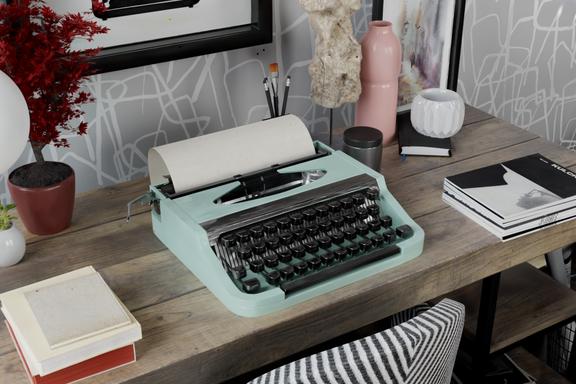} &
  \includegraphics[width=0.20\linewidth]{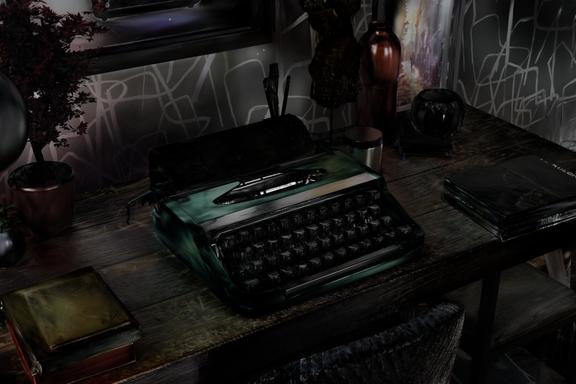} &
  \includegraphics[width=0.20\linewidth]{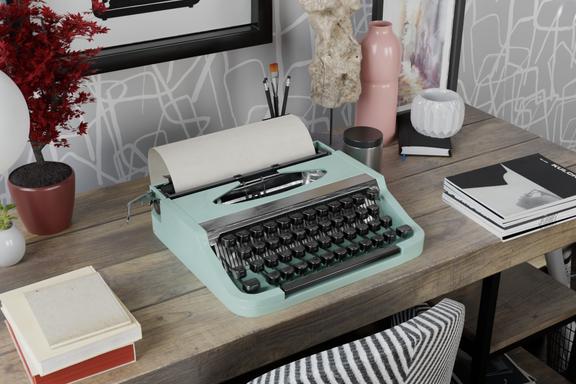} \\
\rotatebox{90}{EnvGS\added{\textsubscript{(g.t. normals)}}} &
  \includegraphics[width=0.20\linewidth]{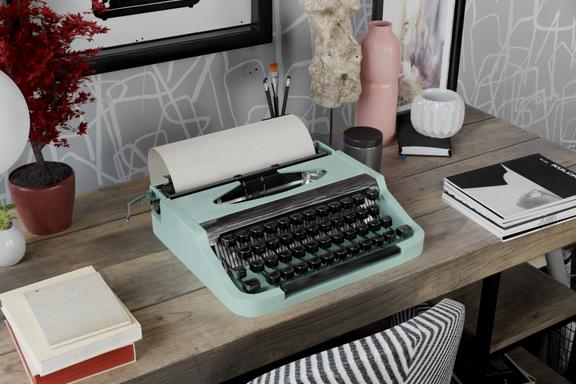} &
  \includegraphics[width=0.20\linewidth]{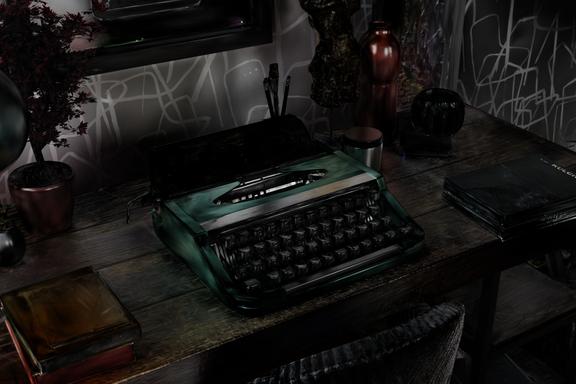} &
  \includegraphics[width=0.20\linewidth]{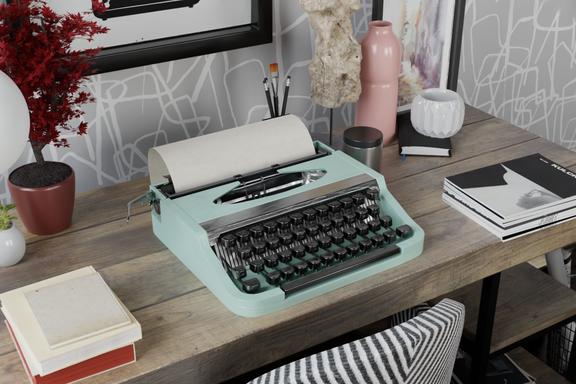} \\
\rotatebox{90}{Ours\added{\textsubscript{(net. inputs)}}} &
  \includegraphics[width=0.20\linewidth]{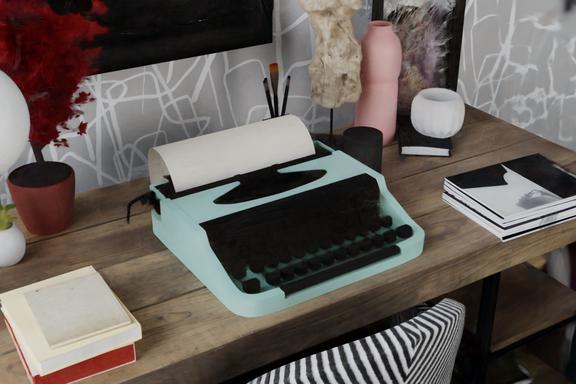} &
  \includegraphics[width=0.20\linewidth]{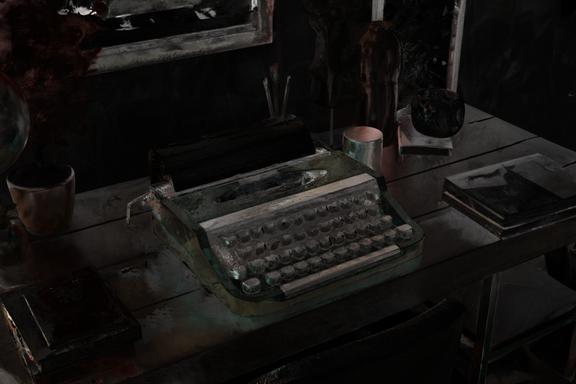} &
  \includegraphics[width=0.20\linewidth]{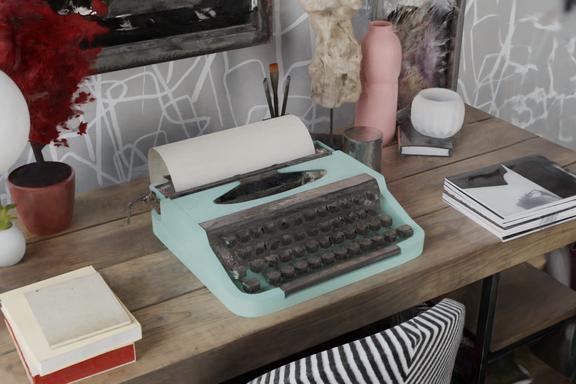} \\
\rotatebox{90}{Ours\added{\textsubscript{(g.t. inputs)}}} &
  \includegraphics[width=0.20\linewidth]{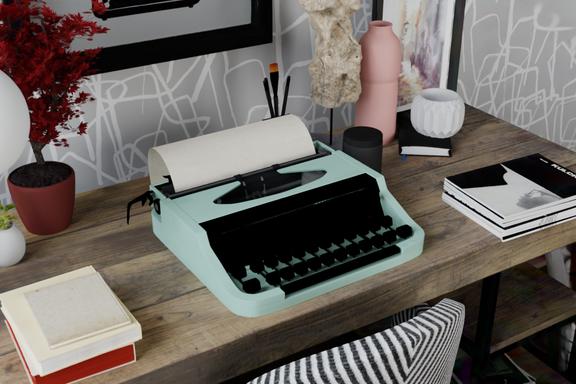} &
  \includegraphics[width=0.20\linewidth]{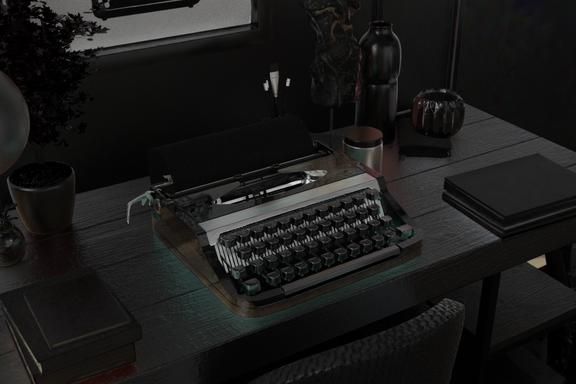} &
  \includegraphics[width=0.20\linewidth]{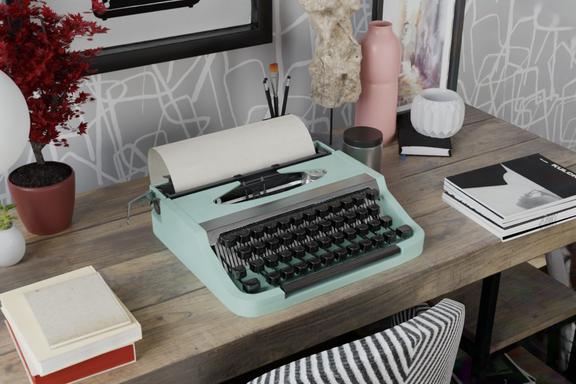} \\
\rotatebox{90}{G.T.} &
  \includegraphics[width=0.20\linewidth]{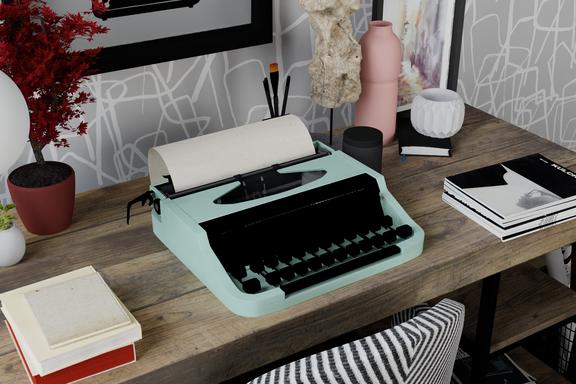} &
  \includegraphics[width=0.20\linewidth]{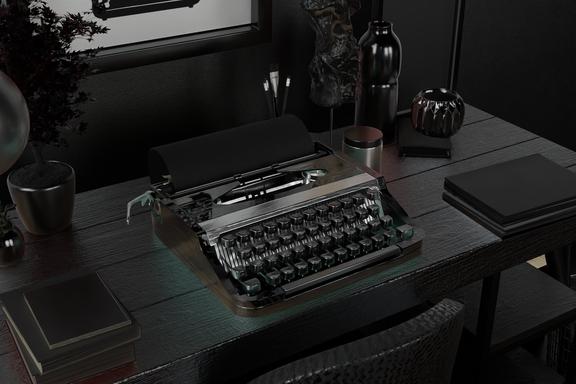} &
  \includegraphics[width=0.20\linewidth]{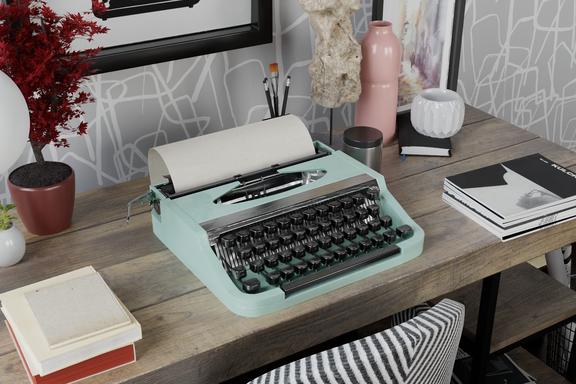} \\
\end{tabular}
\caption{Disentanglement results for the \textsc{Shiny Office} scene.}
\label{tab:image_grid_office}
\end{figure*}

In \cref{tab:image_grid_kitchen,tab:image_grid_livingroom,tab:image_grid_office} we show qualitative results for disentanglement for all
methods on the \textsc{Shiny Kitchen}, \textsc{Shiny Livingroom} and \textsc{Shiny Office} scenes respectively. We see that no other method achieves disentanglement between diffuse
and \added{specular} components. 

\begin{figure*}[h]
    \centering
    \includegraphics[width=0.5\linewidth]{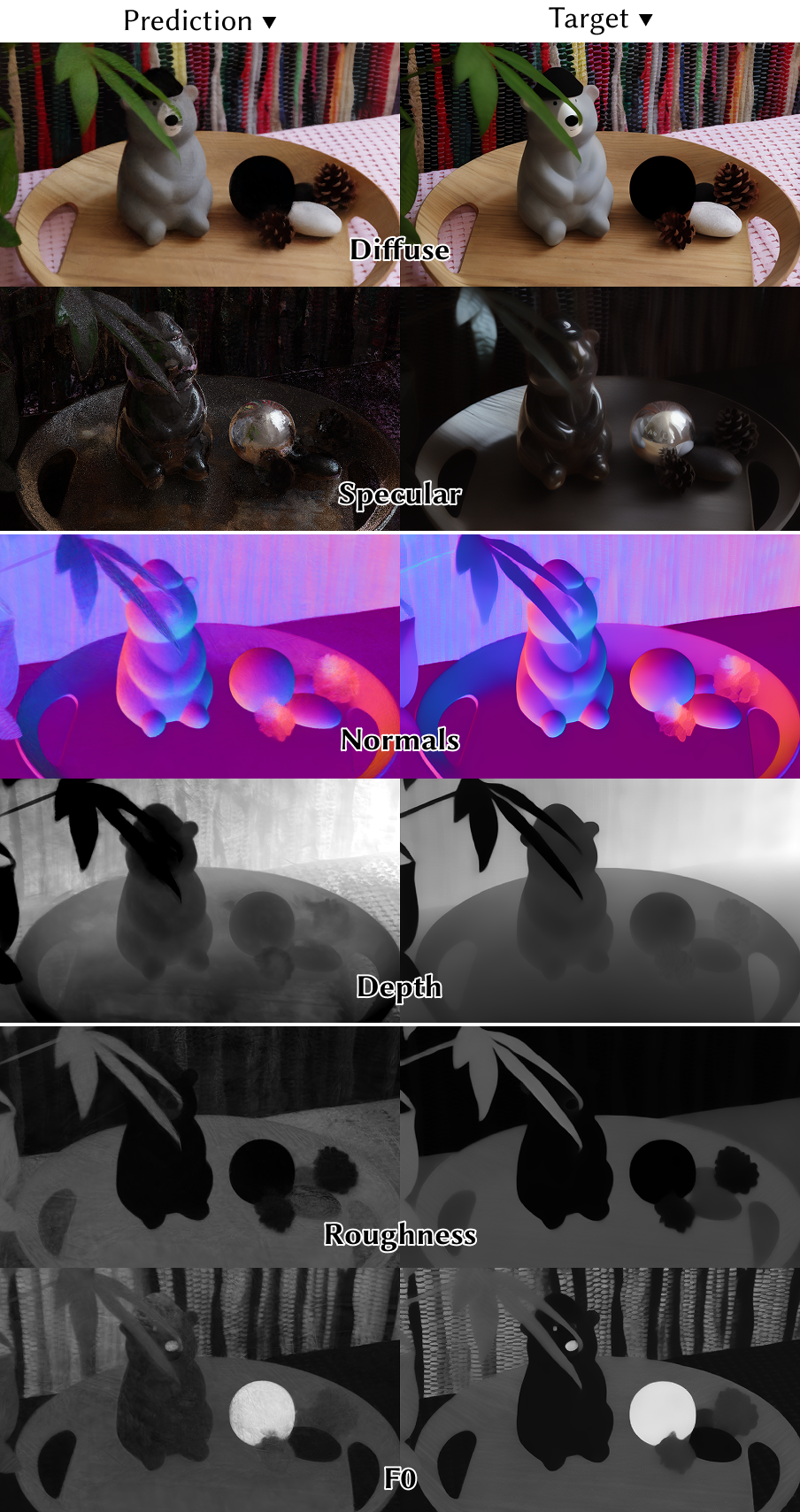}
    \caption{\added{Result of fitting our method (left) to network predicted buffers (right).}}
    \label{fig:real-scene-targets}
\end{figure*}

\added{
\subsection{Real Scene Fit Quality}
\Cref{fig:real-scene-targets} shows our method's fit quality against the network predicted buffers.
}

\begin{table*}
\caption{Disentanglement comparison across all methods.}

\setlength{\tabcolsep}{8.0pt}
\subsubsection*{PSNR}
\begin{tabular}{llllllllllll}
\toprule
 & \multicolumn{3}{l}{Shiny Kitchen} & \multicolumn{3}{l}{Shiny Livingroom} & \multicolumn{3}{l}{Shiny Office} \\
 & \makebox[0pt][l]{\scriptsize Diffuse} & \makebox[0pt][l]{\scriptsize Residual} & \makebox[0pt][l]{\scriptsize Final} & \makebox[0pt][l]{\scriptsize Diffuse} & \makebox[0pt][l]{\scriptsize Residual} & \makebox[0pt][l]{\scriptsize Final} & \makebox[0pt][l]{\scriptsize Diffuse} & \makebox[0pt][l]{\scriptsize Residual} & \makebox[0pt][l]{\scriptsize Final} \\
\midrule
3DGS & 13.11 & 13.17 & 32.83 & 17.01 & 17.35 & \cellcolor{orange!25}32.72 & 19.10 & 15.74 & \cellcolor{orange!25}34.91 \\
2DGS & 13.06 & 13.56 & 32.27 & 16.59 & 16.98 & 30.69 & 18.68 & 15.72 & 33.95 \\
Gaussian Shader & 14.88 & 12.15 & 30.13 & 20.57 & 16.91 & 24.26 & 21.77 & 15.56 & 32.75 \\
3DGS-DR & 10.10 &  8.59 & 32.79 & 16.08 & 12.97 & 31.39 &  9.74 &  8.96 & 34.66 \\
Reflective GS & 13.50 & 13.45 & 32.17 & 20.43 & 20.46 & 29.85 & 20.63 & 17.24 & 33.68 \\
EnvGS (network) & 14.33 & 14.48 & 32.82 & 22.74 & 21.22 & 30.65 & 20.92 & 16.33 & 34.16 \\
Ours (network) & 20.36 & 16.95 & 20.41 & 23.77 & 20.35 & 21.21 & 20.60 & 17.40 & 17.75 \\
EnvGS (optimal) & 14.51 & 15.15 & \cellcolor{orange!25}32.96 & 22.64 & 21.46 & 30.31 & 20.81 & 15.90 & 34.34 \\
Ours (optimal) & \cellcolor{orange!25}33.20 & \cellcolor{orange!25}24.30 & 26.96 & \cellcolor{orange!25}29.68 & \cellcolor{orange!25}26.46 & 26.96 & \cellcolor{orange!25}31.74 & \cellcolor{orange!25}24.48 & 27.54 \\
\bottomrule
\end{tabular}

\subsubsection*{SSIM}
\begin{tabular}{llllllllllll}
\toprule
 & \multicolumn{3}{l}{Shiny Kitchen} & \multicolumn{3}{l}{Shiny Livingroom} & \multicolumn{3}{l}{Shiny Office} \\
 & \makebox[0pt][l]{\scriptsize Diffuse} & \makebox[0pt][l]{\scriptsize Residual} & \makebox[0pt][l]{\scriptsize Final} & \makebox[0pt][l]{\scriptsize Diffuse} & \makebox[0pt][l]{\scriptsize Residual} & \makebox[0pt][l]{\scriptsize Final} & \makebox[0pt][l]{\scriptsize Diffuse} & \makebox[0pt][l]{\scriptsize Residual} & \makebox[0pt][l]{\scriptsize Final} \\
\midrule
3DGS & 0.676 & 0.205 & \cellcolor{orange!25}0.956 & 0.664 & 0.195 & \cellcolor{orange!25}0.934 & 0.797 & 0.169 & \cellcolor{orange!25}0.967 \\
2DGS & 0.658 & 0.189 & 0.946 & 0.640 & 0.187 & 0.910 & 0.780 & 0.128 & 0.961 \\
Gaussian Shader & 0.674 & 0.424 & 0.929 & 0.708 & 0.552 & 0.811 & 0.815 & 0.472 & 0.952 \\
3DGS-DR & 0.396 & 0.336 & 0.955 & 0.634 & 0.520 & 0.925 & 0.400 & 0.324 & 0.964 \\
Reflective GS & 0.691 & 0.351 & 0.946 & 0.757 & 0.553 & 0.909 & 0.814 & 0.458 & 0.957 \\
EnvGS (network) & 0.713 & 0.383 & 0.949 & 0.774 & 0.466 & 0.918 & 0.814 & 0.399 & 0.956 \\
Ours (network) & 0.709 & 0.428 & 0.699 & 0.707 & 0.537 & 0.694 & 0.732 & 0.464 & 0.715 \\
EnvGS (optimal) & 0.715 & 0.435 & 0.951 & 0.769 & 0.453 & 0.917 & 0.816 & 0.402 & 0.957 \\
Ours (optimal) & \cellcolor{orange!25}0.948 & \cellcolor{orange!25}0.790 & 0.888 & \cellcolor{orange!25}0.939 & \cellcolor{orange!25}0.758 & 0.867 & \cellcolor{orange!25}0.943 & \cellcolor{orange!25}0.760 & 0.892 \\
\bottomrule
\end{tabular}

\centering
\label{tab:disentanglement}
\end{table*}

            \begin{table*}[!h]
            \centering
            \caption{Training times}
            \begin{tabular}{llllllllll}
\toprule
 & 3DGS & 2DGS & GShader & 3DGS-DR & ReflGS & EnvGS (net) & EnvGS (opt) & Ours (net) & Ours (opt) \\
\midrule
Shiny Kitchen     & \cellcolor{orange!25}00:17:56 & 00:21:03 & 02:24:26 & 00:44:09 & 01:28:57 & 03:21:47 & 03:34:27 & 00:48:52 & 00:23:16 \\
Shiny Livingroom  & \cellcolor{orange!25}00:20:23 & 00:29:55 & 02:28:28 & 00:45:45 & 01:28:38 & 02:33:45 & 02:40:52 & 00:23:56 & 00:20:48 \\
Shiny Office      & 00:27:13 & 00:25:38 & 02:31:09 & 00:51:41 & 01:43:22 & 03:13:57 & 03:13:21 & 00:37:05 & \cellcolor{orange!25}00:24:13 \\
\bottomrule
\end{tabular}

            \label{tab:time-all}
            \end{table*}

            \begin{table*}[!h]
            \centering
            \caption{Final number of gaussians}
            \begin{tabular}{llllllllll}
\toprule
 & 3DGS & 2DGS & GShader & 3DGS-DR & ReflGS & EnvGS (net) & EnvGS (opt) & Ours (net) & Ours (opt) \\
\midrule
Shiny Kitchen & 511238 & 287530 & 1161931 & 353471 & \cellcolor{orange!25}198884 & 241368 & 250381 & 305755 & 280804 \\
Shiny Livingroom & 920534 & 1807710 & 1938117 & 442109 & 236010 & 396360 & 404594 & \cellcolor{orange!25}175494 & 267530 \\
Shiny Office & 1115487 & 680562 & 1669378 & 458777 & 332547 & 427645 & 438814 & \cellcolor{orange!25}249058 & 368487 \\
\bottomrule
\end{tabular}

            \label{tab:num_gaussians}
            \end{table*}

            \begin{table*}[!h]
            \centering
            \caption{Frames per second}
            \begin{tabular}{llllllllll}
\toprule
 & 3DGS & 2DGS & GShader & 3DGS-DR & ReflGS & EnvGS (net) & EnvGS (opt) & Ours (net) & Ours (opt) \\
\midrule
Shiny Kitchen & \cellcolor{orange!25}126.38 & 94.29 & 24.42 & 73.86 & 33.45 & 6.88 & 6.07 & 7.08 & 16.45 \\
Shiny Livingroom & \cellcolor{orange!25}110.17 & 54.26 & 21.22 & 81.69 & 48.26 & 11.18 & 9.72 & 18.88 & 18.77 \\
Shiny Office & \cellcolor{orange!25}82.11 & 79.89 & 18.07 & 74.15 & 42.43 & 7.99 & 8.17 & 12.18 & 16.59 \\
\bottomrule
\end{tabular}

            \label{tab:fps}
            \end{table*}

\begin{table*}
\centering
\small
\caption{Training times of our raytracer compared to 3DGRT for 7k iterations on MipNeRF scenes on an A6000 card.}
\begin{tabular}{lccccccccc}
\toprule
\scriptsize Downsize (\(\rightarrow\)) & \multicolumn{3}{c}{\( 2 \times \)} & \multicolumn{3}{c}{\( 4 \times \)} & \multicolumn{3}{c}{\( 8 \times \)} \\
\scriptsize Scene (\(\downarrow\)) & 3DGRT & Ours & Speedup & 3DGRT & Ours & Speedup & 3DGRT & Ours & Speedup \\
\midrule
\textsc{bicycle} & 00:20:53 & \cellcolor{orange!20}00:07:39 & 2.73\texttimes{} & 00:05:27 & \cellcolor{orange!20}00:02:21 & 2.32\texttimes{} & 00:02:13 & \cellcolor{orange!20}00:01:03 & 2.11\texttimes{} \\
\textsc{bonsai} & 00:40:13 & \cellcolor{orange!20}00:05:26 & 7.40\texttimes{} & 00:18:38 & \cellcolor{orange!20}00:02:08 & 8.73\texttimes{} & 00:14:59 & \cellcolor{orange!20}00:01:04 & 14.05\texttimes{} \\
\textsc{counter} & 00:26:29 & \cellcolor{orange!20}00:06:42 & 3.95\texttimes{} & 00:09:27 & \cellcolor{orange!20}00:02:39 & 3.57\texttimes{} & 00:05:08 & \cellcolor{orange!20}00:01:16 & 4.05\texttimes{} \\
\textsc{garden} & 00:30:06 & \cellcolor{orange!20}00:09:22 & 3.21\texttimes{} & 00:08:17 & \cellcolor{orange!20}00:03:22 & 2.46\texttimes{} & 00:02:48 & \cellcolor{orange!20}00:01:28 & 1.91\texttimes{} \\
\textsc{kitchen} & 00:56:57 & \cellcolor{orange!20}00:07:26 & 7.66\texttimes{} & 00:19:18 & \cellcolor{orange!20}00:02:54 & 6.66\texttimes{} & 00:08:57 & \cellcolor{orange!20}00:01:21 & 6.63\texttimes{} \\
\textsc{room} & 00:13:00 & \cellcolor{orange!20}00:04:34 & 2.85\texttimes{} & 00:04:31 & \cellcolor{orange!20}00:01:52 & 2.42\texttimes{} & 00:02:27 & \cellcolor{orange!20}00:00:56 & 2.62\texttimes{} \\
\textsc{stump} & 00:18:15 & \cellcolor{orange!20}00:06:51 & 2.66\texttimes{} & 00:04:48 & \cellcolor{orange!20}00:01:59 & 2.42\texttimes{} & 00:01:44 & \cellcolor{orange!20}00:00:52 & 2.00\texttimes{} \\
\bottomrule
\end{tabular}

\end{table*}

\begin{table*}
\centering
\small
\caption{Rendering framerates (in FPS) of our raytracer compared to 3DGRT for MipNeRF scenes at various downsampling levels.}
\begin{tabular}{lccccccccc}
\toprule
\scriptsize Downsize (\(\rightarrow\)) & \multicolumn{3}{c}{\( 2 \times \)} & \multicolumn{3}{c}{\( 4 \times \)} & \multicolumn{3}{c}{\( 8 \times \)} \\
\scriptsize Scene (\(\downarrow\)) & 3DGRT & Ours & Speedup & 3DGRT & Ours & Speedup & 3DGRT & Ours & Speedup \\
\midrule
\textsc{Bicycle} & 29.18 & \cellcolor{orange!20}33.65 & 1.15\texttimes{} & 89.75 & \cellcolor{orange!20}102.00 & 1.14\texttimes{} & 198.97 & \cellcolor{orange!20}230.67 & 1.16\texttimes{} \\
\textsc{Bonsai} & 27.11 & \cellcolor{orange!20}64.09 & 2.36\texttimes{} & 60.76 & \cellcolor{orange!20}165.77 & 2.73\texttimes{} & 78.76 & \cellcolor{orange!20}259.38 & 3.29\texttimes{} \\
\textsc{Counter} & 23.45 & \cellcolor{orange!20}50.76 & 2.16\texttimes{} & 65.44 & \cellcolor{orange!20}145.72 & 2.23\texttimes{} & 116.86 & \cellcolor{orange!20}231.06 & 1.98\texttimes{} \\
\textsc{Garden} & 19.53 & \cellcolor{orange!20}31.53 & 1.61\texttimes{} & 61.07 & \cellcolor{orange!20}103.75 & 1.70\texttimes{} & 160.51 & \cellcolor{orange!20}214.90 & 1.34\texttimes{} \\
\textsc{Kitchen} & 9.10 & \cellcolor{orange!20}56.13 & 6.17\texttimes{} & 26.81 & \cellcolor{orange!20}164.81 & 6.15\texttimes{} & 53.46 & \cellcolor{orange!20}259.23 & 4.85\texttimes{} \\
\textsc{Room} & 36.20 & \cellcolor{orange!20}72.54 & 2.00\texttimes{} & 98.26 & \cellcolor{orange!20}192.30 & 1.96\texttimes{} & 171.48 & \cellcolor{orange!20}318.21 & 1.86\texttimes{} \\
\textsc{Stump} & \cellcolor{orange!20}42.81 & 40.82 & 0.95\texttimes{} & 126.32 & \cellcolor{orange!20}133.18 & 1.05\texttimes{} & 193.59 & \cellcolor{orange!20}227.75 & 1.18\texttimes{} \\
\bottomrule
\end{tabular}

\end{table*}

\begin{table*}
\centering
\small
\caption{PSNR scores on the MipNeRF scenes at various downsampling levels.}
\begin{tabular}{lcccccc}
\toprule
\scriptsize Downsize (\(\rightarrow\)) & \multicolumn{2}{c}{\( 2 \times \)} & \multicolumn{2}{c}{\( 4 \times \)} & \multicolumn{2}{c}{\( 8 \times \)} \\
\scriptsize Scene (\(\downarrow\)) & 3DGRT & Ours & 3DGRT & Ours & 3DGRT & Ours \\
\midrule
\textsc{Bicycle} & \cellcolor{orange!25}21.65 & 20.87 & \cellcolor{orange!25}22.06 & 21.30 & \cellcolor{orange!25}23.13 & 22.76 \\
\textsc{Bonsai} & \cellcolor{orange!25}28.09 & 27.86 & \cellcolor{orange!25}28.47 & 28.10 & 28.29 & \cellcolor{orange!25}28.55 \\
\textsc{Counter} & \cellcolor{orange!25}26.14 & 25.87 & \cellcolor{orange!25}26.30 & 26.05 & 26.19 & \cellcolor{orange!25}26.59 \\
\textsc{Garden} & \cellcolor{orange!25}23.37 & 22.87 & \cellcolor{orange!25}24.22 & 23.58 & \cellcolor{orange!25}25.74 & 25.40 \\
\textsc{Kitchen} & 26.55 & \cellcolor{orange!25}27.19 & 26.91 & \cellcolor{orange!25}27.34 & 27.25 & \cellcolor{orange!25}28.25 \\
\textsc{Room} & \cellcolor{orange!25}28.39 & 28.07 & \cellcolor{orange!25}28.94 & 28.69 & 29.08 & \cellcolor{orange!25}29.35 \\
\textsc{Stump} & \cellcolor{orange!25}23.91 & 22.36 & \cellcolor{orange!25}23.94 & 22.42 & \cellcolor{orange!25}24.19 & 22.88 \\
\bottomrule
\end{tabular}

\end{table*}

\begin{table*}
\centering
\small
\caption{Final number of gaussians compared to 3DGRT for MipNeRF scenes at various downsampling levels.}
\begin{tabular}{lcccccc}
\toprule
\scriptsize Downsize (\(\rightarrow\)) & \multicolumn{2}{c}{\( 2 \times \)} & \multicolumn{2}{c}{\( 4 \times \)} & \multicolumn{2}{c}{\( 8 \times \)} \\
\scriptsize Scene (\(\downarrow\)) & 3DGRT & Ours & 3DGRT & Ours & 3DGRT & Ours \\
\midrule
\textsc{Bicycle} & \cellcolor{orange!20}373956 & 293347 & \cellcolor{orange!20}381338 & 233460 & \cellcolor{orange!20}440262 & 363677 \\
\textsc{Bonsai} & \cellcolor{orange!20}449732 & 441927 & \cellcolor{orange!20}469323 & 427221 & 485899 & \cellcolor{orange!20}527256 \\
\textsc{Counter} & \cellcolor{orange!20}329784 & 272809 & \cellcolor{orange!20}342612 & 272222 & 350056 & \cellcolor{orange!20}384252 \\
\textsc{Garden} & \cellcolor{orange!20}341017 & 205250 & \cellcolor{orange!20}362348 & 219314 & \cellcolor{orange!20}400272 & 266280 \\
\textsc{Kitchen} & \cellcolor{orange!20}417549 & 373662 & \cellcolor{orange!20}428077 & 427069 & 419974 & \cellcolor{orange!20}434215 \\
\textsc{Room} & \cellcolor{orange!20}286533 & 273520 & \cellcolor{orange!20}304513 & 223550 & \cellcolor{orange!20}321132 & 273650 \\
\textsc{Stump} & \cellcolor{orange!20}323825 & 256919 & \cellcolor{orange!20}343281 & 200351 & \cellcolor{orange!20}435707 & 425580 \\
\bottomrule
\end{tabular}

\label{tab:3DGRT}
\end{table*}

\begin{table*}
\centering
\small
\caption{Training times of our raytracer compared to 3DGRT on the Shiny dataset for 7k iterations on an A6000 card.}
\begin{tabular}{lccc}
\toprule
Scene & 3DGRT & Ours & Speedup \\
\midrule
\textsc{Shiny Kitchen} & 00:13:14 & \cellcolor{orange!20}00:02:40 & 4.96\texttimes{} \\
\textsc{Shiny Livingroom} & 00:12:47 & \cellcolor{orange!20}00:02:30 & 5.11\texttimes{} \\
\textsc{Shiny Office} & 00:17:10 & \cellcolor{orange!20}00:02:56 & 5.85\texttimes{} \\
\bottomrule
\end{tabular}

\end{table*}

\begin{table*}
\centering
\small
\caption{Rendering framerates (in FPS) of our raytracer compared to 3DGRT on the Shiny dataset.}
\begin{tabular}{lccc}
\toprule
Scene & 3DGRT & Ours & Speedup \\
\midrule
\textsc{Shiny Kitchen} & 49.52 & \cellcolor{orange!20}181.50 & 3.67\texttimes{} \\
\textsc{Shiny Livingroom} & 62.23 & \cellcolor{orange!20}174.53 & 2.80\texttimes{} \\
\textsc{Shiny Office} & 30.99 & \cellcolor{orange!20}143.00 & 4.61\texttimes{} \\
\bottomrule
\end{tabular}
\end{table*}

\begin{table*}
\centering
\small
\caption{PSNR scores on the Shiny Scenes dataset.}
\begin{tabular}{lcc}
\toprule
Scene & 3DGRT & Ours \\
\midrule
\textsc{Shiny Kitchen} & \cellcolor{orange!25}32.43 & 32.16 \\
\textsc{Shiny Livingroom} & 29.63 & \cellcolor{orange!25}30.13 \\
\textsc{Shiny Office} & 30.76 & \cellcolor{orange!25}31.11 \\
\bottomrule
\end{tabular}
\end{table*}

\begin{table*}
\centering
\small
\caption{Final number of gaussians compared to 3DGRT on the shiny dataset.}
\begin{tabular}{lcc}
\toprule
Scene & 3GDRT & Ours \\
\midrule
\textsc{Shiny Kitchen} & \cellcolor{orange!20}200192 & 137351 \\
\textsc{Shiny Livingroom} & 157768 & \cellcolor{orange!20}160792 \\
\textsc{Shiny Office} & 271238 & \cellcolor{orange!20}286003 \\
\bottomrule
\end{tabular}
\label{tab:shiny}
\end{table*}

\end{document}